\documentclass[twocolumn]{autart}

\usepackage{amsmath,amssymb,amsfonts} 
\usepackage{mathtools}
\usepackage[colorlinks=true, allcolors=blue]{hyperref}
\usepackage{natbib}
\usepackage{textcomp}
\usepackage{bm}
\usepackage{parskip}
\usepackage{siunitx}
\usepackage{cancel}
\usepackage{algorithm}
\usepackage{algpseudocode}
\usepackage{soul}
\usepackage{multirow}

\usepackage{pifont}

\usepackage{lipsum}

\newcommand*{\genbf}[1]{\ifmmode\mathbf{#1}\else\textbf{#1}\fi}
\newcommand{\norm}[1]{\left\lVert#1\right\rVert}

\newcommand{\Phiy}[2]{\bm{\Phi}^{\mathbf{y}}_{\mathbf{#1},\mathbf{#2}}}
\newcommand{\Phiu}[2]{\bm{\Phi}^{\mathbf{u}}_{\mathbf{#1},\mathbf{#2}}}
\newcommand{\Phiyu}[2]{{\bm{\Phi}}_{\mathbf{#1},\mathbf{#2}}}
\newcommand{\tPhiy}[2]{\bm{\Phi}^{\mathbf{y}^\mathrm{o}}_{\mathbf{#1},\mathbf{#2}}}
\newcommand{\tPhiu}[2]{\bm{\Phi}^{\mathbf{u}^\mathrm{o}}_{\mathbf{#1},\mathbf{#2}}}
\newcommand{\tPhiyu}[2]{\bm{\Phi}^\mathrm{o}_{\mathbf{#1},\mathbf{#2}}}
\newcommand{\Psiy}[0]{\bm{\Psi}^{\mathbf{y}}}
\newcommand{\Psiu}[0]{\bm{\Psi}^{\mathbf{u}}}
\newcommand{\tPsiu}[0]{\bm{\Psi}^{\mathbf{u}^\mathrm{o}}}
\newcommand{\tPsiy}[0]{\bm{\Psi}^{\mathbf{y}^\mathrm{o}}}
\newcommand{\Psiyu}[0]{\bm{\Psi}}
\newcommand{\tPsiyu}[0]{\bm{\Psi}^\mathrm{o}}

\newcommand{\Upsilonb}[0]{\bm{\Upsilon}}
\newcommand{\tUpsilonb}[0]{\bm{\Upsilon}^\mathrm{o}}

\newcommand{\CL}[1]{\mathcal{CL}^{#1}}
\newcommand{\hCL}[1]{\widehat{\mathcal{CL}}^{#1}}
\newcommand{\CLp}[1]{\mathcal{CL}_p^{#1}}
\newcommand{\CLtwo}[1]{\mathcal{CL}_2^{#1}}
\newcommand{\hCLp}[1]{\widehat{\mathcal{CL}}_p^{#1}}
\newcommand{\hCLtwo}[1]{\widehat{\mathcal{CL}}_2^{#1}}
\newcommand{\Lp}[0]{\mathcal{L}_p}
\newcommand{\Ltwo}[0]{\mathcal{L}_2}
\newcommand{\Lexp}[0]{\mathcal{L}_{\text{exp}}}

\newcommand{\Emme}[0]{\boldsymbol{\mathcal{M}}}
\newcommand{\Enne}[0]{\boldsymbol{\mathcal{N}}}
\newcommand{\IO}[0]{\begin{bmatrix}\mathbf{I} & \mathbf{0}\end{bmatrix}}
\newcommand{\InO}[1]{\begin{bmatrix}\mathbf{I}_{#1} & \mathbf{0}\end{bmatrix}}

\newcommand{\OI}[0]{\begin{bmatrix}\mathbf{0} & \mathbf{I}\end{bmatrix}}
\newcommand{\OIn}[1]{\begin{bmatrix}\mathbf{0} & \mathbf{I}_{#1}\end{bmatrix}}
\newcommand{\OnegI}[0]{\begin{bmatrix}\mathbf{0} & -\mathbf{I}\end{bmatrix}}
\newcommand{\OnegIn}[1]{\begin{bmatrix}\mathbf{0} & -\mathbf{I}_{#1}\end{bmatrix}}
\newcommand{\OO}[0]{\begin{bmatrix}\mathbf{0} & \mathbf{0}\end{bmatrix}}

\newcommand{\IOv}[0]{(\mathbf{I};\mathbf{0})}
\newcommand{\InOv}[1]{(\mathbf{I}_{#1};\mathbf{0})}

\newcommand{\IOOO}[0]{(\begin{bmatrix}\mathbf{I} & \mathbf{0} \end{bmatrix}; \begin{bmatrix} \mathbf{0} & \mathbf{0}\end{bmatrix})}

\newcommand{\I}[0]{\mathbf{I}}
\newcommand{\Ob}[0]{\mathbf{0}}
\newcommand{\Css}[0]{\mathcal{C}_{ss}}
\newcommand{\Ccs}[0]{\mathcal{C}_{cs}}

\newcommand{\Ccc}[0]{\mathcal{C}_{cc}}
\newcommand{\Cc}[0]{\mathcal{C}_{c}}
\newcommand{\Cs}[0]{\mathcal{C}_{s}}
\newcommand{\Ccscc}[0]{\mathcal{C}_{(cs;cc)}}
\newcommand{\Csscs}[0]{\mathcal{C}_{(ss;cs)}}

\newcommand{\bvec}[2]{({#1};{#2})}

\newcommand{\xb}[0]{\mathbf{x}}
\newcommand{\yb}[0]{\mathbf{y}}
\newcommand{\ub}[0]{\mathbf{u}}
\newcommand{\Gb}[0]{\mathbf{G}}
\newcommand{\Kb}[0]{\mathbf{K}}
\newcommand{\vb}[0]{\mathbf{v}}
\newcommand{\db}[0]{\mathbf{d}}
\newcommand{\tyb}[0]{\mathbf{y}^\mathrm{o}}
\newcommand{\tub}[0]{\mathbf{u}^\mathrm{o}}

\newcommand{\Qb}[0]{\boldsymbol{\mathcal{Q}}}
\newcommand{\betab}[0]{\boldsymbol{\beta}}
\newcommand{\deltab}[0]{\boldsymbol{\delta}}
\newcommand{\omegab}[0]{\boldsymbol{\omega}}
\newcommand{\tomegab}[0]{\tilde{\boldsymbol{\omega}}}

\newcommand{\Ab}[0]{\mathbf{A}}
\newcommand{\Bb}[0]{\mathbf{B}}
\newcommand{\Cb}[0]{\mathbf{C}}

\newcommand{\ab}[0]{\mathbf{a}}
\newcommand{\bb}[0]{\mathbf{b}}

\newcommand{\cmark}{\ding{51}}%
\newcommand{\xmark}{\ding{55}}%

\newcommand{\lpe}[1]{\ell^{#1}}

\let\emptyset\varnothing

\usepackage{tikz}
\usetikzlibrary{arrows}

\newtheorem{proposition}{Proposition}
\newtheorem{theorem}{Theorem}
\newtheorem{corollary}{Corollary}
\newtheorem{remark}{Remark}
\newtheorem{definition}{Definition}

\newtheorem{example}{Example}

\makeatletter
\def\@opargbegintheorem#1#2#3{\trivlist
   \item[]{\bfseries #1\ #2\ (#3)} \itshape}
\makeatother

\let\cite\citep

\begin{document}
	
\begin{frontmatter}
    \title{Parametrizations of All Stable Closed-loop Responses: From Theory to Neural Network Control Design\thanksref{footnoteinfo}}

    \thanks[footnoteinfo]{This work was supported as a part of NCCR Automation, a National Centre of Competence in Research (grant number 51NF40\textunderscore225155), 
    the NECON project (grant number 200021\textunderscore219431) 
    and the RL4NetC Ambizione project (grant number PZ00P2\textunderscore208951), all funded by the Swiss National Science Foundation.}

    \author[EPFL]{Clara L. Galimberti}\ead{clara.galimberti@epfl.ch},
	\author[EPFL]{Luca Furieri}\ead{luca.furieri@epfl.ch},
	\author[EPFL]{Giancarlo Ferrari-Trecate}\ead{giancarlo.ferraritrecate@epfl.ch}
	
	\address[EPFL]{Institute of Mechanical Engineering, \'Ecole Polytechnique F\'ed\'erale de Lausanne, Switzerland}   
	
	
	\begin{abstract}
        The complexity of modern control systems necessitates architectures that achieve high performance while ensuring robust stability, particularly for nonlinear systems. 
        In this work, we tackle the challenge of designing 
        output-feedback controllers to boost the performance of $\ell_p$-stable discrete-time nonlinear systems while preserving closed-loop stability from external disturbances to input and output channels. 
        Leveraging operator theory and neural network representations, we parametrize the achievable closed-loop maps for a given system and propose novel parametrizations of all $\ell_p$-stabilizing controllers, unifying frameworks such as nonlinear Youla parametrization and internal model control. 
        Contributing to a rapidly growing research line, our approach enables unconstrained optimization exclusively over stabilizing controllers and provides sufficient conditions to ensure robustness against model mismatch. 
        Additionally, our methods reveal that stronger notions of stability can be imposed on the closed-loop maps if disturbance realizations are available after one time step. Last, our approaches are compatible with the design of nonlinear distributed controllers.
        Numerical experiments on cooperative robotics demonstrate the flexibility of the proposed framework, allowing cost functions to be freely designed for achieving complex behaviors while preserving stability.
    \end{abstract}
\end{frontmatter}

\section{Introduction}

The characterization of all controllers capable of stabilizing a given system is a cornerstone problem in control theory. For linear time-invariant (LTI) systems, a complete solution is given by the Youla parametrization, which provides the representation of all internally stabilizing LTI controllers based on a system’s doubly coprime factorization~\cite{Youla1976}.
When using this approach, each controller is associated with a transfer function called the Youla parameter, which can be optimized for computing stabilizing regulators that minimize a performance index~\cite{Boyd_Vandenberghe_2004,zhou1998essentials}.
More recently, alternative methods have been proposed, such as System-Level-Synthesis (SLS) \cite{wang2019sls} and the Input-Output Parametrization (IOP) \cite{furieri2019input}, which are equivalent to the Youla parametrization, but offer different numerical stability properties in different control setups \cite{zheng2022system}.
More importantly, these approaches reveal that all controllers that stabilize an LTI system can be directly expressed and implemented in terms of the \emph{closed-loop maps} capturing the relationship between the disturbances affecting the closed-loop system and the control and output variables.

Given the foundational value of the Youla parametrization in linear systems theory, it is not surprising that many works have focused on extensions to the nonlinear setting, both in continuous and discrete time~\cite{desoer1982, anantharam1984stabilization, paice1994cdc, paice1996tac, imura1997, fujimoto2000, fujimoto1998youla}.
The first result was provided in~\citet{desoer1982, anantharam1984stabilization}, where, with reference to $\ell_p$-stability, the authors consider either smooth, stable systems or systems for which a stabilizing controller is available. Subsequent works have aimed at establishing stable kernel representations --- an extension of left coprime factorizations to the nonlinear setting --- for both systems and controllers, with the goal of characterizing all and only stability-preserving controllers for a broader spectrum of nonlinear systems\footnote{For nonlinear systems that are stable or where a stabilizing controller is known.} \cite{paice1994cdc, paice1996tac, fujimoto1998youla, fujimoto2000}. 

Despite their theoretical relevance, to our knowledge, nonlinear Youla parametrizations and stable kernel representations have never found extensive application in the control of real-world systems. The parametrizations presented in the above works involve an operator form that is not well-suited for the numerical optimization of controllers. 
In particular, the controllers in \cite{desoer1982, anantharam1984stabilization, imura1997} are defined in terms of a stable operator --- the nonlinear counterpart of the Youla parameter --- through a relation that includes an operator inverse, which complicates the optimization. 
Additionally, \citet{paice1994cdc, paice1996tac, fujimoto1998youla, fujimoto2000}, employed stable kernel representations and their (pseudo) inverses for modeling stabilizing controllers. However, deriving stable kernel representations of both the system and controller is often challenging, and computing these (pseudo) inverses can be prohibitive.

To provide controller representations more suitable to the numerical solution of optimal control problems, the authors of~\citet{wang2023linear} consider discrete-time stabilizable and detectable LTI systems and parameterize all nonlinear controllers guaranteeing contractive and Lipschitz closed-loop dynamics. These results have recently been extended to nonlinear systems in \citet{barbara2023learning, Furieri22b, furieri2024performance}.
Specifically, \citet{barbara2023learning} addresses time-invariant systems verifying suitable contractivity and Lipschitz assumptions and parameterizes the set of stabilizing controllers that achieve contracting and Lipschitz closed-loop maps. In our previous works~\cite{Furieri22b, furieri2024performance}, we considered $\ell_p$-stable
time-varying discrete-time nonlinear systems and parameterize all and only the state-feedback controllers preserving closed-loop system stability in the $\ell_p$ sense. Moreover, we demonstrated that all these controllers admit an Internal Model Control (IMC) representation \cite{Garcia_Morari_IMC_1982,Economou_Morari_nlIMC_1986} where the regulator includes a copy of the system dynamics and an $\ell_p$-stable operator that can be freely chosen.

To bridge the gap between theoretical results and computational methods for designing optimal control policies, the approaches in \citet{wang2023linear,barbara2023learning, Furieri22b, furieri2024performance} leverage recent results on the representation of stable operators through nonlinear dynamical systems that (i) are freely parameterized and (ii) can embed deep Neural Networks (NN) in their dynamics. 
Property (ii) enables searching within increasingly broad subsets of $\ell_p$-stable operators as the NN depth increases, which is desirable for optimizing highly nonlinear cost functions~\cite{Furieri22b, furieri2024performance}.
The first requirement, instead, concerns the property of only representing $\ell_p$-stable  operators independently of the values of the system parameters.
Examples of freely parametrized models of $\ell_2$-stable operators include Recurrent Equilibrium Networks (RENs) \cite{revay2021RENs}, certain classes of State-Space Models (SSMs) \cite{forgione2021dynonet,gu2022efficiently}, NN parametrizations of Hamiltonian systems~\cite{zakwan2024neural} and Lipschitz-bounded deep networks~\cite{wangDirectParameterizationLipschitzBounded2023a,pauliLipKernelLipschitzBoundedConvolutional2024}.
The primary advantage of free parametrizations is that they allow casting optimal control design into unconstrained optimization problems, which can be efficiently solved with standard gradient descent or its stochastic variants. This is in contrast to other NN controller design approaches~\cite{bonassi2022, damico2023data, gu2022recurrent, deSouza2023event} 
that ensure closed-loop stability by imposing constraints on the controller parameters, hence necessitating the use of more costly and less scalable projected gradient methods or heuristic approximations.

Optimizing within the set of stabilizing controllers provides two key benefits. First, different from Model Predictive Control (MPC), where the cost function is typically designed to represent itself a Lyapunov function for the closed-loop system, our framework allows the cost function to be freely chosen without affecting closed-loop stability.  We remark that dissipativity and Lyapunov arguments still determine the expressivity of the operators we utilize. For instance, RENs \cite{revay2021RENs} inherently constrain the search space to certain closed-loop behaviors dictated by specific Lyapunov functions and dissipativity properties. Nonetheless, the cost function being optimized remains structurally unrestricted and can be chosen independently of stability considerations.  Second, searching in the space of stabilizing controllers enables fail-safe design, meaning that even if the global optimum is not achieved or the optimization process is halted prematurely, the resulting controller will still guarantee stability. This property is especially important because designing NN controllers typically amounts to minimizing nonlinear costs with a very complex optimization landscape.

While the works of \citet{Furieri22b, wang2023linear, barbara2023learning, furieri2024performance} highlight that nonlinear Youla and IMC parametrizations provide a fertile ground for designing stabilizing NN optimal controllers, significant gaps remain, particularly in the nonlinear output-feedback setting.

Firstly, to the authors’ knowledge, the SLS and IOP parametrizations have not been extended to discrete-time nonlinear systems, with the exceptions of \citet{Lu1995,ho2020system,Furieri22b}. 
However, \citet{Lu1995} is limited to input-affine dynamics, and  \citet{ho2020system} and \citet{Furieri22b} tackle the state-feedback case. 
Note that, \citet{ho2020system} characterizes the set of closed-loop maps achievable by some controllers rather than providing a complete description of all stabilizing policies. 
In this work, instead, we show that directly parametrizing the closed-loop maps, offers the following important advantages: (i) it provides a unified methodology for representing $\ell_p$-stabilizing controllers previously derived under different methods such as nonlinear Youla parametrizations and IMC~\cite{desoer1982,Economou_Morari_nlIMC_1986} and (ii) under specific assumptions, it enables parametrizing all controllers that achieve desired closed-loop properties, in addition to $\ell_p$-stability, such as exponential stability, incremental stability, and disturbance localization in distributed systems. 

Secondly, a comprehensive framework that connects operator-theory-based nonlinear Youla parametrizations with computational techniques for designing optimal controllers is currently lacking.

\subsection{Contributions and outline}
\label{sec:contributions}
In this work, we consider discrete-time, time-varying systems specified through a nonlinear non-Markovian input-output mapping,%
\footnote{An operator is Markovian if it can described using a state vector, and the future evolution of the state depends only on its current value and not on the past ones. For instance, common state-space models define Markovian operators. A non-Markovian operator can instead represent a dynamical system where future states depend on the present and past states, making its evolution history-dependent.}
equipped with output-feedback nonlinear dynamic controllers, with the aim of guaranteeing closed-loop $\ell_p$-stability. We study feedback interconnections accounting for both process and measurement disturbances. 

In Section~\ref{sec:achievable_CLM}, we establish necessary and sufficient conditions for general operators to qualify as closed-loop maps and provide non-Markovian state-space models of the controllers that achieve specified closed-loop maps. The latter result is essential for casting optimal controller design into optimization problems.

In Section~\ref{sec:stable_CLM}, we provide the first main result of the paper. We consider $\ell_p$-stable systems and parametrize all and only the achievable stable closed-loop maps, along with their corresponding stability-preserving controllers. This representation, which relies on a unique free operator in $\mathcal{L}_p$, shows that any stabilizing controller inherently possesses an IMC structure. Therefore, IMC architectures are not merely sufficient for stability (as shown in \citet{Economou_Morari_nlIMC_1986} for continuous-time systems) but also necessary. Furthermore, our framework allows recovering the nonlinear Youla parametrization presented in \citet{desoer1982}. 
We also demonstrate that, in the case of interconnected systems, the proposed parametrizations are naturally suited for designing distributed controllers and provide insights into the required relationship between the physical coupling graph of the system and the communication network topology.

In Section~\ref{sec:robust_analysis}, we analyze the robustness of the proposed controllers when there is a mismatch between the actual system and the model used within the IMC controller. This enables the application of our parametrization in scenarios where only an approximate system description is available, such as models derived from simplified physical principles or data-driven approaches.

When controlling real-world systems, achieving closed-loop $\ell_p$-stability is often insufficient. Typically, there is no direct relationship, beyond stability, between the characteristics of the nonlinear Youla parameter and those inherited by the closed-loop maps. 
We however show that various desirable properties, such as exponential and incremental stability, can be attained by design when input disturbances are measurable or can be reconstructed with a one-step delay. 
Moreover, in a distributed setting, we show how to localize the effect of disturbances  --- i.e., prevent disturbances in one local system from influencing distant subsystems --- which is one of the key features enabled by SLS \cite{anderson2019system}. 

The last key contribution of the paper is provided in Section~\ref{sec:NOC_RENs}, where we illustrate how to bridge the gap between the proposed parametrizations and the actual computation of controllers for boosting the performance of stable nonlinear systems. Specifically, we address Nonlinear Optimal Control (NOC) problems, showing how to design controllers by searching within the set of stability-preserving control policies.
To this purpose, we leverage the available methods for representing $\mathcal{L}_2$ operators through freely parametrized nonlinear systems embedding NNs. Numerical experiments demonstrating the effectiveness of the proposed approaches are presented in Section~\ref{sec:numerical}, using examples from cooperative mobile robotics. Specifically, we show that robots pre-stabilized around a target point can be equipped with performance-boosting controllers to enable collision avoidance with obstacles and other robots while maintaining a stable target-reaching behavior.

The appendices collect the proofs of the theorems and the propositions of this work, as well as implementation details of the simulations.

\subsection{Notation}
\label{sec:notation}

The set of all sequences $\mathbf{x} = (x_0,x_1,x_2,\ldots)$, where $x_t \in \mathbb{R}^n$ for all $t = 0,1,\dots$, is denoted as $\ell^n$. 
Moreover,  $\mathbf{x}$ belongs to $\ell_p^n \subset \ell^n$ with $p \in \mathbb{N}$ if $\norm{\mathbf{x}}_p = \left(\sum_{t=0}^\infty |x_t|^p\right)^{\frac{1}{p}} < \infty$, where $|\cdot|$ denotes
any norm of interest. 
We say that $\xb \in \ell^n_\infty$ if $\operatorname{sup}_{t}|x_t|< \infty$. 
We highlight that in the literature, the set $\ell^n$ is sometimes referred to as the \emph{extended} $\ell^n_p$ space, for all $p\in\mathbb{N}\cup \{+\infty\}$, and it is denoted with $\ell^n_e$.
We use the notation $x_{j:i}$ for the truncation $(x_i,x_{i+1},\ldots,x_{j})$ of $\mathbf{x}$, when $j\geq i$. 
If $j < i$, $x_{j:i}$ represents the empty set ($\emptyset$).
We define $\Ob = (0,0,\dots)$.
For two vectors $x_0\in\mathbb{R}^n$ and $y_0\in\mathbb{R}^m$, we denote $(x_0;y_0) = [x_0^\top y_0^\top]^\top\in \mathbb{R}^{n+m}$.
The element-wise concatenation of two sequences $\xb \in \lpe{m}$ and $\yb\in\lpe{r}$ is defined as  $(\xb ; \yb)  =  ((x_0;y_0),(x_1;y_1),\dots) \in \lpe{m+r}$.
Truncated element-wise concatenated sequences are denoted as $(x ; y)_{j:i} = (x_{j:i} ; y_{j:i}) = ((x_i;y_i), (x_{i+1};y_{i+1}), \dots, (x_j;y_j))$. 
By convention, when the truncated sequence lengths do not match, we pad with zeros, e.g., 
\begin{equation*}
    (x_{j-1:i} ; y_{j:i}) = ((x_{j-1:i},0) ; y_{j:i}) = (x_{j:i} ; y_{j:i})\,,
\end{equation*}
where we have defined $x_j=0$.
The sum of truncated sequences of equal dimension and length is defined element-wise.
Next, we introduce relevant properties of operators $\mathbf{A}:\lpe{n} \rightarrow \lpe{m}$ over sequences.
The identity operator from $\lpe{n}$ to $\lpe{n}$ is $\I_n$ and we simply write $\I$ when the dimension $n$ is clear from the context.
We also use the simplified notation $\Ab\ub$ for referring to $\Ab(\ub)$.

\paragraph*{Causality}
An operator $\mathbf{A}:\lpe{n} \rightarrow \lpe{m}$ is said to be \emph{causal}
if 
\begin{equation*}
    \mathbf{A}(\mathbf{u}) = (A_0(u_0),A_1(u_{1:0}),\ldots,A_t(u_{t:0}),\ldots) \,.
\end{equation*}
If in addition $A_t$  is constant with respect to $u_t$ for any $t$, i.e., $A_t(u_{t:0}) = A_t(u_{t-1:0}, \bar{u}_t)$, $\forall \bar{u}_t$, then $\mathbf{A}$ is said to be \emph{strictly-causal}. This means that the output of 
$A_t$ depends only on the past values of the input up to time $t-1$.
The set of all causal (strictly-causal) operators from $\lpe{n}$ to $\lpe{m}$ is $\Cc(\lpe{n},\lpe{m})$ ($\Cs(\lpe{n},\lpe{m})$).
Note that, if $\Ab\in\Cs$, then, by convention, $A_0(u_{-1:0}) = A_0(\emptyset)$ and we assume that $ A_0(\emptyset)$ is a vector in $\mathbb{R}^m$.
An operator $\mathbf{A}:\lpe{n}\times\lpe{r} \rightarrow \lpe{m}$ is \textit{causal---strictly-causal} (also denoted as $\mathbf{A} \in \Ccs(\lpe{n}\times\lpe{r},\lpe{m})$) if it is causal on its first argument and strictly-causal on its second argument, i.e., if for any $t$, it holds $A_t(u_{t:0}; z_{t:0}) = A_t(u_{t:0}; z_{t-1:0}) = A_t((u_{t:0});(z_{t-1:0}, 0))$.
Similarly, the sets $\Ccc(\lpe{n}\times\lpe{r},\lpe{m})$ and $\Css(\lpe{n}\times\lpe{r},\lpe{m})$ comprise operators that are causal or strictly-causal, respectively, in both inputs.
Note that, by convention, if $\Ab\in\Ccs(\lpe{n_1}\times\lpe{n_2}, \lpe{m})$ and $\Bb\in\Css(\lpe{n_1}\times\lpe{n_2}, \lpe{m})$, then, for any $u_0\in\mathbb{R}^{n_1}$, $A_0(u_0;z_{-1:0}) = A_0(u_0;0)$ and $B_0(u_{-1:0};z_{-1:0}) = B_0(\emptyset)$ and we assume that $B_0(\emptyset)$ is a vector in  $\mathbb{R}^{m}$.
Given operators $\mathbf{A}\in \Cc(\lpe{m}, \lpe{s})$, $\mathbf{B}\in \Ccs(\lpe{n}\times\lpe{r}, \lpe{m})$ and $\mathbf{C}\in \Cc(\lpe{n}, \lpe{m})$, 
it is easy to see that the composed operators $\mathbf{A}\mathbf{B}$ and $\mathbf{A}\mathbf{C}$ verify  $\mathbf{A}\mathbf{B} \in \Ccs(\lpe{n}\times\lpe{r}, \lpe{s})$ and $\mathbf{A}\mathbf{C} \in \Cc(\lpe{n}, \lpe{s})$.

\paragraph*{Truncation}
We use the notation $A_{j:i}(u_{j:0})$ to refer to 
\begin{equation*}
    (A_i(u_{i:0}),A_{i+1}(u_{i+1:0}),\ldots,A_j(u_{j:0})) \,.
\end{equation*}

\paragraph*{Sums and products}
The sums and products of operators are defined as 
$(\Ab + \Bb)\ub = \Ab\ub+\Bb\ub$ and $(\Ab\Bb)\ub = \Ab(\Bb\ub)$.
Note that, in general, for operators,  multiplication is not commutative and is only left-distributed over the sum (but not right-distributed), i.e., $(\Ab + \Bb)\Cb = \Ab\Cb + \Bb\Cb$ but $\Cb(\Ab + \Bb) \neq \Cb\Ab + \Cb\Bb$. The latter formula follows from the properties of composition of functions. Indeed, for $a(\cdot)$, $b(\cdot):\mathbb{R}^n\mapsto \mathbb{R}^m $, and $c(\cdot):\mathbb{R}^m\mapsto \mathbb{R}^p $, one has $ c(a(x)+b(x))\neq c(a(x))+c(b(x))$.

\paragraph*{Combined operators}
For two operators 
$\mathbf{A} \in \Cc (\lpe{n}, \lpe{r})$ and $\mathbf{B} \in \Cc(\lpe{n}, \lpe{m})$, 
we define 
$(\mathbf{A} ; \mathbf{B})$ as follows: for any $\ub\in\lpe{n}$, $(\mathbf{A} ; \mathbf{B}) \ub = \left(\mathbf{A} (\ub) ; \mathbf{B} (\ub)\right)$. 
Then, it follows that $(\mathbf{A} ; \mathbf{B}) \ub\in \lpe{r+m}$ and
\begin{equation*}
    (\mathbf{A} ; \mathbf{B}) \ub \hspace{-1pt}=\hspace{-1pt}  ((A_0(u_0); B_0(u_0)), (A_1(u_{1:0}); B_1(u_{1:0})), \dots)  \,.
\end{equation*}
We say that $(\mathbf{A} ; \mathbf{B}) \in \Ccscc(\lpe{n}\times\lpe{r},\lpe{m}\times\lpe{s})$ if $\mathbf{A}\in \Ccs(\lpe{n}\times\lpe{r}, \lpe{m})$ and $\mathbf{B}\in \Ccc(\lpe{n}\times\lpe{r}, \lpe{s})$. 
Similarly, $(\mathbf{A} ; \mathbf{B}) \in \Csscs(\lpe{n}\times\lpe{r},\lpe{m}\times\lpe{s})$ means that $\mathbf{A}\in \Css(\lpe{n}\times\lpe{r}, \lpe{m})$ and $\mathbf{B}\in \Ccs(\lpe{n}\times\lpe{r}, \lpe{s})$. 
Given operators 
$\mathbf{A}_1:\lpe{n_1} \rightarrow \lpe{m}$,
$\mathbf{A}_2:\lpe{n_2} \rightarrow \lpe{m}$ and
$\mathbf{A}_3:\lpe{n_3} \rightarrow \lpe{m}$,
and sequences $\ub_1\in\lpe{n_1}$, $\ub_2\in\lpe{n_2}$ and $\ub_3\in\lpe{n_3}$, 
we define the matrix-reminiscent notation
$\begin{bmatrix} \Ab_1 & \Ab_2 \end{bmatrix} (\ub_1;\ub_2) = (\Ab_1\ub_1 + \Ab_2\ub_2) \in\lpe{m}$.
Moreover, 
\begin{equation*}
    \begin{bmatrix} \Ab_1 & \Ab_2 & \Ab_3 \end{bmatrix} \hspace{-3pt} (\ub_1;\ub_2;\ub_3) \hspace{-1pt}=\hspace{-1pt} \begin{bmatrix} \begin{bmatrix}\Ab_1 & \Ab_2 \end{bmatrix} & \hspace{-3pt}\Ab_3 \end{bmatrix} \hspace{-3pt}((\ub_1;\ub_2);\ub_3) .
\end{equation*}

\section{Operator representation of dynamical systems and achievable closed-loop maps}
\label{sec:achievable_CLM}

This section formally introduces the system models we consider and the closed-loop map operators.
With these definitions, we derive necessary and sufficient conditions for operators to be closed-loop maps of a given system.
Finally, we provide an explicit formulation for implementing the controller that achieves those maps.

\subsection{System model}\label{sec:sys_model}

We consider discrete-time time-varying nonlinear closed-loop systems described by
\begin{subequations}
\label{eq:system_control_dyn}
	\begin{align}
		\label{eq:system_dyn}
		y_{t} &= G_t(u_{t-1:0})+v_t\,, \\
		\label{eq:control_dyn}
		u_{t} &= K_t(y_{t:0}) + d_t\,,
	\end{align}
\end{subequations}
where $t= 0,1,2,\ldots$ is the time index, $y_t \in \mathbb{R}^r$ and $u_t \in \mathbb{R}^m$  denote the system output and the control input, and $v_t \in \mathbb{R}^r$ and $d_t \in \mathbb{R}^m$ are the measurement and process noises. 
We consider disturbances with support $\mathbb{V}_t\subseteq\mathbb{R}^r$ and $\mathbb{D}_t\subseteq\mathbb{R}^m$ following potentially unknown distributions $\mathcal{D}^v_t$ and $\mathcal{D}^d_t$, that is, $v_t \in\mathbb{V}_t$, $d_t \in\mathbb{D}_t$ and $v_t \sim \mathcal{D}^v_t$, $d_t \sim \mathcal{D}^d_t$ for every $t=0,1,\dots$.
The functions $G_t : \mathbb{R}^{m\times t} \rightarrow \mathbb{R}^{r}$, with $G_0(u_{-1:0}) = G_0(\emptyset) \in\mathbb{R}^r$, characterize the open-loop input-output system behavior.
Similarly, functions $K_t : \mathbb{R}^{r\times (t+1)} \rightarrow \mathbb{R}^{m}$ represent a causal dynamical output feedback controller applied to the system.

In \emph{operator form}, system \eqref{eq:system_control_dyn} is represented by
\begin{subequations}
    \label{eq:system_control}
	\begin{align}
		\label{eq:system}
		\yb &= \Gb \ub + \vb\,, \\
		\label{eq:control}
		\ub &= \Kb \yb + \db\,,
	\end{align}
\end{subequations}
where $\Gb : \lpe{m} \rightarrow \lpe{r}$ is a strictly-causal operator and 
$\Kb:\lpe{r} \rightarrow \lpe{m}$ is a causal operator, 
i.e.  $\Gb\in\Cs$ and $\Kb \in \Cc$.%
\footnote{
The strict-causality of the operator $\Gb$ is a standard assumption in control systems since the control input $u_t$ can be computed only after receiving the measurements $(y_0,y_1,\dots,y_t)$, and only affects future measurements $(y_{t+1},y_{t+2},\dots)$.}
Equivalently, 
\begin{subequations}
	\begin{align*}
		\Gb\ub &= (G_0(\emptyset),G_1(u_0),\ldots,G_t(u_{t-1:0}),\ldots)\,, \\
		\Kb\yb &= (K_0(y_0), K_1(y_{1:0}) \dots, K_t(y_{t:0}), \dots)\,.
	\end{align*}
\end{subequations}
For later use, we also define the noiseless version of the input and output signals as $\tyb = \Gb \ub$ and $\tub = \Kb \yb$ which verify $\tyb = \yb-\vb$ and $\tub = \ub-\db$.
Moreover, for the free evolution of the system, we will use the notation 
\begin{equation}
    \label{eq:y_free}
    \yb^{\text{free}} = \Gb(\Ob)\,.
\end{equation}

\begin{figure}
	\centering
	\includegraphics{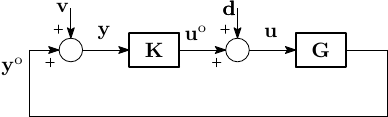}
	\caption{Block diagram of the closed-loop system~\eqref{eq:system_control}.}
	\label{fig:blockdiagram_o}
\end{figure}

One of the key advantages of operator theory \cite{zames1966input} is its ability to model dynamical systems using algebraic relations, similar to transfer functions for LTI models. This, in turn, enables the representation of system interconnections through block diagrams, as illustrated in Figure~\ref{fig:blockdiagram_o} for the closed-loop system \eqref{eq:system_control}.

\begin{example}[From state-space to operator models]
\label{ex:nonlin}
Consider the strictly causal time-invariant dynamical system
\begin{subequations}
 \label{eq:NLsys}
 \begin{align}
        x_{t+1} &= f(x_t, u_t) \,, \quad x_t\in\mathbb{R}^n \,,\label{eq:NLsys_state}\\
        y_t &= h(x_t) \label{eq:NLsys_out}\,,\\
        x_0 &= \bar{x}\,, \label{eq:NLsys_init}
    \end{align}
    \end{subequations}
    along with the causal controller
\begin{subequations}
 \label{eq:NLcontr}
    \begin{align}
        \xi_{t+1} &= f_K(\xi_t, y_t)\,, \quad \xi_t\in\mathbb{R}^{n_K} \,, \label{eq:NLcontr_state}\\
        u_t &= h_K(\xi_t, y_t)\,, \label{eq:NLcontr_out}\\
        \xi_0 &= \bar{\xi} \,. \label{eq:NLcontr_init}
    \end{align}
\end{subequations}
By recursive substitution, one obtains that the functions $G_t$ in \eqref{eq:system_dyn} are given by 
   \begin{align*}
        G_0(\emptyset) &= h(\bar x) \,, \\
        G_1(u_0) &= h(f(\bar x, u_0)) \,, \\
        G_2(u_{1:0}) &= h(f(f(\bar x, u_0), u_1)) \,,\\
        \vdots
    \end{align*}
Similarly, $K_t$ in  \eqref{eq:control_dyn} are given by 
   \begin{align*}
        K_0(y_0) &= h_K(\bar \xi, y_0) \,, \\
        K_1(y_{1:0}) &= h_K(f_K(\bar \xi, y_0), y_1) \,, \\
        K_2(y_{2:0}) &= h_K(f_K(f_K(\bar \xi, y_0), y_1), y_2) \,, \\
        \vdots
    \end{align*}
These formulae reveal that $\Gb$ (respectively, $\Kb$) depends on the specific initial state $\bar x\in\mathbb{R}^n$ (respectively, $\bar \xi\in\mathbb{R}^{n_K}$).\footnote{This is coherent with the definition of input-output operators in the literature, see, e.g., \citet{zames1966input,paice1996tac,fujimoto1998youla}.}
Moreover, $G_0(\emptyset)$ only depends on the initial state of the system. The nonstrict causality of \eqref{eq:NLcontr} allows one also to consider static controllers solely defined by the output equation $ u_t = h_K(y_t)$. 

The systems \eqref{eq:NLsys} and \eqref{eq:NLcontr} are Markovian. However, models \eqref{eq:system_dyn} and \eqref{eq:control_dyn} can also capture systems and controllers that do not enjoy this property.
\end{example}

\begin{example}[LTI systems and relations with transfer functions]
    \label{ex:LTI_1}
    Assume \eqref{eq:NLsys} is an LTI system, i.e.,
    \begin{align*}
        x_{t+1} &= A x_t + B u_t \,,\\
        y_t &= C x_t + v_t\,,\\
        x_0& =\bar{x}\,.
    \end{align*}
    The function $G_t$ in \eqref{eq:system_dyn} is given by
    \begin{align*}
        G_0(\emptyset) &= C \bar{x} \,, \\
        G_t(u_{t-1:0}) &= CA^t\bar{x} + \sum_{\tau=0}^{t-1}CA^{t-\tau-1}B u_\tau\,, \quad t=1,\dots \,,
    \end{align*}
    and model \eqref{eq:system} becomes
    \begin{equation}
        \label{eq:system_dyn_LTI}
        \begin{bmatrix}
            y_0 \\
            y_1 \\
            y_2 \\
            \vdots
        \end{bmatrix}
        =
        \begin{bmatrix}
            C \\
            CA \\
            CA^2 \\
            \vdots
        \end{bmatrix}
        \bar{x}
        +
        \begin{bmatrix}
            0 & 0 & 0 & \cdots \\
            CB & 0 & 0 & \cdots \\
            CAB & CB & 0 & \cdots \\
            \vdots & \vdots & \vdots & \ddots 
        \end{bmatrix}
        \begin{bmatrix}
            u_0 \\
            u_1 \\
            u_2 \\
            \vdots
        \end{bmatrix} 
        +
        \begin{bmatrix}
            v_0 \\
            v_1 \\
            v_2 \\
            \vdots
        \end{bmatrix}
        \,,
    \end{equation}
    showing that $\Gb$ is an affine operator and $G_0 = (C ; CA ; CA^2; \dots)\bar{x}$.
    If $\bar{x}=0$, $\Gb$ is linear and can be equivalently represented by the transfer matrix in the $z$-domain
    \begin{equation}
        \label{eq:G_linear_TF}
        \Gb(z) = C (zI-A)^{-1} B + D \,,
    \end{equation}
    i.e., \eqref{eq:system} is the same as 
    \begin{equation*}
        \mathcal{Z}(\yb) = \Gb(z) \mathcal{Z}(\ub) + \mathcal{Z}(\vb) \,,
    \end{equation*}
    where $\mathcal{Z}(\cdot)$ is the Z-transform operator.
\end{example}

\subsection{Closed-loop maps}
\label{sec:CL_maps}
Assume we are given a system $\Gb$ and a controller $\Kb$.
Due to the strict causality of $\Gb$, 
for any pair of disturbances $\vb$ and $\db$, the closed-loop \eqref{eq:system_control} produces unique sequences $\yb$ and $\ub$ (see also \eqref{eq:system_control_dyn}).
Moreover, due to causality of $\Kb$, we can define {unique causal maps} from disturbances $(\vb;\db)$ to $(\yb;\ub)$.

\begin{definition}[Closed-loop maps]\label{def:closed_loop_maps}
	The closed-loop map $\Phiyu{G}{K} \in \Cc(\lpe{r} \times \lpe{m}, \lpe{r} \times \lpe{m})$ 
	is the unique operator that satisfies 
	$(\yb; \ub) = \Phiyu{G}{K} (\vb; \db)$ for all the sequences $\vb, \db, \yb, \ub$ obtained from the interconnection of the system $\Gb$ with a controller $\Kb$ as per \eqref{eq:system_control}.
	Moreover, the partial maps $(\vb; \db) \mapsto \yb$  and $(\vb; \db) \mapsto \ub$ will be denoted with $\Phiy{G}{K}$ and $\Phiu{G}{K}$, respectively.
\end{definition}

We also define the closed-loop operator $\tPhiyu{G}{K} := \Phiyu{G}{K} - \I$. This operator maps $(\vb;\db) \mapsto (\tyb;\tub)$, since
\begin{align*}
    \tPhiyu{G}{K} (\vb;\db) 
    &= \left(\Phiyu{G}{K} - \I \right) (\vb;\db) \,, \\
    &= \Phiyu{G}{K} (\vb;\db) - (\vb;\db) \,, \\
    &= (\yb; \ub) - (\vb;\db) \,, \\
    &= (\tyb;\tub)\,.
\end{align*}
Note that this relationship is also captured in the block diagram of Figure~\ref{fig:blockdiagram_o}.
Analogously, we define 
\begin{equation*}
	\tPhiy{G}{K}:= \Phiy{G}{K} - \InO{r} \text{ and } \tPhiu{G}{K}:= \Phiu{G}{K} - \OIn{m} \,.
\end{equation*}
Since there is a one-to-one relationship between the operators ($\Phiyu{G}{K}$, $\Phiy{G}{K}$, $\Phiu{G}{K}$) and ($\tPhiyu{G}{K}$,$\tPhiy{G}{K}$, $\tPhiu{G}{K}$),
any formula involving the ``$^{\mathrm{o}}$'' operators can be written in terms of the others. 
We will, however, use both sets of operators to have a more compact notation and improve readability.

Next, we introduce two key properties of the closed-loop map $\Phiyu{\Gb}{\Kb}$.

\begin{proposition}\label{prop:phi_invertible}
	If $\Gb\in\Cs$ and $\Kb\in\Cc$, the closed-loop operator $\Phiyu{\Gb}{\Kb}$ is invertible and satisfies $\Phiyu{\Gb}{\Kb} ^{-1} \in \Ccscc$.\footnote{Note that the terms \emph{invertible} and \emph{bijective} are equivalent in the context of the map $\Phiyu{\Gb}{\Kb}$. Since $\Phiyu{\Gb}{\Kb}$ is invertible, it establishes a one-to-one correspondence --- i.e., a bijective relationship --- between $(\vb;\db)$ and $(\yb;\ub)$.}
\end{proposition}
\begin{pf}
	From \eqref{eq:system_control}, one has
	$\vb = \yb - \Gb\ub$ and 
	$\db = \ub - \Kb\yb$
	which is the mapping  $(\yb ; \ub) \mapsto (\vb ; \db)$. Moreover, due to $\Gb\in\Cs$ and $\Kb\in\Cc$, we have that $\left((\yb ; \ub) \mapsto \vb \right) \in \Ccs$ and $\left((\yb ; \ub) \mapsto \db \right) \in \Cc$. Thus, $\Phiyu{\Gb}{\Kb} ^{-1} \in \Ccscc$.
\end{pf}

\begin{proposition}\label{prop:phi_achievable}
	Given $\Gb\in\Cs$ and $\Kb\in\Cc$, the closed-loop operator $\Phiyu{\Gb}{\Kb} = (\Phiy{\Gb}{\Kb}; \Phiu{\Gb}{\Kb}) = (\tPhiy{\Gb}{\Kb}; \tPhiu{\Gb}{\Kb}) + \I$ satisfies
	\begin{align}
		&\qquad~ \tPhiu{\Gb}{\Kb}  \in \Ccs\,, \label{eq:SLSo1_phi}\\
		&\qquad~ \tPhiy{\Gb}{\Kb} = \Gb \Phiu{\Gb}{\Kb} \,, \label{eq:SLSo2_phi} \\
		&\qquad~ \tPhiu{\Gb}{\Kb} = \tPhiu{\Gb}{\Kb} \Phiyu{\Gb}{\Kb}^{-1} \InOv{r} \Phiy{\Gb}{\Kb} \label{eq:SLSo3_phi}
		\,.
	\end{align}
	Moreover, the controller $\Kb$ verifies
	\begin{equation}
		\label{eq:SLSoK_phi}
		\Kb = \tPhiu{\Gb}{\Kb} \Phiyu{\Gb}{\Kb}^{-1} \InOv{r} \,.
	\end{equation}
\end{proposition}
The proof can be found in Appendix~\ref{app:proof_phi_achievable}. 

Proposition~\ref{prop:phi_achievable} gives necessary conditions that any closed-loop operator must satisfy. 
In particular, \eqref{eq:SLSo1_phi} is a causality condition 
on the map $(\vb;\db)\mapsto\tub$,
\eqref{eq:SLSo2_phi} follows from the system dynamics \eqref{eq:system}, and the right-hand side of \eqref{eq:SLSo3_phi} guarantee consistency with the relation between $(\vb,\db)$ and $\tub$ provided by the closed-loop diagram in Figure~\ref{fig:blockdiagram_o}. 
Moreover, \eqref{eq:SLSo3_phi} and \eqref{eq:SLSoK_phi} give the relation between closed-loop operators and the controller dynamics \eqref{eq:control}.

\begin{figure}
	\centering
	\includegraphics{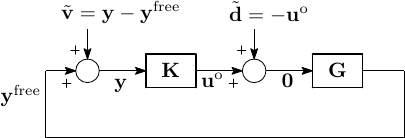}
	\caption{Block diagram of the closed-loop system~\eqref{eq:system_control} for the input sequence $(\tilde{\vb};\tilde{\db})$ as defined in Remark~\ref{rmk:BD_beta_delta}.}
	\label{fig:BD_beta_delta}
\end{figure}

\begin{remark} \label{rmk:BD_beta_delta}
    Equation~\eqref{eq:SLSo3_phi} reveals that the output of the map $\tPhiu{\Gb}{\Kb}$ is the same when the input is $(\vb;\db)$ or $\Phiyu{\Gb}{\Kb}^{-1} \InOv{r} \Phiy{\Gb}{\Kb}(\vb;\db) $. 
    We show that 
    $\Phiyu{\Gb}{\Kb}^{-1} \InOv{r} \Phiy{\Gb}{\Kb}(\vb;\db)$ 
    is the signal 
    $(\tilde{\vb}; \tilde{\db}) = (\yb-\yb^{\text{free}} ; -\tub)$.
    Indeed, one has
    $\Phiyu{\Gb}{\Kb}^{-1} \InOv{r} \Phiy{\Gb}{\Kb}(\vb;\db)
    =
    \Phiyu{\Gb}{\Kb}^{-1} \InOv{r} \yb
    =
    \Phiyu{\Gb}{\Kb}^{-1} (\yb;\Ob)
    $.
    The last expression gives the disturbances $(\tilde{\vb};\tilde{\db})$ in Figure~\ref{fig:BD_beta_delta} such that $\yb$ is still the input to $\Kb$ and $\ub=\Ob$ is the input to $\Gb$.
    From the latter condition, we have $\tilde{\db} = -\tub$. 
    This implies $\tyb = \Gb(\Ob)=\yb^{\text{free}}$ and, therefore, $\tilde{\vb} = \yb - \yb^{\text{free}}$, where $\yb^{\text{free}}$ is defined in \eqref{eq:y_free}.
    Notice, however, that $\tPhiy{\Gb}{\Kb}(\vb;\db) \neq \tPhiy{\Gb}{\Kb}(\tilde{\vb}; \tilde{\db})$.
\end{remark}

\begin{example}[Continuation from Example~\ref{ex:LTI_1}]
\label{ex:closed_loop_map_LTI}
    In the same setting of Example~\ref{ex:LTI_1}, assume that $u_t = K y_t + d_t$. Therefore, the map $\Phiu{\Gb}{\Kb}$ is given by
    \begin{multline}
        \label{eq:LTI_2}
        \begin{bmatrix}
            u_0 \\
            u_1 \\
            \vdots
        \end{bmatrix}
        =
        \begin{bmatrix}
            KC \\
            KAC + KCBKC \\
            \vdots
        \end{bmatrix}
        \bar{x} \\
        +
        \begin{bmatrix}
            K & \mathcal{P} & 0 & 0 & \cdots \\
            KCBK & KCB & K & \mathcal{P} & \cdots \\
            \vdots & \vdots & \vdots & \vdots & \ddots
        \end{bmatrix}
        \begin{bmatrix}
            v_0 \\
            d_0 \\
            v_1 \\
            d_1 \\
            \vdots
        \end{bmatrix},
    \end{multline}
    for $\mathcal{P}=I_m$, highlighting its nonlinear dependence of $\Phiu{\Gb}{\Kb}$ on the matrix gain $K$.
    Similarly, since $\begin{bmatrix} \Ob & \I_m \end{bmatrix} (\vb;\db) = \db$, one has that $\tPhiu{\Gb}{\Kb} = \Phiu{\Gb}{\Kb} - \begin{bmatrix} \Ob & \I_m \end{bmatrix}$ is given by \eqref{eq:LTI_2} for $\mathcal{P}=0$.
\end{example}

\subsection{Parametrization of all achievable closed-loop maps}
We consider the case where the controller $\Kb$ is not specified a priori. 
We characterize the set of all possible closed-loop maps for a given system $\Gb$ that are achieved by some causal controller.
Similar to~\cite{ho2020system}, which focuses on state-feedback controllers,
we say that $\Psiyu = (\Psiy;\Psiu) : \lpe{r+m} \rightarrow \lpe{r+m}$ is a closed-loop map of $\Gb$ if there exists a controller $\Kb'\in\Cc$ such that $\Psiyu = \Phiyu{\Gb}{\Kb'}$.
For later use, we also define
\begin{equation*}
	\tPsiy:= \Psiy - \InO{r} \text{ and } \tPsiu:= \Psiu - \OIn{m} \,.
\end{equation*}

\begin{definition}[Achievable closed-loop maps]\label{def:achievable_CLM}
	\label{def:achievable_CL_maps}
 For a system $\Gb$,
	the set of all closed-loop maps achievable by \eqref{eq:system_control} is 
	\begin{align}
		\label{eq:Phi_ach}
		\CL{\Gb} 
		&= \{\Phiyu{G}{K}|~ \Kb \in \Cc \}\,.
	\end{align}
\end{definition}

In view of Proposition~\ref{prop:phi_achievable}, the following conditions are necessary to have $\Psiyu \in \CL{\Gb}$:
\begin{subequations}
\label{eq:SLSo}
\begin{align}
	\tPsiu  &\in \Ccs\,, \label{eq:SLSo1}\\
	\tPsiy &= \Gb \Psiu \,,\label{eq:SLSo2} \\
	\tPsiu &= \tPsiu \Psiyu^{-1} \InOv{r} \Psiy \,. \label{eq:SLSo3} 
\end{align}
\end{subequations}

Our first main result is to show that \eqref{eq:SLSo} are also sufficient conditions, hence providing an equivalent characterization of $\CL{\Gb}$.
In the spirit of IOP and SLS \cite{furieri2019input, wang2019sls, ho2020system}, the key idea is to remove from \eqref{eq:Phi_ach}
the explicit dependency on the controller operator $\Kb$.

A crucial step for proving the sufficiency is to show that $\Psiyu^{-1}$ always exists.
This fact follows from the next result, which is an extension of Proposition III.1 of \citet{ho2020system}
and whose proof can be found in Appendix~\ref{app:proof_invertibility}.
\begin{proposition}\label{prop:invertibility}
	Consider an operator $\Psiyu = (\Psiy;\Psiu) = (\tPsiy;\tPsiu) + \I$. 
	If 
    $\,\tPsiy \in \Css$ 
    and 
    $\,\tPsiu \in \Ccs$, 
    then 
    $\Psiyu^{-1} \in \Ccscc$ exists and 
    $(\vb;\db) = \Psiyu^{-1} (\yb;\ub)$ satisfies, for $t=0, 1, \dots$, the recursive formulae
    \begin{subequations}
    \label{eq:prop_inv_eq}
	\begin{align}
		v_t &= y_t - \Psi^{y^\mathrm{o}}_{t}(v_{t-1:0},d_{t-1:0})\,, \label{eq:prop_inv_eq1}\\
		d_t &= u_t - \Psi^{u^\mathrm{o}}_{t}(v_{t:0},d_{t-1:0})\,. \label{eq:prop_inv_eq2}
	\end{align}
    \end{subequations}
\end{proposition}
Proposition~\ref{prop:invertibility} requires that $\tPsiy \in \Css$ and $\tPsiu \in \Ccs$. Under the stronger assumption that $\tPsiu \in \Css$,
the equations in \eqref{eq:prop_inv_eq} can be merged into the following one
\begin{equation*}
    (v_t;d_t) = (y_t ; u_t) - \Psi^{\mathrm{o}}_t (v_{t-1:0},d_{t-1:0}).
\end{equation*}
Therefore, one obtains the following result matching Proposition III.1 of \citet{ho2020system}.
\begin{corollary}\label{cor:invertibility_ho}
	Consider an operator $\Upsilonb \in \Cc(\lpe{n}, \lpe{n})$ such that $\tUpsilonb := \Upsilonb-\I \in \Cs$. Then $\Upsilonb^{-1} \in \Cc$ exists and $\bb = \Upsilonb^{-1} \ab$ satisfies, for $t=0,1,\dots$,
	\begin{align*}
		b_t &= a_t - \Upsilon^\mathrm{o}_{t}(b_{t-1:0})\,.
	\end{align*}
\end{corollary}

We now show that \eqref{eq:SLSo} are also sufficient conditions.
\begin{theorem}\label{th:achiev_of}
    For a system $\Gb$,
	the set of all closed-loop maps achievable by \eqref{eq:system_control} can be represented as
	\begin{align*}
		\CL{\Gb} = \{\Psiyu = (\Psiy; \Psiu) = (\tPsiy; \tPsiu) + \I ~~| ~~
		 \eqref{eq:SLSo}
		\}\,.
	\end{align*}
	Moreover, the unique controller achieving $\Psiyu$ is given by
	\begin{equation}
		\label{eq:SLSoK}
		\Kb = \tPsiu \Psiyu^{-1} \InOv{r} \,.
	\end{equation}
\end{theorem}
The proof can be found in Appendix~\ref{app:proof_th_achiev_of}. 

To illustrate the value of Theorem~\ref{th:achiev_of}, several comments are in order.
First, compared to \eqref{eq:Phi_ach} where closed-loop maps are parametrized through $\Kb$, Theorem~\ref{th:achiev_of} presents an alternative parametrization of $\CL{\Gb}$ where constraints are directly over achievable closed-loop maps. 
Moreover, \eqref{eq:SLSoK} provides an explicit method for computing the controller associated with a given closed-loop map $\Psiyu$. 

Second,
Theorem~\ref{th:achiev_of} is a generalization of the input-output parametrization provided in~\citet{furieri2019input}, which considers LTI systems and controllers.
In the LTI case, the operators $\Psiyu$, $\Gb$ and $\Kb$ in Theorem~\ref{th:achiev_of} are equivalently represented by transfer matrices. 
In Appendix~\ref{app:generalization}, we show that for LTI systems Theorem~\ref{th:achiev_of} coincides with Theorem~1 of \citet{furieri2019input}.

Third, \eqref{eq:SLSo2} gives an explicit formula to obtain the operator $\Psiy$ from $\Psiu$. 
However, the operator $\Psiu$ cannot be freely chosen as it must satisfy \eqref{eq:SLSo3}.
The meaning of \eqref{eq:SLSo3} becomes clear when combining it with the controller \eqref{eq:SLSoK}. Indeed, the closed-loop maps are related by the dynamics of the controller as per  $\tPsiu = \Kb \Psiy$, (see also the relation between $\yb$ and $\tub$ in Figure~\ref{fig:blockdiagram_o}).

Next, we provide an alternative representation of closed-loop maps that overcomes the problem of choosing $\Psiu$ such that \eqref{eq:SLSo3} is satisfied. To this aim, we leverage the recursive implementation of an operator inverse established in Proposition~\ref{prop:invertibility}.

\subsection{Alternative parametrization and recursive implementation of achievable closed-loop maps}
\label{sec:recursive_implementation}

In this section, we give an equivalent characterization of the set $\CL{\Gb}$ that relies on a unique operator $\Emme$.
This result is the cornerstone that, for a given plant $\Gb$, allows exploring the set of all possible closed-loop maps by choosing only the ``free parameter'' $\Emme$.

\begin{theorem}\label{th:SLS_M_2}
	Given a system $\Gb$ and the feedback architecture \eqref{eq:system_control}, the following statements are true:
	\begin{enumerate}
		\item 
		For any $\Emme \in \Ccs(\lpe{r} \times \lpe{m} , \lpe{m})$, the operator $\Psiyu$ verifying%
        \footnote{Recall that, by definition, we have $\Psiyu = (\Psiy;\Psiu) = (\tPsiy;\tPsiu) + \I$.}
        \begin{subequations}
        \label{eq:SLSo_psi_2}
		\begin{align}
			\tPsiu 
			&= 
			\Emme
            \bvec{\Upsilonb_{2a}}{\Upsilonb_{2b}}
            \Upsilonb_1
			\,,
			\label{eq:SLSo_psiu_2}
			\\
			\tPsiy
			&= 
			\Gb \Psiu\,,
			\label{eq:SLSo_psiy_2}
		\end{align}
        \end{subequations}
        with 
        \begin{align*}
            \Upsilonb_{2a} &= \begin{bmatrix}\I_r & \Ob & \Ob \end{bmatrix} + \Gb \begin{bmatrix}\Ob & \I_m & \I_m \end{bmatrix} -\Gb \begin{bmatrix}\Ob & \Ob & \Ob \end{bmatrix} \,, \\ 
            \Upsilonb_{2b} &= - \begin{bmatrix}\Ob & \Ob & \I_m \end{bmatrix} \,, \\ 
            \Upsilonb_1 &= \bvec{\I_{r+m}}{\tPsiu}\,,
        \end{align*}
		satisfies $\Psiyu \in \CL{\Gb}$.
		\item 
		For any $\Psiyu \in\CL{\Gb}$, there exists a unique operator $\Emme \in \Ccs(\lpe{r} \times \lpe{m} , \lpe{m})$ such that
		$\Psiyu$ can be computed through \eqref{eq:SLSo_psi_2}. 
	\end{enumerate}
\end{theorem}
The proof can be found in Appendix~\ref{app:proof_SLS_M}.
It is worth highlighting that the term $\Gb\begin{bmatrix}\Ob & \Ob & \Ob \end{bmatrix}$ --- equal to $\Gb(\Ob)$, and independent of the input sequence --- represents the free response of the system
acting as a constant bias term. 
Let us clarify the notation  in~\eqref{eq:SLSo_psiu_2}. The operators on both sides act on a sequence $(\vb;\db) \in \lpe{r+m}$.
Applying $\Upsilonb_1$ to $(\vb;\db)$ gives the sequence $(\vb;\db;\tPsiu(\vb;\db))$, which is the input to operator $(\Upsilonb_{2a};\Upsilonb_{2b})$. 
Then, using the notation of Section~\ref{sec:notation}, one has 
\begin{multline*}
    (\Upsilonb_{2a};\Upsilonb_{2b}) (\vb;\db;\tPsiu(\vb;\db)) = \\ \left(\Upsilonb_{2a} (\vb;\db;\tPsiu(\vb;\db)) ; \Upsilonb_{2b} (\vb;\db;\tPsiu(\vb;\db)) \right)\,.
\end{multline*}

\begin{proposition}[Recursive implementation of $\Emme$]
\label{prop:rec_implememntation_M}
	Given a system $\Gb$ and the feedback architecture \eqref{eq:system_control}, 
    the output of the controller $\tub = \Kb \yb$, where $\Kb$ is defined by \eqref{eq:SLSoK} and \eqref{eq:SLSo_psi_2}, can be computed recursively through the equation
    \begin{equation}
    	\label{eq:K_implementation}
    	u^\mathrm{o}_t =  \mathcal{M}_t(y_{t:0}  - y^{\text{free}}_{t:0} ; - u^\mathrm{o}_{t-1:0})\,.
    \end{equation}
\end{proposition}
The proof can be found in Appendix~\ref{app:proof_prop_recursivity_M}.
Note that formula~\eqref{eq:K_implementation} circumvents the problem of computing the inverse of $\Psiyu$ appearing in~\eqref{eq:SLSoK}.
We also highlight that \eqref{eq:K_implementation} represents a non-Markovian\footnote{This is because, at time $t$, the state $u^{\mathrm{o}}_t$ depends on its complete history, i.e., on $u^{\mathrm{o}}_{t-1:0}$.} dynamical system with input $\yb -\yb^{\text{free}}$ and internal state $\tub$.

\section{Parametrization of stable closed-loop maps}
\label{sec:stable_CLM}

While Theorem~\ref{th:SLS_M_2} gives a parametrization of all and only achievable closed-loop maps for a given plant, 
it does not say anything about the stability of the closed-loop system. 
Note that constraining $\Emme$ to be a stable operator is not enough to guarantee closed-loop stability. 
Indeed, 
the recursive implementation \eqref{eq:K_implementation} shows that $\Emme$ operates in a feedback loop as some of its inputs correspond to previous output values.
Even in the SISO LTI case,  a stable transfer matrix $\Emme$ can generate an unstable transfer matrix $\tPsiu$.%
\footnote{In the LTI setting, the operator $\tPsiu$ represents the transfer matrix formed by the following two blocks: the complementary sensitivity transfer function matrix, mapping  $\db\mapsto\tub$, and the noise sensitivity transfer function matrix, mapping $\vb\mapsto\tub$.}

In the sequel, we will focus on the following notions of stability.

\begin{definition}
\label{def:ifg_stab}
	An operator $\Ab\in\Cc(\lpe{m},\lpe{r})$ is termed:
	\begin{itemize}
		\item 
		$\ell_p$-stable (denoted as $\Ab\in\Lp$), if $\Ab(\xb) \in \ell^r_p$ for all $\xb\in\ell^m_p$;
		\item 
		incrementally finite gain $\ell_p$-stable (denoted as i.f.g. $\ell_p$-stable), if there exists ${\gamma}\in [0,+\infty)$ such that for any $\xb_1,\xb_2\in\ell_p^n$, 
        one has 
        \begin{equation}
        \label{eq_IFG}
            \norm{\Ab\xb_1 - \Ab\xb_2}_p \leq {\gamma} \norm{\xb_1 - \xb_2}_p \,.
        \end{equation}
        Moreover, $\gamma$ is called the incremental $\ell_p$-gain of ${\Ab}$.
	\end{itemize}
\end{definition}
Note that i.f.g.\ $\ell_p$-stability is a notion of smoothness~\cite{vanderSchaft2017}.
Moreover, the i.f.g.\ $\ell_p$-stability of  $\Ab$ does not imply that $\Ab\in\Lp$.
A notable exception is when $\Ab(\mathbf{0})=\mathbf{0}$ because, by setting $ \xb_2=\mathbf{0}$ in \eqref{eq_IFG}, one obtains $ \norm{\Ab\xb_1}_p \leq {\gamma} \norm{\xb_1}_p$, which implies $\ell_p$-stability.

\begin{remark} [Static stable operators]
\label{rmk:static_operators}
    Consider a function $a:\mathbb{R}^n\mapsto \mathbb{R}^m$ and the associated operator 
    \begin{equation}
        \label{eq:static_op}
        \Ab \xb=(a(x_0), a(x_1), \ldots).
    \end{equation} From Definition \ref{def:ifg_stab} it is easy to see that $\Ab\in\mathcal{L}_p$ if 
    \begin{equation}
    \label{eq:sublinear_growth}
        |a(x)|<\eta|x|,~\forall x\in \mathbb{R}^n\,,
    \end{equation} 
    for some $\eta>0$.  Moreover, any function $a$ verifying the Lipschitz bound 
    \begin{equation}
    \label{eq:lip_bound}
        |a(x)-a(\tilde x)|\leq \gamma |x-\tilde x|,~\forall x,\tilde x\in \mathbb{R}^n\,,
    \end{equation}
    for some $\gamma>0$, defines an i.f.g.\ $\ell_p$-stable operator with incremental gain $\gamma$.
\end{remark}

\begin{example}[Continuation from Example~\ref{ex:LTI_1}]
    Consider the same setting of Example~\ref{ex:LTI_1} and let us analyze the $\ell_p$-stability and i.f.g. $\ell_p$-stability for $p=2$.
    In this case, both definitions coincide and amount requiring that the transfer matrix $\Gb(z)$ is asymptotically stable.  
\end{example}

The subset of $\CL{\Gb}$ 
of achievable and stable closed-loop maps is given by 
\begin{equation}
	\label{eq:Phi_ach_stab}
	\CLp{\Gb} = \{\Phiyu{G}{K}|~\mathbf{K} \in \Cc \quad\text{and}\quad \Phiyu{G}{K}\in\Lp \}\,.
\end{equation}
In the LTI case, 
imposing $\Phiyu{G}{K}\in\Lp$
is not difficult \cite{furieri2019input} 
because \eqref{eq:SLSo} gives rise to affine relations between
the four input-output operators that characterize the closed-loop maps $\vb \mapsto \yb$, $\vb \mapsto \ub$,  $\db \mapsto \yb$, and $\db \mapsto \ub$. 
Thus, imposing stability is equivalent to restricting these maps to be proper stable transfer matrices satisfying mutual affine relationships.
However, in the nonlinear case, it is not straightforward to impose $\Phiyu{G}{K}\in\Lp$ while satisfying \eqref{eq:SLSo}.

Next, we address this issue by parametrizing $\CLp{\Gb}$, and particularly $\Emme$, in terms of a free operator $\Qb\in\Lp$.
We first consider, in Section~\ref{sec:stable_plant}, the simpler case where the plant $\Gb$ satisfies  $\Gb\in\Lp$ and is i.f.g. $\ell_p$-stable.
Then, we introduce three extensions:
\begin{itemize}
	\item
	In Section~\ref{sec:extension_base_controller}, we consider the case of unstable systems for which a base stabilizing controller $\Kb'(\yb)$ is available. The goal will be to show how to describe all stabilizing controllers as a function of $\Kb'(\yb)$ and $\Qb$.
	\item
	In Section~\ref{sec:robust_analysis}, we provide robustness guarantees for the case where there is some model mismatch between the true plant and the available model for parametrizing closed-loop maps.
	\item 
	In Section~\ref{sec:extension_distributed}, we extend the parametrization to distributed control architectures for interconnected systems where each local controller can only receive information from a subset of subsystems.
\end{itemize}

\subsection{The case of stable plants} \label{sec:stable_plant}
Our goal is to parametrize the set $\CLp{\Gb}$ when $\Gb$ is i.f.g. $\ell_p$-stable and $\Gb\in\Lp$.
To this purpose, we will represent the operator $\boldsymbol{\mathcal{M}}$ in Theorem~\ref{th:SLS_M_2} by using an auxiliary operator $\Qb$ and the system model.  

\begin{theorem}\label{th:SLS_stable}
	Given an i.f.g. $\ell_p$-stable plant $\Gb$, such that $\Gb\in\Lp$,
    each $\Psiyu \in\CLp{\Gb}$ can be obtained
	by selecting $\Emme$ in Theorem~\ref{th:SLS_M_2} as:
	\begin{equation}\label{eq:M_from_Q}
		\Emme = \Qb \left( \InO{r} - \Gb \OnegIn{m} + \Gb \OO \right)\,,
	\end{equation}
	for a suitable $\Qb:\lpe{r}\rightarrow\lpe{m}\in\Lp$.
	Moreover, 
    for any $\Qb\in\Lp$, $\Emme$ in \eqref{eq:M_from_Q} defines a closed-loop map $\Psiyu \in\CLp{\Gb}$ 
    through \eqref{eq:SLSo_psi_2}.
\end{theorem} 
The proof can be found in Appendix~\ref{app:proof_SLS_stable}.
Moreover, Figure~\ref{fig:blockdiagram_MWS} presents a block diagram 
with the implementation of $\tPsiu$ stemming from \eqref{eq:M_from_Q}, and using \eqref{eq:SLSo_psiu_2}. 

\begin{figure}
	\centering
	\includegraphics{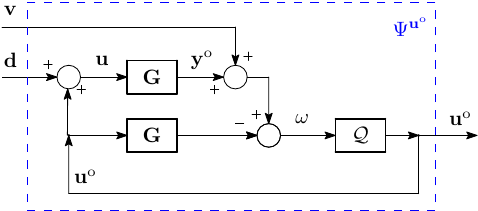}
	\caption{Block diagram implementation of the stable operator $\tPsiu$ using the free operator $\Qb\in\Lp$. Note that $\tPsiy$ can then be obtained as $\Gb\Psiu$. }
	\label{fig:blockdiagram_MWS}
\end{figure}

The next proposition shows the dynamical system implementing the controller $\Kb$ associated with a given $\Qb\in\Lp$.
This result is the cornerstone bridging the theoretical framework provided by Theorems \ref{th:SLS_M_2} and \ref{th:SLS_stable} with the neural network control design methods described in Section~\ref{sec:NOC_RENs}.
\begin{proposition}[Recursive implementation of the controller]
\label{prop:rec_implemenatation_KQ}
    Given an i.f.g. $\ell_p$-stable plant $\Gb$, such that $\Gb\in\Lp$, and any $\Qb\in\Lp$, a non-Markovian recursive implementation for the controller $\Kb$ achieving the closed-loop maps described in Theorem~\ref{th:SLS_stable}%
    \footnote{Equivalently, when $\Kb$ is given by \eqref{eq:SLSoK}, \eqref{eq:SLSo_psi_2} and \eqref{eq:M_from_Q}.}
    is given by
    \begin{subequations}
    \label{eq:recursive_implementation_omega}
    \begin{align}
    \label{eq:recursive_implementation_omega_1}
	\omega_t &= y_{t} - G_{t}( \mathcal{Q}_{t-1:0}(\omega_{t-1:0}))\,, \\
    \label{eq:recursive_implementation_omega_2}
    u^\mathrm{o}_t &=  \mathcal{Q}_t(\omega_{t:0}) \,.
    \end{align}
    \end{subequations}
\end{proposition}
The proof can be found in Appendix~\ref{app:proof_rec_implementation_KQ}.

\begin{figure}
	\centering
	\includegraphics{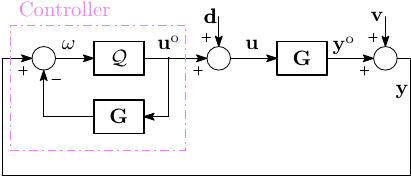}
	\caption{Block diagram implementation of the closed-loop system using the controller form of~\eqref{eq:controller_youla}.}
	\label{fig:blockdiagram_CL_stable}
\end{figure}

\begin{remark}[Relation with \citet{desoer1982}]
\label{rmk:cmp_desoer}
The work of \citet{desoer1982} provides a characterization of all 
stabilizing controllers in Figure~\ref{fig:blockdiagram_o} for an i.f.g. $\ell_p$-stable plant $\Gb$, hence establishing a nonlinear version of the classic Youla parametrization. Specifically, \citet{desoer1982} show that any stabilizing controller can be written as 
\begin{equation}
	\label{eq:controller_youla}
	\Kb = \Qb (\Gb \Qb + \I)^{-1}\,,
\end{equation}
for a suitable $\Qb\in\Lp$.
We show that \eqref{eq:recursive_implementation_omega} provides \eqref{eq:controller_youla}.
Consider the composed operator $\Gb \Qb$ appearing in~\eqref{eq:recursive_implementation_omega_1}. Since $\Gb\in\Cs$, we have that $\Gb \Qb \in\Cs$.
Thus, we can use Corollary~\ref{cor:invertibility_ho} to rewrite~\eqref{eq:recursive_implementation_omega_1} in its operator form as
$\omegab = \left(\Gb\Qb + \I\right)^{-1}\yb$.
Hence, the controller that achieves the closed-loop maps $\Psiyu$ of Theorem~\ref{th:SLS_stable} is given by \eqref{eq:controller_youla}.
However, we notice that Theorems \ref{th:SLS_M_2} and \ref{th:SLS_stable} do not characterize stabilizing controllers only but also the associated closed-loop maps. Moreover, different from Proposition~\ref{prop:rec_implemenatation_KQ}, \citet{desoer1982} do not provide any recursive method for computing control actions.
For a concise comparison with these and other related works, we refer the reader to Table~\ref{tab:comparison}.
\end{remark}
Figure~\ref{fig:blockdiagram_CL_stable} shows the block diagram of the closed-loop of a system $\Gb$ with the controller~\eqref{eq:recursive_implementation_omega} (or equivalently \eqref{eq:controller_youla}).

\begin{remark}
	Since every stabilizing controller contains a model (or copy) of the plant, the controller architecture can be regarded through the lens of IMC in the nonlinear setting~\cite{Garcia_Morari_IMC_1982, Economou_Morari_nlIMC_1986}.\footnote{The reader can compare the schemes in Figure~\ref{fig:blockdiagram_CL_stable} with the one in Figure~3 of \citet{Economou_Morari_nlIMC_1986}.}
	As in IMC, the controller implementation relies on the knowledge of the system model. 
	Hence, a mismatch between the true system and the model $\Gb$ may compromise stability. This issue is further analyzed in Section~\ref{sec:robust_analysis}.
\end{remark}

\begin{remark}
\label{rmk:history}
	According to \eqref{eq:recursive_implementation_omega_2}, for obtaining $u_t$ one needs to store the whole past of the variable $\omegab$, leading to intractable computations as $t$ grows. However, this issue can be circumvented by restricting the attention to Markovian controllers in the form \eqref{eq:NLcontr}. Examples of controllers of this type embedding NNs are provided in Section~\ref{sec:intro_REN}.
\end{remark}

\subsection{Plants equipped with a stabilizing controller}
\label{sec:extension_base_controller}

We extend the base case described in Section~\ref{sec:stable_plant} to plants that can be unstable but for which at least one stabilizing policy is known.
More specifically,
given the stabilizing controller $\mathbf{K}'(\mathbf{y})$, 
with $\mathbf{K}' \in \mathcal{L}_{p}$,
for the i.f.g. $\ell_p$-stable system $\mathbf{G}$, 
we show that \emph{all other} stabilizing control policies can be represented as 
\begin{equation}
	\label{eq:stabilizing_input}
	\tub = \Kb(\yb) = \mathbf{K}'(\mathbf{y})+\Qb(\tomegab)\,,
\end{equation}
where the operator $\Qb\in \mathcal{L}_{p}$ is a free parameter and $\tomegab$ is given by 
\begin{equation}\label{eq:omega_prestabilized}
    \tomegab = \yb-\Gb\tub \,.
\end{equation}
Furthermore, if $\mathbf{K}'$ is stabilizing but not itself a stable operator, the control policy \eqref{eq:stabilizing_input} still describes closed-loop maps 
$(\Psiy;\Psiu) \in \CLp{\Gb}$
although the parametrization is not complete.

Thanks to the causality of $\mathbf{K}'(\mathbf{y})$ and $\Qb(\tomegab)$,  \eqref{eq:stabilizing_input} defines an overall control policy $\Kb$ that maps $\yb$ into $\tub$.
Note that, 
the recursive implementation of
$\tomegab$ 
stemming from \eqref{eq:stabilizing_input} and \eqref{eq:omega_prestabilized}
is slightly different from the one  given in \eqref{eq:recursive_implementation_omega} for $\omegab$ 
since, in this case, the action of the base controller is also considered.

Similar to the works of \citet{zhou1998essentials, anantharam1984stabilization}, the role of the base controller $\mathbf{K}'$ is to appropriately stabilize the system, which allows us to define a set of ``stable coordinates'' and then freely optimize over~$\Qb$.

More formally, in the sequel, we use the following properties for our base controller in the statement of our results.
\begin{definition}[(Strongly) $\ell_p$-stabilizing i.f.g. controller]
	\label{def:stabilizing}
	Given an i.f.g. $\ell_p$-stable system $\Gb$,
	we say that a controller $\mathbf{K}'$ is 
	\begin{enumerate}
		\item 
		$\ell_p$-stabilizing i.f.g. for $\mathbf{G}$, if $\Phiyu{\Gb}{\Kb'} : (\mathbf{v}; \mathbf{d}) \mapsto (\mathbf{y};\mathbf{u})$  lies in $\mathcal{L}_p$ and $\Phiyu{\Gb}{\Kb'}$ is i.f.g. $\ell_p$-stable.
		\item 
		strongly $\ell_p$-stabilizing i.f.g. for $\Gb$, if it is i.f.g. $\ell_p$-stabilizing, and, in addition, $\mathbf{K}'\in \mathcal{L}_{p}$.
	\end{enumerate}
\end{definition} 
While this paper does not deal with the computation of a base controller, we refer the interested reader to \citet{Koelewijn_2021} and references therein for modern methods to design  $\ell_p$-stabilizing i.f.g. controllers in discrete-time.

The next theorem goes beyond Theorem~\ref{th:SLS_stable} and parameterizes all achievable closed-loop maps for classes of systems that can be stabilized through a base controller.
\begin{theorem}\label{th:stabilizing_SLS}
	Consider the closed-loop system \eqref{eq:system_control} with $\Kb$ as in \eqref{eq:stabilizing_input} where $\Gb$  is i.f.g. $\ell_p$-stable.
	\begin{enumerate}
		\item Assume that $\mathbf{K}'$ is an $\ell_p$-stabilizing i.f.g. controller. 
		Then,  $\Phiyu{\Gb}{\Kb} \in \CLp{\Gb}$ for every $\Qb \in \Lp$.
		\item If, in addition, $\mathbf{K}'$ is a strongly  $\ell_p$-stabilizing i.f.g. controller, then, for any $(\Psiy;\Psiu) \in \CLp{\Gb}$, there exists $\Qb \in \Lp$ such that the control policy \eqref{eq:stabilizing_input}
		achieves the closed-loop maps $\Psiyu$, i.e., $(\Phiy{\Gb}{\Kb};\Phiu{\Gb}{\Kb}) = (\Psiy;\Psiu)$.
	\end{enumerate}
\end{theorem}
The proof can be found in Appendix~\ref{app:proof_stabilizing}.

\begin{remark}\label{rmk_cmp_Fujimoto}
\citet{fujimoto1998youla} and \citet{paice1996tac} develop Youla parametrizations for nonlinear systems using observer-based kernel representations, which generalize nonlinear left coprime factorizations \cite{paice1994stable}.
Note that, unlike \citet{fujimoto1998youla}, where a stable kernel representation for both the system and controller is needed,
point~2 of  Theorem~\ref{th:stabilizing_SLS} relies only on a property of the base controller for obtaining a parametrization of all other stabilizing controllers.
This new characterization lends itself to the design of algorithms that search in the set of stabilizing controllers (see Section~\ref{sec:NOC_RENs}).
It is also worth highlighting that it is generally not easy to obtain \emph{stable} kernel representations when the nonlinear system is not stable by itself.
Table~\ref{tab:comparison} provides a synthetic comparison between our approach and related methods.
\end{remark}

\begin{figure}
	\centering
	\includegraphics[width=0.95\linewidth]{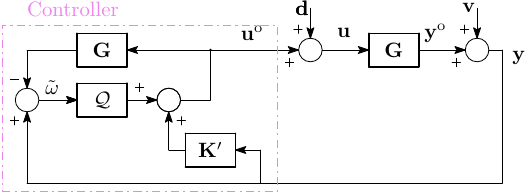}
	\caption{Block diagram implementation of the closed-loop system using the controller form of~\eqref{eq:K_stable}.}
	\label{fig:blockdiagram_CL_with_stabK}
\end{figure}

The next proposition provides an implementation of the controller $\Kb$ associated with $\Kb'$ and a given $\Qb\in\Lp$.
As for Proposition~\ref{prop:rec_implemenatation_KQ}, this result allows translating Theorem~\ref{th:stabilizing_SLS} into concrete procedures for the design of neural network controllers --- see Section~\ref{sec:NOC_RENs}.
\begin{proposition}[Recursive implementation of the controller with $\Kb'$]
\label{prop:rec_implemenatation_KQ_with_base}
    Given an i.f.g. $\ell_p$-stable plant $\Gb$, a controller $\Kb'$ satisfying point (1) of Theorem~\ref{th:stabilizing_SLS} and any $\Qb\in\Lp$,
    a non-Markovian recursive implementation for the controller $\Kb$ in \eqref{eq:stabilizing_input} is given by
    \begin{subequations}
    \label{eq:K_stable}
    \begin{align}
        u^\mathrm{o}_t &=  \mathcal{Q}_t(\tilde{\omega}_{t:0}) + K'_t(y_{t:0})\,, \label{eq:K_stable_output} \\
        \tilde{\omega}_t &= y_t - G_t(\mathcal{Q}_{t-1:0} (\tilde{\omega}_{t-1:0}) + K'_{t-1:0} (y_{t-1:0}) )\,. \label{eq:K_stable_omega}
    \end{align}
    \end{subequations}
\end{proposition}
The proof can be found in Appendix~\ref{app:proof_rec_implementation_KQ_base}.
Note that \eqref{eq:K_stable} is a dynamical system with input $\yb$, output $\tub$ and internal state $\tomegab$. 
The block diagram of the closed-loop is shown in Figure~\ref{fig:blockdiagram_CL_with_stabK}.

\subsection{Robustness against model mismatch}\label{sec:robust_analysis}
As highlighted in the previous sections, the controller implementation needs a copy of the plant model. 
However, in practice, the system model might not be exact, leading to a mismatch between the true plant and the dynamics used in the controller.

\begin{figure}
	\centering
    \includegraphics{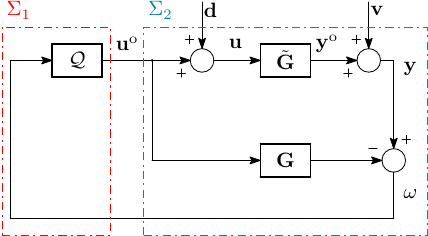}
	\caption{Scheme of the closed-loop system with plant-model mismatch.}
	\label{fig:robust}
\end{figure}

In this section, we analyze under which conditions the stability result in Theorem~\ref{th:SLS_stable} holds in the presence of a model mismatch. We denote with $\tilde{\Gb}$ the true plant dynamics, and with $\Gb$ the modeled dynamics, see Figure~\ref{fig:robust}.
The next proposition characterizes robust closed-loop stability.

\begin{proposition}\label{prop:robust}
	Given an i.f.g. $\ell_p$-stable plant $\tilde{\Gb}$, such that $\tilde{\Gb}\in\Lp$, and its nominal model is $\Gb$, the following statements hold:
	\begin{enumerate}
		\item 
		If $(\tilde{\Gb} - \Gb)\in\Lp$, then there exists $\gamma_Q \in \mathbb{R}^+$, which is the finite gain of operator $\Qb$, such that the closed-loop system in Figure~\ref{fig:robust} is stable.
		\item 
		If $(\tilde{\Gb} - \Gb)\ub\in\ell_p$ for any $\ub$, 
		then the closed-loop of Figure~\ref{fig:robust} is stable for any choice of $\Qb \in \Lp$.
	\end{enumerate}
\end{proposition}
The proof can be found in Appendix~\ref{app:proof_robust}.
Note that point~2 of Proposition~\ref{prop:robust} relies on a stronger assumption compared to point~1.
In point~1, one assumes that for any $\ub\in\ell_p$, we have that $(\tilde{\Gb} - \Gb) \ub \in \ell_p$.
However, in point~2, one assumes that the latter holds even when $\ub \notin \ell_p$.

The result of Proposition~\ref{prop:robust} is in the spirit of the small-gain theorem~\cite{zames1966input}, but it is not a direct consequence of it since we are not applying the small-gain condition to the closed-loop formed by the system $\Gb$ and the controller $\Kb$. 
Rather, the small-gain condition is evaluated over the auxiliary closed-loop interconnection shown in Figure~\ref{fig:robust} formed by: the operator $\Sigma_1$, which consists only of $\Qb$, and the operator $\Sigma_2$ which combines $\tilde{\Gb}$ and $\Gb$.
This setup allows obtaining weaker stability conditions compared to the use of the small gain theorem with blocks $\Gb$ and $\Kb$.
Moreover, we highlight that the proof of point~1 is constructive. 
This means that it provides a formula for choosing the gain $\gamma_Q$ as a function of the finite gain $\gamma_{\Delta}$ of the operator $(\tilde{\Gb}-\Gb)$. In particular, one needs $\gamma_Q < \gamma_{\Delta}^{-1}$. 
It is worth noticing that in practical applications, $\gamma_{\Delta}$ would probably be unknown. However, Proposition~\ref{prop:robust} suggests decreasing $\gamma_Q$ until closed-loop stability is eventually obtained.

We highlight that the model of the system as per $\Gb$ contains the information of the initial condition. 
Thus, one of the sources of model mismatch could be the difference between the initial condition of the true system $\tilde{\Gb}$ and the model $\Gb$.

\begin{remark}[Relations with existing results]
\label{rmk_cmp_robust}
    In the work of \citet{desoer1982}, a robustness analysis is performed to assess whether closed-loop stability is preserved under a plant perturbation $\boldsymbol{\Delta}$. However, this analysis is limited to linear stable parameters $\boldsymbol{\mathcal{Q}}$.
    
    The assumption in point~2 is 
    similar to the detectability assumption in~\citet{fujimoto2000} for the stable kernel representations  (see Definition~14 in~\citet{fujimoto2000}).
	Moreover, \citet{barbara2023learning} also consider the assumption of point~2 (see the assumption A3 of Section~II.B  and Definition~1 (Contraction), in~\citet{barbara2023learning}). However, they focus their analysis on the case where the model mismatch is due to differences in the initial conditions since the state-space model of the system is known for the control design. 
    
    Furthermore, we highlight that our results are consistent with those of \citet{desoer1984simultaneous}, which address the conditions under which a controller can simultaneously stabilize two distinct plants.  However, unlike \citet{desoer1984simultaneous}, our Proposition~\ref{prop:robust} emphasizes the conditions on the operator $\boldsymbol{\mathcal{Q}}$ required for robustness against model mismatch. This is particularly important for optimization purposes, specifically for searching within the set of stabilizing controllers (see Section~\ref{sec:NOC_RENs}).
\end{remark}
 
\begin{example}[Robustness w.r.t. a change of the initial conditions in LTI systems]
Assume the same setting of Example~\ref{ex:LTI_1} and let $\Gb$ and $\tilde{\Gb}$ be the operators generated by the same asymptotically stable LTI models when starting from $x_0 = \bar{x}$ and ${x}_0 = \tilde{x}$, respectively. 
Since the initial state affects only $\Gb(\Ob)$ and $\tilde{\Gb}(\Ob)$ we have, for any 
$\Gb(\tub) = \Gb (\ub) - \Gb(\Ob) + \tilde{\Gb}(\Ob)$
and hence,
\begin{equation}
    \label{eq:robust_LTI}
    (\tilde{\Gb} - \Gb) \ub = \tilde{\Gb}(\Ob) - \Gb(\Ob) 
    = 
    \begin{bmatrix}
        C \\
        CA \\
        CA^2 \\
        \vdots
    \end{bmatrix}
    (\tilde{x} - \bar{x})\,,
\end{equation}
where the last equality follows from \eqref{eq:system_dyn_LTI}.
The asymptotic stability of the system implies that the sequence in the right-hand side of \eqref{eq:robust_LTI} belongs to $\ell_p$ for any $\ub$. Therefore, by point (2) of Proposition~\ref{prop:robust}, the closed-loop system in Figure~\ref{fig:robust} is stable for any $\Qb\in\Lp$.
\end{example}

\subsection{Design of distributed controllers}
\label{sec:extension_distributed}

We consider now a system composed of $M$ interconnected subsystems, each equipped with a local controller.
By splitting the signals components of the closed-loop \eqref{eq:system_control},
$y_t$, $u_t$, $v_t$ and $d_t$ into $M$ vectors of suitable dimensions,
the dynamics  of subsystem $i$ and its associated controller are
\begin{align*}
	y_t^i &= G_t^i(u_{t-1:0}) + v_t^i \,, \\
	u_t^i &= K_t^i(y_{t:0}) + d_t^i \,,
\end{align*}
for $t=0,1,\dots$, or, in the signal space,
\begin{subequations}
\label{eq:split_y_u}
\begin{align}
	\mathbf{y}^i &= \mathbf{G}^i(\mathbf{u}) + \mathbf{v}^i \,, \label{eq:split_y}\\
	\mathbf{u}^i &= \mathbf{K}^i(\mathbf{y}) + \mathbf{d}^i \,, \label{eq:split_u}
\end{align}
\end{subequations}
for $i=1,\dots,M$.

Next, we consider the case where the input of each local subsystem (respectively controller) can only depend on the control variables $u^j_t$ (respectively the system outputs $y^j_t$) produced by some other controllers (respectively subsystems), denoted as
\textit{neighbors}.
To this aim, for an operator $\Ab$ composed by $M$ subsystems,
we introduce a directed graph with binary adjacency matrix $\mathcal{D}(\mathbf{A})\in\{0,1\}^{M\times M}$
where node $i$ represents the subsystem $i$ and a 
$1$ in  position $(i,j)$, i.e., $[\mathcal{D}(\mathbf{A})]_{i,j} = 1$, indicates that subsystem $i$ can send information to subsystem $j$.
By convention, $[\mathcal{D}(\mathbf{A})]_{i,i} = 1$, for all $i=1,\dots,M$.

For binary matrices $S_1\in\{0,1\}^{M\times M}$ and $S_2\in\{0,1\}^{M\times M}$, we adopt the following notation:
$S_1 + S_2 \in\{0,1\}^{M\times M}$ 
(respectively, $S_1  S_2 \in\{0,1\}^{M\times M}$)
is a binary matrix having a zero entry in position $(i,j)$ 
if and only if 
$S_1 + S_2$ 
(respectively, $S_1 S_2$) has a zero entry in position $(i,j)$.
Moreover, $S_1\leq S_2$ denotes that $[S_1]_{i,j} \leq [S_2]_{i,j}$ for all $i,j\in\{1,\ldots,M\}$. To provide an interpretation of these definitions, let $S_1$ and $S_2$ be the binary adjacency matrices of the directed graphs $\mathcal{G}_1=(V,\mathcal{E}_1)$, and $\mathcal{G}_2=(V,\mathcal{E}_2)$, respectively, where $V=\{1,\ldots,M\}$. Then, $S_1+S_2$ is the adjacency matrix of the union graph $\mathcal{G}=(V,\mathcal{E}_1\cup \mathcal{E}_2)$
and $[S_1 S_2]_{i,j}=1$ if and only if there is a two-hops path from node $i$ to node $j$ by crossing first an edge $(i, i')$ of $\mathcal{G}_2$ and then the edge $(i', j)$ in $\mathcal{G}_1$. Finally, $S_1\leq S_2$ corresponds to  $\mathcal{E}_1\subseteq \mathcal{E}_2$.

We indicate the set of in-neighbors of node $i$ as
\begin{equation}
\label{eq:in_neighbors}
	\mathcal{N}^-_A(i) = \{j\in\{1,\dots,M\} ~|~ [\mathcal{D}(\mathbf{A})]_{j,i}=1 \} 
    \,,
\end{equation}
and the set of out-neighbors as
\begin{equation}
	\mathcal{N}^+_A(i) = \{j\in\{1,\dots,M\} ~|~ [\mathcal{D}(\mathbf{A})]_{i,j}=1 \} 
    \,.
\end{equation}
Moreover, 
we define $\bar{\mathcal{N}}^-_A(i) = \mathcal{N}^-_A(i) \setminus\{i\}$ and $\bar{\mathcal{N}}^+_A(i) = \mathcal{N}^+_A(i) \setminus\{i\}$.
The \textit{interconnection structure} of an operator $\Ab$ is then captured by the \textit{sparsity pattern} of the matrix $\mathcal{D}(\Ab)$.

\begin{figure}
    \centering
    \includegraphics[width=0.99\linewidth]{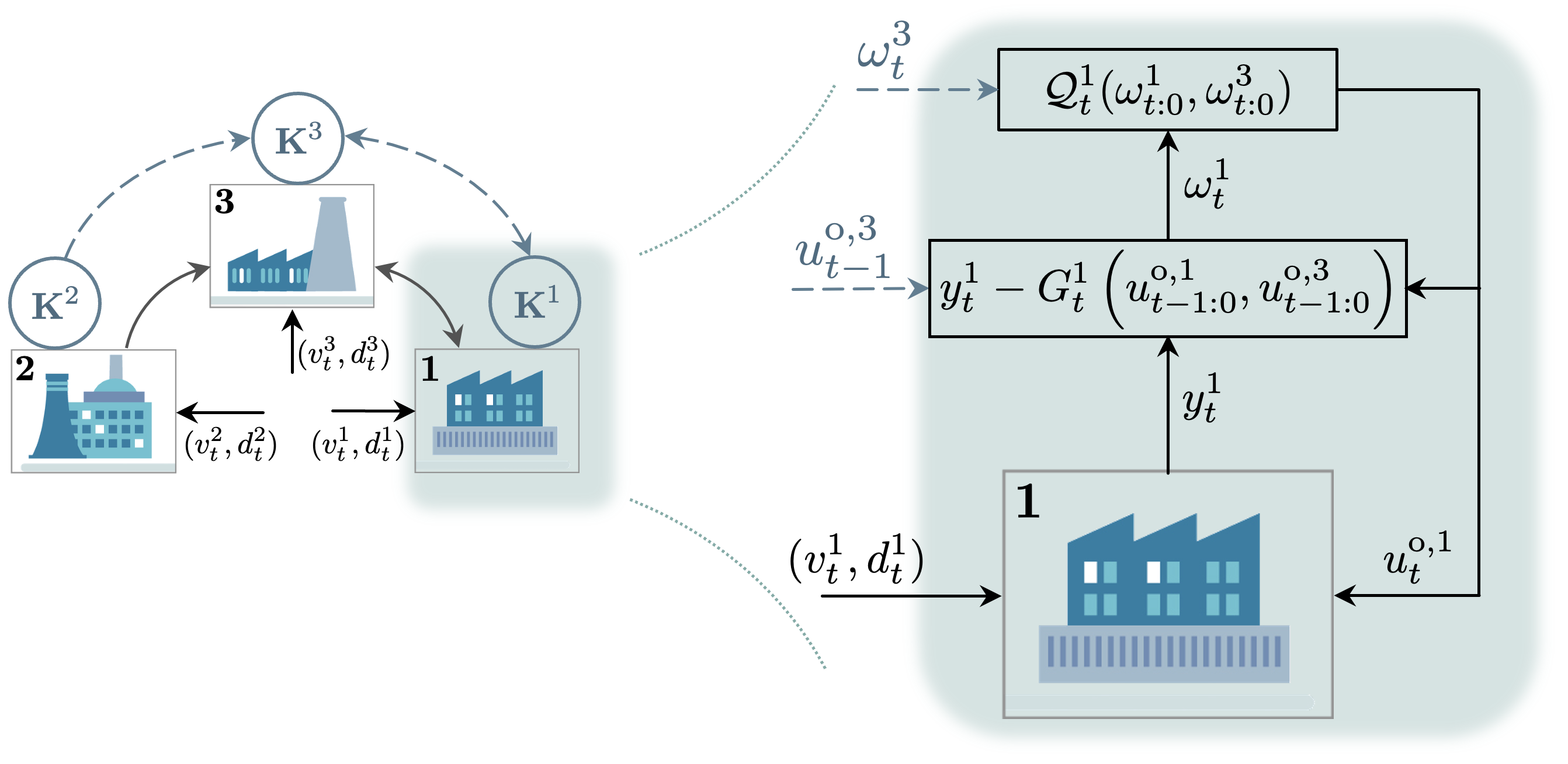}
    \caption{\textit{Left panel}: example of the interconnected system \eqref {eq:split_y_neighbors} with the associated distributed controller \eqref{eq:split_u_neighbors}. The continuous arrows define the coupling among systems and the adjacency matrix  $\mathcal{D}(\Gb)$. The dashed arrows represent the communication topology between local controllers, which is equivalently captured by the matrix $\mathcal{D}(\Kb)$.
    \textit{Right panel}: graphical representation of the steps needed for computing the control action $u_t^{o,1}$ for subsystem $1$ according to Algorithm~\ref{alg:dist_implementation}.}
    \label{fig:distributed_example}
\end{figure}

The adjacency matrix $\mathcal{D}(\mathbf{G})$ is given by the coupling topology between subsystems, i.e., $[\mathcal{D}(\Gb)]_{i,j} = 1$ means that $u^i$ influences the evolution of $y^j$. Instead, $\mathcal{D}(\mathbf{K})$ can be chosen 
for designing a distributed controller with a prescribed communication topology.
We can  write 
$\mathbf{G}^i\left(\{\mathbf{u}^j\}_{j\in\mathcal{N}^-_G(i)}\right)$ --- respectively 
$\mathbf{K}^i\left(\{\mathbf{y}^j\}_{j\in\mathcal{N}^-_K(i)}\right)$ --- for highlighting that the operator depends only on the subset of the inputs indexed by $\mathcal{N}^-_G(i)$ --- respectively $\mathcal{N}^-_K(i)$. 
Thus, the closed-loop system dynamics~\eqref{eq:split_y_u} in a distributed setting is given by:
\begin{subequations}
    \label{eq:split_y_u_neighbors}
	\begin{align}
		\label{eq:split_y_neighbors}
		\mathbf{y}^i = \mathbf{G}^i\left(\{\mathbf{u}^j\}_{j\in\mathcal{N}^-_G(i)}\right) + \mathbf{v}^i \,, \\
		\label{eq:split_u_neighbors}
	\mathbf{u}^i = \mathbf{K}^i\left(\{\mathbf{y}^j\}_{j\in\mathcal{N}^-_K(i)}\right) + \mathbf{d}^i \,.
	\end{align}
\end{subequations}
An illustrative example of the model \eqref{eq:split_y_u_neighbors} is given in Figure~\ref{fig:distributed_example}.

We now provide a procedure for designing $\ell_p$-stabilizing controllers which comply with a given maximal admissible communication topology.

\begin{proposition}
\label{prop:distributed}
    Consider an i.f.g. $\ell_p$-stable system $\Gb$ composed of $M$ interconnected subsystems $\Gb^i$, $i=1,2,\dots,M$ as in~\eqref{eq:split_y_neighbors} where each $\Gb^i\in \Lp$ and is i.f.g. $\ell_p$-stable. Let $\mathcal{T}\in \{0,1\}^{M\times M}$ denote the adjacency matrix of the densest admissible communication graph for the controller $\Kb$ and assume $ \mathcal{D}(\Gb)\leq\mathcal{T}$. Assume also that at most two rounds of communication are allowed per time step between local controllers, and 
    consider an operator $\Qb$ consisting
    of $M$ operators $\Qb^i \in \Lp$, connected through a communication graph represented by the adjacency matrix $ \mathcal{D}(\Qb)$, and verifying $\mathcal{D}(\Qb) \leq \mathcal{T}$. Then, the controller given by \eqref{eq:controller_youla} and implemented as described in Algorithm~\ref{alg:dist_implementation} verifies 
    \begin{align}
        \label{eq:sparsity_controller_distributed}
        \mathcal{D}(\Kb) &= \mathcal{D}(\Qb) + \mathcal{D}(\Gb) \leq \mathcal{T} \,,
        \\
        \label{eq:sparsity_controller_distributed_GK}
        \mathcal{D}(\Kb) &\geq \mathcal{D}(\Gb)\,,
    \end{align}
    and guarantees that the closed-loop system is $\ell_p$-stable.
\end{proposition}
The proof of Proposition~\ref{prop:distributed} can be found in \ref{app:proof_prop_distributed}.

\begin{algorithm}[!htb]
	\caption{Distributed controller implementation - Algorithm for agent $i$}\label{alg:dist_implementation}
	\begin{algorithmic}[1]
		\Require 
        An i.f.g.\ $\ell_p$-stable system $\mathbf{G}$ split in $M$ subsystems $\Gb^i$ such that $\mathbf{G}^i\in\Lp$ and $\Gb^i$ are i.f.g. $\ell_p$-stable.
        A densest admissible control communication topology $\mathcal{T}\leq \mathcal{D}(\Gb)$.
		An operator $\Qb$ split in $M$ subsystems $\Qb^i\in\Lp$ such that $\mathcal{D}(\Qb)\leq \mathcal{T}$.
		\For{$t=0,1,\dots$}
		\State Measure (and store) the local output $y_t^i$.
		\State Reconstruct (and store) the signal $\omega_t^i$ as per
		\begin{equation}
            \label{eq:omega_algorithm}
			\omega_t^i = y_t^i - G_t^i\left(\{u_{t-1:0}^{\mathrm{o},k}\}_{k\in\mathcal{N}^-_G(i)}\right)\,.
		\end{equation}
		\State Send $\omega_{t}^i$ to $\bar{\mathcal{N}}^+_{\mathcal{Q}}(i)$ i.e., the out-neighbors of subsystem $i$ according to $\mathcal{D}(\boldsymbol{\mathcal{Q}})$.
		\State Receive (and store) $\omega_{t}^j$ from $j\in\bar{\mathcal{N}}^-_{\mathcal{Q}}(i)$.
		\State Calculate (and store) the $i^{\text{th}}$ output as per 
		\begin{equation}\label{eq:algorithm_u}
			u_t^{\mathrm{o},i} = \mathcal{Q}^i_t\left(\{\omega_{t:0}^k\}_{k\in\mathcal{N}^-_{\mathcal{Q}}(i)}\right) \,.
		\end{equation}
		\State Send $u_{t}^{\mathrm{o},i}$ to $\bar{\mathcal{N}}^+_{G}(i)$ i.e., the out-neighbors of subsystem $i$ according to $\mathcal{D}(\mathbf{G})$.
		\State Receive (and store) $u_{t}^{\mathrm{o},j}$ from $j\in\bar{\mathcal{N}}^-_G(i)$.
		\EndFor
	\end{algorithmic}
\end{algorithm}

The controller implementation in Algorithm~\ref{alg:dist_implementation}
requires two rounds of computations and communications within each sampling interval.
Indeed, after computing locally the variables $\omega_t^i$ in \eqref{eq:omega_algorithm}, these signals are transmitted according to the communication topology described by $\mathcal{D}(\boldsymbol{\mathcal{Q}})$.
Thanks to the information received from the in-neighbors, i.e., $\omega^j_t$ from $j\in\bar{\mathcal{N}}^-_{\mathcal{Q}}(i)$,
each local controller can  compute the local output, $u_t^{\mathrm{o},i}$.
These are then transmitted to the neighbors defined by the adjacency matrix $\mathcal{D}(\mathbf{G})$, which mirrors the physical interconnection of the subsystems. We highlight that this round of communication is always possible because Proposition~\ref{prop:distributed} guarantees $ \mathcal{D}(\Gb)\leq\mathcal{D}(\Kb)$.

Let us also provide a more specific example by considering the networked control system in Figure~\ref{fig:distributed_example} and setting $\mathcal{D}(\Gb) = \mathcal{T}$, and $\mathcal{D}(\Kb) = \mathcal{D}(\Qb)= \mathcal{T}$. 
Let us focus on the subsystem and controller $i=1$. 
At time $t$, 
in order to compute $u^{\mathrm{o},1}_t =\mathcal{Q}_t^1(\omega_{t:0}^1,\omega_{t:0}^3)$,
the controller must calculate $\omega_t^1$, and 
receive $\omega_{t}^3$.
To calculate $\omega_t^1$, the local controller 
must evaluate the local dynamics 
$G_t^1(u^{\mathrm{o},1}_{t-1:0} , u^{\mathrm{o},3}_{t-1:0} )$; 
which in turn requires first receiving the signal $u^{\mathrm{o},3}_{t-1}$.
Note that the value of $\omega_{t}^3$ must first be computed locally by the controller at location 3 before it can be transmitted to controller 1.
The right panel of Figure~\ref{fig:distributed_example} provides a graphical representation of these computational steps.

In several applications, however, at most one communication round is allowed per time step. In this case, one can use the following more restrictive results for building a distributed stability-preserving controller $\Kb$ complying with a pre-specified maximal admissible communication topology.

\begin{corollary}
    \label{cor:distributed} Let $\Gb$ and $\mathcal{T}$ be defined as in
     Proposition~\ref{prop:distributed}. 
    Assume that at most one communication round per time step is allowed and that $\Qb$ in \eqref{eq:controller_youla} is decentralized, i.e. it consists of $M$ operators $\Qb^i \in \Lp$ and verifies $\mathcal{D}(\Qb) =I_M$. Moreover, let $\Kb$ be the controller given by \eqref{eq:controller_youla} and implemented as in Algorithm~\ref{alg:dist_implementation} where \eqref{eq:algorithm_u} is replaced by 
$u_t^{\mathrm{o},i} = \mathcal{Q}^i_t\left(\omega_{t:0}^i\right)$. Then, $\Kb$ verifies 
    \begin{equation}
        \mathcal{D}(\Kb) = \mathcal{D}(\Gb) \leq \mathcal{T}\,,
    \end{equation}
    and guarantees that the closed-loop system is $\ell_p$-stable.
\end{corollary}
The proof follows from Proposition~\ref{prop:distributed} by setting $\mathcal{D}(\Qb) = \I_M$.

\section{Closed-loop maps parametrizations in presence of measurable disturbances}\label{sec:measured_disturbances}

The control architecture in Figure~\ref{fig:blockdiagram_CL_with_stabK} hinges on the reconstruction of the sequence $\omegab$ which, according to \eqref{eq:omega_prestabilized}, is a nonlinear combination of $\yb$ and $\tub$; thus implicitly, a function of the
the disturbances $\vb$ and $\db$.
We now consider the scenario where the input disturbance $\db$ can be measured (or it represents a known reference signal), implying that the controller $\Kb$ can internally reconstruct the input signal $\ub$.

Compared to the previous section, we increase the information available to the controller at a given time instant, which requires additional sensors or system knowledge.
On the other hand, this setup will allow enforcing not only $\Lp$ stability on the closed-loop maps but also stronger properties such as Lipschitzness and exponential stability.
Moreover, it enables the design of a control scheme for the isolation of disturbances in a distributed setting (see Section~\ref{sec:disturbance_localization}).

\begin{figure}
	\centering
	\includegraphics{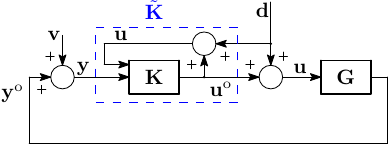}
    \caption{Scheme of the closed-loop system with disturbance measurements considered in Section~\ref{sec:measured_disturbances}.}
	
	\label{fig:blockdiagram_io_d}
\end{figure}

\begin{figure}
	\centering
	\includegraphics{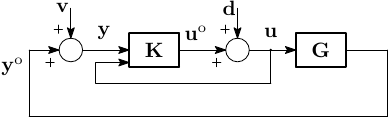}
	\caption{Equivalent scheme of the closed-loop system in Figure~\ref{fig:blockdiagram_io} with disturbance measurements.}
	\label{fig:blockdiagram_io}
\end{figure}

In this section, we consider controllers of the form $\tilde{\Kb}(\yb;\db)\in\Ccs$, allowing the control law to depend upon the outputs of the system as well as previous values of the disturbance $\db$, as shown in Figure~\ref{fig:blockdiagram_io_d}.
Equivalently, and with a slight abuse of notation compared to the previous sections, we can identify the controller with the operator $\Kb(\yb;\ub)\in\Ccs$ appearing in Figure~\ref{fig:blockdiagram_io_d}. Indeed, as it is clear from the block diagram, there is a one-to-one mapping between the operators $\tilde{\Kb}$ and $\Kb$.
This is further highlighted by the closed-loop system in Figure \ref{fig:blockdiagram_io}, which is an equivalent representation of the diagram in Figure~\ref{fig:blockdiagram_io_d}.
Note that if $\db=\Ob$, the block diagram reduces to the one in Figure~\ref{fig:blockdiagram_o}.

The definition of the achievable closed-loop maps, as per Definition~\ref{def:achievable_CLM} remains the same, except that model \eqref{eq:system_control} is replaced by 
\begin{subequations}
\label{eq:system_control_Kyu}
	\begin{align}
		\label{eq:system_Kyu}
		\yb &= \Gb(\ub) + \vb \,, \\
		\label{eq:K_control_yu}
		\ub &= \Kb(\yb;\ub) + \db\,, \quad \Kb\in\Ccs\,.
	\end{align}
\end{subequations}


The next theorem provides a parametrization of all closed-loop maps 
that can be achieved by using the control architecture \eqref{eq:system_control_Kyu} shown in Figure~\ref{fig:blockdiagram_io}.

\begin{theorem}
	\label{th:achievability_measured_disturbances}
	For a system $\Gb$, the set of all closed-loop maps achievable by~\eqref{eq:system_control_Kyu} can be represented as
    \begin{subequations}
     \label{eq:SLSio}
	\begin{align}
		\hCL{\Gb} = & \{\Psiyu = (\Psiy; \Psiu) = (\tPsiy; \tPsiu) + \I ~~| \\
		&\quad \tPsiu  \in \Ccs\,, \label{eq:SLSio1}\\
		&\quad \tPsiy = \Gb \Psiu \label{eq:SLSio2} 
		\}\,.
	\end{align}
    \end{subequations}
	Moreover, 
    \begin{equation}
        \label{eq:K_io}
        \Kb = \tPsiu \Psiyu^{-1}
    \end{equation}
    is the unique controller achieving the closed-loop map $\Psiyu$.
\end{theorem}
The proof can be found in Appendix~\ref{app:proof_achievability_measured_disturbances}.
It is worth noticing that due to the 
use of $\db$ in $\Kb$, the achievability constraints for the closed-loop maps are simplified compared to Theorem~\ref{th:achiev_of}.
Indeed, the constraint \eqref{eq:SLSo3} is not needed anymore.
This observation has two key implications. First, one can directly use the operator $\tPsiu\in\Ccs$ for describing all closed-loop maps without the need for computing operator inverses (see Theorem~\ref{th:SLS_M_2}). Second, this setting allows enforcing additional properties on the set of closed-loop maps (see Sections \ref{sec:exp_stability} and \ref{sec:disturbance_localization}).

Similar to the case of the controller in Proposition~\ref{th:achiev_of}, 
in order to obtain $\Kb$ in~\eqref{eq:K_io},
we need the inverse of $\Psiyu$.
We rely again on  Proposition~\ref{prop:invertibility} to avoid this calculation and obtain a recursive method for computing the control variable.
The resulting implementation of the controller is given in the next proposition and, similarly to Propositions~\ref{prop:rec_implemenatation_KQ} and \ref{prop:rec_implemenatation_KQ_with_base}, paves the way to the design of neural network controllers (see Section~\ref{sec:NOC_RENs}).
\begin{proposition}[Recursive implementation of the controller \eqref{eq:K_io}]
\label{prop:rec_implemenatation_Kio}
    Given a system $\Gb$ and the feedback architecture \eqref{eq:system_control_Kyu}, the output $\tub = \Kb(\yb;\ub)$, with $\Kb$ as in~\eqref{eq:K_io}, can be computed recursively through the equations
    \begin{subequations}
    \label{eq:rec_impl_disturbances}
    \begin{align}
    	\delta_{t-1} &= u_{t-1} - {\Psi}^{u^{\mathrm{o}}}_{t-1}(\beta_{t-1:0}, \delta_{t-2:0}) \,, \label{eq:rec_impl_disturbances_1} \\
    	\beta_t &= y_t - {\Psi}^{y^{\mathrm{o}}}_t(\beta_{t-1:0}, \delta_{t-1:0}) \,, \label{eq:rec_impl_disturbances_2}\\
    	u^{\mathrm{o}}_t &=  {\Psi}^{u^{\mathrm{o}}}_t(\beta_{t:0}, \delta_{t-1:0}) \,. \label{eq:rec_impl_disturbances_3}
    \end{align}
    \end{subequations}
\end{proposition}
The proof can be found in Appendix~\ref{app:proof_rec_implementation_Kio}.
Note that \eqref{eq:rec_impl_disturbances} is a non-Markovian dynamical system with input $(\yb;\ub)$, output $\tub$ and internal state $(\boldsymbol{\beta}; \boldsymbol{\delta})$.

Having parametrized all possible achievable closed-loop maps, we now focus our attention on those that are stable; in particular, we would like to parametrize the following set:
\begin{equation}
	\label{eq:CLp_measured_disturbances}
	\hCLp{\Gb} = \{\Psiyu\in\hCL{\Gb} ~|~\Psiyu\in\Lp \}\,.
\end{equation}
When the plant is itself a stable operator, the next theorem shows that one can use $\tPsiu\in\Lp$ as a free parameter.

\begin{theorem}
	\label{th:stability_measured_disturbances}
	For a system $\Gb\in\Lp$,  
	the set $\hCLp{\Gb}$ in \eqref{eq:CLp_measured_disturbances} can be written as
    \begin{subequations}
    \label{eq:SLSio_stable} 
	\begin{align}
		\hCLp{\Gb} = & \{\Psiyu = (\Psiy; \Psiu) = (\tPsiy; \tPsiu) + \I ~~| \\
		&\quad \tPsiu \in\Ccs\,, ~~ \tPsiu  \in \Lp \,,   \\
		&\quad  \tPsiy = \Gb\Psiu 
		\}\,.
	\end{align}
    \end{subequations}
\end{theorem}
The proof can be found in Appendix~\ref{app:proof_stability_measured_disturbances} and relies on the fact that the composition of $\Lp$ operators remains in $\Lp$.

Next, we show how to impose further constraints on the closed-loop system.
In particular, we will focus on methods for guaranteeing exponential stability or i.f.g. $\ell_p$-stability of the closed-loop maps. 
Finally, in Section~\ref{sec:disturbance_localization}, we show how to design $\Psiyu$ that allows for disturbances isolation when working in a distributed setting.

\subsection{Exponential and i.f.g. \texorpdfstring{$\ell_p$}{lp}-stability of closed-loop maps}
\label{sec:exp_stability}
First, we define exponentially stable signals and operators. Then, we show that it is straightforward to guarantee closed-loop  exponential stability by imposing 
the same requirement to the plant $\Gb$ and the free parameter
$\tPsiu$.

\begin{definition}[Exponentially decaying sequence]
	A sequence $\xb\in\lpe{m}$ is exponentially decaying 
    if there exists $k>0$ and $\alpha \in (0, 1)$ such that
	\begin{equation*}
		\lvert x_t\rvert \leq k\alpha^{t}  \,,
	\end{equation*}
	for all $t=0,1,\dots$.
	The set of exponentially decaying signals is denoted with $\ell^m_{exp}$.
\end{definition}

\begin{definition}[Exponentially stable operator]
	An operator $\Ab:\lpe{n} \rightarrow \lpe{m}$ is exponentially stable if for all $\xb\in\ell_{exp}^n$ we have that $\Ab\xb\in\ell_{\text{exp}}^m$.
	The set of exponentially stable operators is denoted with $\Lexp$.
\end{definition}
For the sake of completeness, in Appendix~\ref{app:exponential_stability}, we show that exponentially stable operators are closed under the summation and composition.

The next theorem characterizes the set of all exponentially stable achievable closed-loop maps for a system $\Gb\in\Lexp$, i.e., we provide a parametrization of
\begin{equation}\label{eq:set_exponentially_stable_short}
    \CL{\Gb}_{\text{exp}} = \{ \Psiyu \in \hCL{\Gb} ~|~ \Psiyu\in\Lexp \} \,.
\end{equation}

\begin{theorem}
	\label{th:exp_stab}
    For a system $\Gb\in\Lexp$, 
	the set 
    $\CL{\Gb}_{\text{exp}}$ in \eqref{eq:set_exponentially_stable_short}
    can be written as
    \begin{subequations}
    \label{eq:SLSio_exp_stable}
	\begin{align}
		\CL{\Gb}_{\text{exp}} = & \{\Psiyu = (\Psiy; \Psiu) = (\tPsiy; \tPsiu) + \I ~~| \label{eq:SLSio_exp_stable_1} \\
		&\quad \tPsiu  \in \Ccs\,,  ~\tPsiu  \in \Lexp\,, \label{eq:SLSio_exp_stable_2}\\
		&\quad \tPsiy =  \Gb \Psiu  \label{eq:SLSio_exp_stable_3} 
		\}\,.
	\end{align}
    \end{subequations}
\end{theorem}
The proof is given in Appendix~\ref{app:proof_exp_stab}.

We can also impose the Lipschitzness of the closed-loop maps, or more generally, i.f.g. $\ell_p$-stability. 
The next theorem characterizes the set of all i.f.g. $\ell_p$-stable achievable closed-loop maps for an i.f.g. $\ell_p$-stable plant $\Gb$, i.e., we provide a parametrization of the set
\begin{equation}\label{eq:set_ifg_stable_short}
    \CL{\Gb}_{i.f.g} = \{ \Psiyu \in \hCL{\Gb} ~|~ \Psiyu\text{ is } i.f.g.\ \ell_p\text{-stable} \} \,.
\end{equation}

\begin{theorem}
	\label{th:lipschitzness}
    For an i.f.g. $\ell_p$-stable system $\Gb$,
	the set 
    $\CL{\Gb}_{i.f.g}$ in \eqref{eq:set_ifg_stable_short}
    can be written as
    \begin{subequations}
    \label{eq:SLSio_ifg_stable}
	\begin{align}
		\CL{\Gb}_{i.f.g} = & \{\Psiyu = (\Psiy; \Psiu) = (\tPsiy; \tPsiu) + \I ~~| \\
		& \tPsiu  \in \Ccs\,,  ~\tPsiu \text{ is i.f.g. } \ell_p\text{-stable}\,,    \\
		& \tPsiy =  \Gb \Psiu  
		\}\,.
	\end{align}
    \end{subequations}
\end{theorem}
The proof is provided in Appendix~\ref{app:proof_lipschitzness}.

The parametrizations of Theorems~\ref{th:exp_stab} and \ref{th:lipschitzness} highlight that, 
given a system $\Gb$ with specific stability properties (such as exponentially stable, or i.f.g. $\ell_p$-stability),
it is possible to construct achievable closed-loop maps $(\tPsiy;\tPsiu)$ for that system while preserving these characteristics.
This can be obtained by freely selecting an operator $\tPsiu$ that shares the same stability properties.

\subsection{Disturbance localization}
\label{sec:disturbance_localization}

In a distributed setting, as the one introduced in Section~\ref{sec:extension_distributed}, disturbance localization allows confining the effect of disturbances to a local set of subsystems, preventing that subsystems located \emph{far away} from the disturbance generation point are affected.
For example, given a graph of interconnected subsystems as in Section~\ref{sec:extension_distributed}, we seek closed-loop maps ensuring that disturbances entering at location $i$ can only affect system $i$ and its \emph{direct} out-neighbors.
For LTI systems, disturbance localization has been well studied under the SLS framework in~\citet{wang2019sls, wang2018separable}. 
In this section, we analyze the same property in the nonlinear setting, 
i.e., we characterize a subset of the achievable closed-loop maps that allows the isolation of disturbances.

Let us assume, as in Section~\ref{sec:extension_distributed}, that the system $\Gb$ is composed of $M$ subsystems, 
and can be written as in~\eqref{eq:split_y_neighbors}.
Moreover, we adopt the same notation for binary matrices used in Section~\ref{sec:extension_distributed}. Specifically,
we assume that the system $\Gb$ has a sparsity pattern given by the adjacency matrix $\mathcal{D}(\mathbf{G})$.
The next proposition shows that the sparsity pattern of the mapping $\Psiy$ from disturbances to outputs is shaped by the sparsity of the system and the sparsity of the free parameter $\Psiu$ (which is the map from disturbances to inputs).

\begin{proposition}
	\label{prop:disturbance_localization}
	Consider a system $\Gb$ whose sparsity is characterized by $\mathcal{D}(\Gb) \in \{0,1\}^{M\times M}$, 
	and a matrix $S_u \in \{0,1\}^{M\times M}$. Assume $[\mathcal{D}(\Gb)]_{i,i} = 1$ and $[S_u]_{i,i} = 1$ for $i=1,2,\dots,M$.
	Then any operator $\tPsiu \in \Ccs$ such that $\mathcal{D}(\tPsiu) = S_u$ guarantees that the sparsity pattern of the 
	closed-loop maps $\Psiu$ and $\Psiy$, have the following properties:
	\begin{itemize}
		\item 
		$\mathcal{D}(\Psiu) = S_u$,
		\item 
		$\mathcal{D}(\Psiy) = S_u  \mathcal{D}(\Gb)$.
	\end{itemize}
\end{proposition}
The proof is given in Appendix~\ref{app:proof_disturbance_localization}.

\section{Parametrization of stable operators for nonlinear optimal control problems}
\label{sec:NOC_RENs}
In this section, we show how the parametrizations of stabilizing controllers provided in Sections~\ref{sec:stable_CLM} and \ref{sec:measured_disturbances} can be used for addressing output-feedback NOC problems. 
Moreover, we will show that there are classes of deep neural networks (DNNs) that allow implementing the stable operators $\Qb$ and $\tPsiu$ and act as ``degrees of freedom'' in the controller parametrizations.

Our goal is to synthesize a control policy for a given system $\Gb$, 
such that $\Psiyu\in\CLp{\Gb}$ (or $\Psiyu\in\hCLp{\Gb}$), i.e., closed-loop $\ell_p$-stability is enforced as a \textit{hard} constraint.
Moreover, we aim at devising a fail-safe design procedure, meaning that closed-loop stability holds for any policy generated at every step of the optimization process. To address this requirement, we employ the parametrizations introduced in Sections~\ref{sec:stable_CLM} and \ref{sec:measured_disturbances}.
The second condition involves minimizing a loss function:
\begin{equation}
    \label{eq:cost_J}
    J = \mathbb{E}_{v_{T:0}, d_{T:0}} \left[ L(y_{T:0}, u_{T:0}) \right] \,,
\end{equation}
where $L$ is a piece-wise differentiable function such that $L(y_{T:0}, u_{T:0})\geq 0$ for any input.\footnote{Another common approach is to use $\max_{v_{T:0},d_{T:0}}[\cdot]$ instead of the expectation, while assuming that disturbances belong to a bounded set. Other useful choices  include $\operatorname{Var}_{v_{T:0},d_{T:0}}[\cdot]$, $\operatorname{CVAR}_{v_{T:0},d_{T:0}}[\cdot]$, and weighted combinations of all the above. As explained in Section \ref{subs:mapping_NOC}, the proposed framework only requires the ability to approximate the chosen operator --- used to remove the effect of disturbances from the cost --- by performing multiple experiments.}
Unlike the stability constraint, this optimization objective is treated as a \textit{soft} constraint, as standard in DNN training. 
We do not expect gradient-based methods to achieve a globally optimal solution for all disturbance sequences, as such guarantees are generally unattainable for problems beyond Linear Quadratic Gaussian (LQG) control --- which enjoy convexity of the cost and linearity of the optimal policies~\cite{tang2021analysis,furieri2020first}.

We are now ready to formulate the NOC problem as:
\begin{alignat}{3}
\operatorname{NOC:}~~~&\min_{\mathbf{K}(\cdot)}&& \quad \mathbb{E}_{v_{T:0}, d_{T:0}}\left[L(y_{T:0},u_{T:0})\right] \label{eq:NOC_problem}\\
&~\operatorname{s.t.}~~ && \text{closed-loop dynamics}\,, \nonumber \\
&~~&& (\Phiy{G}{K};\Phiu{G}{K}) \in \Lp\,, \nonumber
\end{alignat}
where the closed-loop dynamics are either \eqref{eq:system_control} or \eqref{eq:system_control_Kyu}.
Searching over the space of stabilizing control policies leads to intractable optimization problems in general.
Here, similar to the SLS approach \cite{anderson2019system, ho2020system, Furieri22b}, the idea is to circumvent the difficulty of characterizing stabilizing controllers by instead directly designing stable closed-loop maps making use of the parametrizations presented in Sections~\ref{sec:stable_CLM} and \ref{sec:measured_disturbances}.
Specifically, we can equivalently rewrite the NOC problem 
by searching over operators $(\Psiy;\Psiu)$ that are stable closed-loop maps achieved by $\Gb$, i.e., searching either in the set \eqref{eq:Phi_ach_stab} or in \eqref{eq:CLp_measured_disturbances}, depending on the considered feedback architecture (\eqref{eq:system_control} or \eqref{eq:system_control_Kyu}, respectively).
Then, one has
$y_t = \Psi^y_t(v_{t:0}, d_{t-1:0})$ and
$u_t = \Psi^u_t(v_{t:0}, d_{t:0})$,
and the NOC problem can be written as
\begin{alignat}{3}
\operatorname{N-SLS}_{\Psiyu}: ~&\min_{(\Psiy,\Psiu)}&& \quad \mathbb{E}_{v_{T:0}, d_{T:0}}\left[L(y_{T:0},u_{T:0})\right] \nonumber \\
&~~~\operatorname{s.t.}~~ && y_t = \Psi^y_t(v_{t:0}, d_{t-1:0})\,, \nonumber \\
& ~~ && u_t = \Psi^u_t(v_{t:0}, d_{t:0})\,, \forall t = 0,1,\ldots \,,\nonumber
\end{alignat}
with the additional constraint $(\Psiy;\Psiu)\in \CLp{\Gb}$ if considering the feedback architecture \eqref{eq:system_control},  or $(\Psiy;\Psiu)\in \hCLp{\Gb}$ for \eqref{eq:system_control_Kyu}.
As remarked in Sections~\ref{sec:achievable_CLM} and \ref{sec:measured_disturbances}, this last constraint of the
N-SLS$_{\Psiyu}$ problem
cannot be directly used in computations, and the next step is to get rid of it by using Theorem~\ref{th:stabilizing_SLS} and Theorem~\ref{th:stability_measured_disturbances}.

More precisely, 
for the case of the feedback architecture \eqref{eq:system_control}, when a base controller is available, 
we can use the parametrization of Theorem~\ref{th:stabilizing_SLS} and the recursive formulae \eqref{eq:K_stable} to write the
N-SLS$_{\Psiyu}$ problem
as
\begin{alignat}{3}
\operatorname{N-SLS}_{\Qb}: ~&\min_{\Qb\in\Lp}&& \quad \mathbb{E}_{v_{T:0}, d_{T:0}}\left[L(y_{T:0},u_{T:0})\right] \nonumber\\
&~~~\operatorname{s.t.}~~ && y_t = G_t(u_{t-1:0}) + v_t\,, \nonumber \\
& ~~ && \tilde{\omega}_t = y_t - G_t(u^\mathrm{o}_{t-1:0}) \,, \nonumber \\
& ~~ && u_t = \mathcal{Q}_t(\tilde{\omega}_{t:0}) + K'_t(y_{t:0}) + d_t\,, \nonumber \\
& ~~ && ~~~~~ \forall t = 0,1,\ldots  \nonumber
\end{alignat}
Similarly, when using the feedback architecture \eqref{eq:system_control_Kyu}, 
one can use Theorem~\ref{th:stability_measured_disturbances} to rewrite 
the
N-SLS$_{\Psiyu}$ problem
as
\begin{alignat}{3}
\operatorname{N-SLS}_{\tPsiu}: ~&\min_{\tPsiu\in\Lp}&& \quad \mathbb{E}_{v_{T:0}, d_{T:0}}\left[L(y_{T:0},u_{T:0})\right] \nonumber\\
&~~~\operatorname{s.t.}~~ && \tPsiy = \Gb\Psiu\,, \nonumber \\
& ~~ && y_t = \Psi^y_t(v_{t:0},d_{t-1:0}) \,,\nonumber \\
& ~~ && u_t = \Psi^u_t(v_{t:0},d_{t:0}) \,, \forall t = 0,1,\ldots \nonumber
\end{alignat}
Solving the N-SLS$_{\Qb}$ and N-SLS$_{\tPsiu}$ problems depends on our ability to search in the set of $\Lp$ operators, and, practically, on how effectively we can calculate the expectation.
For a tractable implementation, we shift to using finite-dimensional Markovian operators and approximate the expected value by using samples of $v_{T:0}, d_{T:0}$.

\begin{remark}
Note that $\Lp$ operators can be discontinuous functions of their inputs. This property allows the proposed method to accommodate discontinuous stabilizing feedback policies, which may be necessary for optimality in certain contexts, such as in the continuous-time Brockett integrator \cite{astolfi1998discontinuous}. This observation highlights an interesting direction for future work, that is, parametrizing discontinuous operators in $\Lp$. However, in this work, we focus on existing parametrizations of stable dynamical operators that are continuous functions of their inputs.
\end{remark}

When linear systems are considered, one can search over Finite Impulse Response (FIR) transfer matrices, expressed as $\mathbf{M} = \sum_{i=0}^N M_iz^{-i} \in \mathcal{TF}_s$, where $\mathcal{TF}_s$ stands for the space of all stable transfer matrices. By optimizing over the real matrices $M_i$, progressively less conservative solutions can be achieved by increasing the FIR order $N$.
In the nonlinear case, \citet{KimPatronBraatz18,revay2021RENs,martinelli2023unconstrained} 
have recently introduced finite dimensional DNN approximations for certain classes of nonlinear $\Ltwo$ operators. 
In the next section, we briefly review the nonlinear models proposed by~\citet{revay2021RENs} that can be used to freely parametrize subsets of  Markovian operators in $\Ltwo$.
Moreover, since these systems embed arbitrarily deep NNs, they are flexible tools for representing $\Ltwo$ operators.
This observation is corroborated by the examples in Section~\ref{sec:numerical}.

\subsection{Brief introduction to RENs for parametrizing \texorpdfstring{$\Ltwo$}{L2} operators}
\label{sec:intro_REN}

The effectiveness of our approaches (i.e., of solving the N-SLS$_{\Qb}$ and N-SLS$_{\tPsiu}$ problems) hinges on the ability to parametrize $\Lp$ operators. A major challenge lies in the fact that the space $\Lp$ is infinite-dimensional. Consequently, practical implementations typically involve restricting the search to subsets of $\Lp$ characterized by a finite number of parameters.

RENs, as introduced by \citet{revay2021RENs}, are finite-dimensional Markovian approximators of nonlinear operators. 
An operator $\Enne$ is a REN if the relationship $\hat{\yb} = \Enne\hat{\ub}$ is generated by the following dynamical system:
\begin{subequations}
\label{eq:REN}
\begin{equation}
	\label{eq:REN_linear_part}
	\begin{bmatrix}
		{\xi}_{t+1} \\
		\zeta_{t} \\
		\hat{u}_{t}
	\end{bmatrix}
	= 
	\overbrace{{\begin{bmatrix}
				A & B_1 & B_2 \\
				C_1 & D_{11} & D_{12} \\
				C_2 & D_{21} & D_{22}
	\end{bmatrix}}}^{\hat{W}}
	\begin{bmatrix}
		\xi_t \\
		w_t \\
		\hat{y}_t
	\end{bmatrix}
	+ 
	\overbrace{\begin{bmatrix}
			0 \\
			0 \\
			b_t
	\end{bmatrix}}^{\hat{b_t}}\,,
\end{equation}
\begin{equation}
	\label{eq:REN_nonlinear_part}
	w_t = \sigma ( \zeta_t)\,,
\end{equation}
\end{subequations}
where $\xi_t\in\mathbb{R}^{q_1}$, $\zeta_t,w_t \in\mathbb{R}^{q_2}$, $\hat{u}_t\in\mathbb{R}^{q_{out}}$, $\hat{y}_t\in\mathbb{R}^{q_{in}}$, the activation function $\sigma:\mathbb{R}\rightarrow\mathbb{R}$ is applied element-wise, and with initial condition $\xi_0\in\mathbb{R}^{q_1}$.
Further, $\sigma(\cdot)$ must be piecewise differentiable and verify $0 \leq \frac{\sigma(z) - \sigma(\tilde{z})}{z - \tilde{z}} \leq 1, \quad \forall z, \tilde{z} \in \mathbb{R}, \, z \neq \tilde{z}$.
This implies  $\sigma(0)=0$ and  $\frac{d \sigma}{d z}\in [0, 1]$.
The vector $\hat{b}_t$ represents the bias term of the REN architecture. 
Different from \citet{revay2021RENs}, where it is assumed that the bias is a time-invariant trainable vector, here we allow it to be time-varying, as far as it is an $\ell_2$ sequence.
For instance, in one of the experiments in Section~\ref{sec:numerical}, we set it to be a trainable sequence for $t=0,\ldots,T$, and $b_t=0$, for $t>T$.
As noted by \citet{revay2021RENs}, RENs include many existing DNN architectures. In general, RENs define deep equilibrium network models due to the implicit relationships 
between the signals involved in \eqref{eq:REN}. 
By restricting $D_{11}$ to be strictly lower-triangular, $\zeta_t$ in \eqref{eq:REN} can be computed explicitly, thus significantly speeding up computations. 
To exemplify the modeling power offered by RENs, as shown in \citet{furieri2024performance}, one can utilize \eqref{eq:REN} to construct nonlinear systems of the form:
\begin{align*}
    \xi_{t+1} &= \hat A \xi_{t}+\hat B \,\text{NN}^{\xi} (\xi_{t},\hat{y}_{t}) \,,\\  
    \hat u_t &= \hat C\xi_t + \hat D \,\text{NN}^{\hat u}(\xi_{t},\hat{y}_{t})\,,
\end{align*}
where $\hat{A}$, $\hat{B}$, $\hat{C}$, $\hat{D}$ are arbitrary matrices of suitable dimensions and $NN^\star$, $\star\in\{\xi, \hat{u}\}$, are neural networks of depth $L$ defined by the layer equations
\begin{align*}
    \tilde{z}_{0,t}^\star &= (\xi_{t},\hat{u}_t), \\
    \tilde{z}_{k+1,t}^\star &= \sigma(W_k^\star \tilde{z}_{k,t}^\star+b_k^\star),\quad k=0,\ldots, L-1 \,,
\end{align*}
where $W_k^\star$ and $b_k^\star$ are weights and biases, respectively, and $\tilde z_{L,t}^\star$ is the NN output.

Next, we address the problem that, for an arbitrary choice of $(\hat{W},\hat{b})$, the map $\Enne$ induced by \eqref{eq:REN} may not lie in $\Lp$. 
To this purpose, we introduce the following properties.
\begin{definition}[\citet{revay2021RENs}]
    \label{def:contractivity}
    For the system \eqref{eq:REN} let $\xi_t^\star$ and $\hat u_t^\star$ be the state and output trajectories generated by $\xi_0^\star$ and $\hat y_t^\star$, where $\star\in\{a,b\}$. The system 
    \begin{enumerate}
        \item is contractive with rate $\alpha \in (0,1)$ if there is $\kappa>0$ such that for any $\xi_0^a$ and $\xi_0^b$ and $\hat{y}^a_t=\hat{y}^b_t$ one has
        \begin{equation*}
            |\xi_t^a- \xi_t^b|_2\leq\kappa \alpha ^t |\xi_0^a- \xi_0^b|_2 \,,
        \end{equation*}
        where $|\cdot|_2$ is the 2-norm.
        \item has an incremental $\ell_2$-gain $\beta>0$ if, for all $T>0$
        \begin{equation}
        \label{eq:iL2-gain_time}
            \sum_{t=0}^{T}- \frac{1}{\beta}|\hat u_t^a-\hat u_t^b|_2^2 +\beta  | \hat{y}^a_t-\hat{y}^b_t|_2^2 \geq \,,
    -d(\xi_0^a,\xi_o^b)
\end{equation}
for some function $d(\xi, \tilde \xi) \geq 0$ with $d(\xi, \xi) = 0$.
    \end{enumerate}
\end{definition}
Note that \eqref{eq:iL2-gain_time} implies that 
\begin{equation}
\label{eq:REN_iL2_gain}
    \|\hat{\ub}^a-\hat{\ub}^b\|_2\leq \beta^2\|\hat{\yb}^a-\hat{\yb}^b\|_2+\beta d(\xi_0^a,\xi_o^b)
\end{equation} 
and, therefore, when the initial state of the controller is fixed, i.e. $\xi_0^a=\xi_o^b=\bar{\xi}_0$, \eqref{eq:REN} defines an operator $\hat\yb\mapsto \hat \ub$ that, according to Definition \ref{def:ifg_stab}, is i.f.g.\ $\ell_2$-stable with incremental $\ell_2$-gain $\beta^2$.

The breakthrough of~\citet{revay2021RENs} is to provide explicit smooth mappings
$\Theta_c,\Theta_{i\ell_2}:\mathbb{R}^d \rightarrow \mathbb{R}^{(q_1+q_2+q_{out})\times(q_1+q_2+q_{in})}$ from unconstrained optimization parameters
$\theta\in \mathbb{R}^{d}$ 
to the matrix 
$\hat{W}$
defining \eqref{eq:REN}, 
such that, for any $\theta$,  (i) $\Theta_c(\theta)$ generates a REN \eqref{eq:REN} that  is  contractive with prescribed rate $\alpha$ and 
 has a finite, albeit \textit{a priori} unknown, incremental $\ell_2$-gain. (ii) $\Theta_{i\ell_2}$ generates a REN \eqref{eq:REN} which is contractive with prescribed rate $\alpha$ and finite incremental $\ell_2$-gain $\beta$. 
Since no constraints are imposed on $\theta$ we call $\Theta_c$ and $\Theta_{i\ell_2}$ \textit{free} parameterizations.\footnote{We defer the reader to \citet{revay2021RENs} for the explicit definitions of these mappings.}

Next, we show that RENs parametrized by $\Theta_c$ or $\Theta_{i\ell_2}$ define $\mathcal{L}_2$ operators.
\begin{proposition}
\label{prop:REN_L2}
Assume that the REN  \eqref{eq:REN} has an incremental $\ell_2$-gain $\beta>0$. If $\hat\bb\in\ell_2$, then, the map $\hat{\yb}\mapsto\hat\ub$ induced by \eqref{eq:REN} is in $\mathcal{L}_2$.  
\end{proposition}
\begin{pf}
Since $\sigma(0)=0$, when $\hat y_t=0$ one has that $\xi_t=0$, $w_t=0$, $\forall t\geq 0$ verify \eqref{eq:REN}. Moreover, the corresponding output is $\hat u_t=\hat b_t$. Setting $\xi_0 ^b=0$ and $\hat y_t^b=0$ in \eqref{eq:iL2-gain_time},  one obtains from \eqref{eq:REN_iL2_gain} that 
\begin{equation}
\label{eq:REN_L2_proof}
    \|\hat{\ub}^a-\hat \bb\|_2\leq \beta^2\|\hat{\yb}^a\|_2+\beta d(0,\xi_o^b)
\end{equation} 
Therefore $\hat \yb^a\in\ell_2$ implies $\ub^a-\hat \bb \in\ell_2$ and, hence, $\ub^a\in \ell_2$.
\end{pf}

Proposition \ref{prop:REN_L2} implies that the free parameterizations $\Theta_c$ and $\Theta_{i\ell_2}$ can be used to describe the stable operators $\Qb$  or $\tPsiu$ needed in the N-SLS$_{\Qb}$ and N-SLS$_{\tPsiu}$ problems, respectively.
\begin{remark}
\label{rmk:other_L2_NN}
    As highlighted in the introduction, besides RENs there are other dynamical models providing free parameterizations on $\mathcal{L}_2$ operators, e.g., some classes of 
    SSMs \cite{forgione2021dynonet, gu2022efficiently} 
    and NN architectures based on port-Hamiltonian systems \cite{zakwan2024neural}. Moreover any freely parametrized family of static functions $\mathcal{A}=\{a_\theta:\mathbb{R}^m\mapsto\mathbb{R}^r,\,\theta\in\mathbb{R}^d, \, a_\theta(0)=0\}$ verifying  \eqref{eq:sublinear_growth} or \eqref{eq:lip_bound} uniformly in $\theta$ induces a static $\mathcal{L}_p$ operator through \eqref{eq:static_op}. An example of such a family is provided by the Lipschitz-bounded deep NNs proposed by \citet{wangDirectParameterizationLipschitzBounded2023a} and \citet{pauliLipKernelLipschitzBoundedConvolutional2024} that verify \eqref{eq:lip_bound} for a prescribed Lipschitz constant $\gamma>0$.
\end{remark}

\subsection{Mapping NOC to an unconstrained optimization problem}
\label{subs:mapping_NOC}

We use RENs models for parametrizing $\Lp$ operators in the N-SLS$_{\Qb}$ and N-SLS$_{\tPsiu}$ problems, which allows turning them into unconstrained optimization
problems over $\theta\in\mathbb{R}^d$.
The next issue to be addressed is the computation of the
average in the  N-SLS$_{\Qb}$ and N-SLS$_{\tPsiu}$ problems which, as noticed before, is generally intractable. 
This problem is usually circumvented by approximating the
exact average with its empirical counterpart obtained using a finite set of samples drawn from the distributions $\mathcal{D}^v_{T:0}$ and $\mathcal{D}^d_{T:0}$.
For the feedback architecture~\eqref{eq:system_control}, when a base controller is available, the N-SLS$_{\Qb}$ problem reads as
\begin{subequations}\label{eq:noc_Ky}
\begin{alignat}{3}
	\quad &\min_{\theta \in \mathbb{R}^d}&& \frac{1}{S}\sum_{s=1}^S 
	L(y_{T:0}^s, u_{T:0}^s)
	\label{eq:loss}\\
	&\operatorname{s.t.}~~ && 
	y_{t}^s = \tilde{G}_t(u_{t-1:0}^{\mathrm{o},s} + d_{t-1:0}^s) + v_t^s \,,
	\label{eq:dynamics_s}\\
	&
	\begin{bmatrix}\xi_{t+1}^s\\\zeta_t^s\\u_t^{\mathrm{o},s}\end{bmatrix} 
	&& \hspace{-3pt}=\hspace{-3pt} 
	\begin{bmatrix}0\\0\\K'_t(y_{t:0}^s)\end{bmatrix}
	\hspace{-3pt}+\hspace{-3pt} 
    \Theta(\theta) \hspace{-1pt}
	\begin{bmatrix}\xi_{t}^s\\\sigma(\zeta_t^s)\\y_{t}^s - G_t(u_{t-1:0}^{\mathrm{o},s}) \end{bmatrix}
	\hspace{-3pt}+\hspace{-3pt}
    \begin{bmatrix} 0 \\ 0 \\ b_t \end{bmatrix} 
	\hspace{-2pt},
	\label{eq:OutputSLS_controller}\\
	&~&& t = 0,1,\ldots,T, \qquad \xi_{0} = 0\,.\label{eq:OutputSLS_end}
\end{alignat}
\end{subequations}
where $\theta$ is a free parameter and, hereafter, $\Theta(\cdot)$ is a placeholder for either $\Theta_c(\cdot)$ or $\Theta_{i\ell_2}(\cdot)$. Therefore \eqref{eq:OutputSLS_controller} is a REN.
In the above problem, $\Kb'$ is an i.f.g. $\ell_p$-stabilizing base controller,
and $\{ v^s_{T:0}, d^s_{T:0} \}_{s=1}^S$ is a given training set of $S$ sampled disturbances.
The cost function \eqref{eq:loss} is defined as the sample average of the loss evaluated over the training set, and the system dynamics are imposed through~\eqref{eq:dynamics_s} for every $(v^s_{T:0};d^s_{T:0})$, $s=1,\dots,S$.
The relationship \eqref{eq:OutputSLS_controller}-\eqref{eq:OutputSLS_end} define a control sequence $\mathbf{u}^{\mathrm{o},s} = \Kb'(\yb^s) + \Enne_\theta(\yb^s-\Gb(\mathbf{u}^{\mathrm{o},s}))$, where $\Enne_\theta\in\Ltwo$ for every $\theta$. As a result, each value of $\theta\in\mathbb{R}^d$ yields closed-loop maps $(\Psiy;\Psiu)\in\CLtwo{\Gb}$.

For the case of the feedback architecture \eqref{eq:system_control_Kyu}, the optimization problem N-SLS$_{\tPsiu}$ reads as
\begin{subequations}\label{eq:noc_Kyu}
\begin{alignat}{3}
	\quad &\min_{\theta \in \mathbb{R}^d}&& \frac{1}{S}\sum_{s=1}^S  
	L(y_{T;0}^s, u_{T;0}^s)
	\label{eq:loss_Kyu}\\
	&\operatorname{s.t.}~~ && 
	y_{t}^s = \tilde{G}_t(u_{t-1:0}^{\mathrm{o},s} + d_{t-1:0}^s) + v_t^s \,,
	\label{eq:dynamics_s_Kyu}\\
	&
	\begin{bmatrix}\xi_{t+1}^s\\w_t^s\\u_t^{\mathrm{o},s}\end{bmatrix} 
	&&=
	\Theta(\theta)
	\begin{bmatrix}\xi_{t}^s\\\sigma(w_t^s)\\(y_{t}^s ; u_{t-1}^s) \end{bmatrix}
	+ \begin{bmatrix} 0 \\ 0 \\ b_t \end{bmatrix} 
	,
	\label{eq:OutputSLS_controller_Kyu}\\
	&~&& t = 0,1,\ldots,T, ~~~~ \xi_{0} = 0\,.\label{eq:OutputSLS_end_Kyu}
\end{alignat}
\end{subequations}
In this case, the relationship \eqref{eq:OutputSLS_controller_Kyu}-\eqref{eq:OutputSLS_end_Kyu} define a control sequence 
$\mathbf{u}^{\mathrm{o},s} = \Enne_\theta(\yb^s ; \ub^s)$, where 
$\Enne_\theta\in \Ccs$ and $\Enne_\theta\in\Ltwo$ for every $\theta$. 
As a result, each value of $\theta\in\mathbb{R}^d$ yields closed-loop maps $(\Psiy;\Psiu)\in\hCLtwo{\Gb}$.

Note that in both cases, any $\theta\in\mathbb{R}^d$ parametrizes closed-loop maps that are achievable for system $\Gb$. 
This key property enables using unconstrained gradient-descent algorithms for optimizing over $\theta$. Thus, the NOC problem is now equivalent to the \emph{training} of a DNN.

\begin{remark}
We emphasize that using gradient descent to solve \eqref{eq:noc_Ky} (or \eqref{eq:noc_Kyu}) is not equivalent to direct policy gradient methods, such as those proposed by \citet{fazel2018global} for LQR control. In standard policy gradient approaches, the control policy $\mathbf{K}$ is directly optimized through gradient descent — this simply becomes a matrix $K$ in the LQR setting of \citet{fazel2018global}. While \citet{fazel2018global} establish global optimality guarantees, careful step-size tuning is required to avoid destabilizing policies during optimization. Instead, our approach optimizes over parameters $\theta$ that always define a stable operator linked to the closed-loop behavior. This ensures that every iterate corresponds to a stabilizing policy via a nonlinear transformation, irrespective of the chosen step size.
While such a reformulation is also possible in the LQR case --- using the standard Youla or SLS parametrizations --- it leads to a convex optimization landscape, making convex programming a more favorable alternative to direct gradient descent.
\end{remark}

We remark that the class of all $\ell_2$-stable REN operators is significantly more restrictive than the class of all operators in $\Ltwo$. 
Indeed, in \citet{wang2023linear}, it is shown that RENs given by \eqref{eq:REN} with full, time-invariant vector $\hat b_t=\bar b\in\mathbb{R}^{q_1+q_2+q_{out}}$ are universal approximations of contractive nonlinear systems with incremental $\ell_2$-gain.
Consequently, learning exclusively within the set of $\ell_2$-stable REN operators may limit the applicability of the completeness result presented in  point~2 of Theorem~\ref{th:stabilizing_SLS}.
This is why in the learning problem \eqref{eq:noc_Ky}, 
we allow $\Kb'$ being  i.f.g. $\ell_p$-stabilizing, but not necessarily strongly  i.f.g. $\ell_p$-stabilizing.

Based on the above discussion, an important takeaway is that developing finite-dimensional approximations of $\Lp$ operators that are as large as possible is a crucial step toward the computation of globally optimal solutions to NOC problems.

\section{Numerical experiments}
\label{sec:numerical}
\allowdisplaybreaks

In this section, we illustrate 
through the formulations N-SLS$_{\Qb}$ and N-SLS$_{\tPsiu}$ how to address NOC problems while using RENs to represent $\Ltwo$ operators.
As remarked in Section~\ref{sec:NOC_RENs}, the goal is to minimize
an empirical average of the cost evaluated over sampled trajectories of noise realizations.
Through the machine learning lenses, this is an unsupervised learning problem where the input data corresponds to the initial conditions of the system and disturbance trajectories.
In the sequel, we use the terms ``control design'' and ``training'' interchangeably.

We implement the learning problem \eqref{eq:noc_Ky} using PyTorch and train the resulting DNNs with ADAM, a stochastic gradient descent method. 
The code to reproduce the examples is available at \url{https://github.com/DecodEPFL/outputSLS}.

\subsection{Simulation setup}
We consider point-mass robots. The position of robot $i$ is $x^i_{1,t} \in \mathbb{R}^2$ and its velocity is $x^i_{2,t} \in \mathbb{R}^2$. The robots are affected by nonlinear drag forces (e.g., air or water resistance).
The discrete-time model of robot $i$ of mass $m_i\in \mathbb{R}^+$ is 
\begin{subequations}
\label{eq:mechanical_system}
\begin{align}
	\label{eq:mechanical_system_state}
	\begin{bmatrix}
    x^i_{1,t+1}\\
    x^i_{2,t+1}
    \end{bmatrix} 
    &=  
    \begin{bmatrix}
    x^i_{1,t}\\
    x^i_{2,t}
    \end{bmatrix} 
    + 
    T_s
    \begin{bmatrix}
    x^i_{2,t}\\
    \frac{1}{m^i}
    \left(-C^i(x^i_{2,t}) + F^i_{t}\right)
    \end{bmatrix}\,, 
    \\
    \label{eq:mechanical_system_output}
    y^i_t &= x^i_{1,t}\,,
\end{align}
\end{subequations}
where $y^i_t \in \mathbb{R}^2$ is the output of robot $i$, $F^i_t \in \mathbb{R}^2$ denotes the force control input, $T_s>0$ is the sampling time and $C^i:\mathbb{R}^2\rightarrow \mathbb{R}^2$ is a \emph{drag function}.
In our case, we set $C^i(x_2^i) = b^i_1 x_2^i - b^i_2 \tanh(x_2^i)$ for some $0< b^i_2 < b^i_1$~\cite{falkovich2011fluid}.

For robot $i$, consider a base controller $\nu^i_t = \bar{K}^i(\bar{x}^i_1-x^i_{1,t})$ with $\bar{K}^i = \operatorname{diag}(\bar{k}^i_1,\bar{k}^i_2)$ and $\bar{k}^i_1,\bar{k}^i_2>0$ for reaching the predefined target position $\bar{x}^i_1\in \mathbb{R}^2$. 
One can easily verify that the base controller is strongly $\ell_2$-stabilizing.
Then, the input to each robot is given by $F_t^i=\nu_t^i+u_t^i$, where $u_t^i$ denotes the control input over which we optimize.

We model a set of two robots~\eqref{eq:mechanical_system} by defining an overall state $x_t \in \mathbb{R}^{8}$, input $u_t \in \mathbb{R}^{4}$ and output $y_t \in \mathbb{R}^{4}$.
The initial condition of the system, $x_0$, is fixed a priori.
Two scenarios are considered, involving the coordination of the two robots in the $xy$-plane to complete a given task while avoiding obstacles and preventing collisions between them.

The task of scenario  \texttt{corridor}, shown in Figure~\ref{fig:corridor}, consists in coordinating the passage of the two robots through the narrow valley to reach their respective endpoints designated with `$\star$'.
The nominal system model used in the controller considers a fixed initial condition $x_0$, whose position ($y_0$) is indicated with  `$\times$' in Figure~\ref{fig:corridor}, and the initial velocity is set to zero. 
The real agents forming the \textit{true system} start instead from zero velocity and random initial positions sampled from a normal distribution with mean $y_0$ and variance $0.2$. In Figure~\ref{fig:corridor}, the training data is marked with `$\circ$'.
In this simulation, all the disturbances have been set to zero.
By means of Proposition~\ref{prop:robust}, the closed-loop system is robust to the differences in the initial condition since it is stabilized by a base controller that is i.f.g. $\ell_2$-stabilizing (see discussion in Section~\ref{sec:robust_analysis}).

The task of scenario \texttt{waypoint-tracking},  in Figure~\ref{fig:waypoints}, is to visit the waypoints $g_{a}$, $g_{b}$, $g_{c}$ in a prescribed order, given by
$g_{b}$, $g_{a}$ and $g_{c}$
for the \textit{blue} robot, and
$g_{c}$, $g_{b}$ and $g_{a}$
for the \textit{orange} robot.
In this scenario, the initial conditions are fixed while the disturbances consist of i.i.d. samples from a Gaussian distribution with zero mean and standard deviation of~$0.1$.%
\footnote{The \texttt{corridor} and \texttt{waypoint-tracking} benchmarks are motivated by the examples in \citet{onken2021neural, li2017reinforcement}.}

For the  \texttt{corridor} scenario,
we use in \eqref{eq:loss} the cost function
\begin{align}
	L(y_{T:0},u_{T:0}) &= \sum_{t=0}^T l(y_t,u_t) \nonumber \\
	&= \sum_{t=0}^T \left( l_\text{traj}(y_t,u_t) + l_\text{ca}(y_t)+ l_\text{obs}(y_t) \right) \,,  \label{eq:loss_CA}
\end{align}
where
$l_\text{traj}(y_t,u_t)$ is a quadratic function penalizing the distance of agents from their target position and the control energy;
$l_{\text{ca}}(y_t)$ penalizes collisions between agents,
and;
$l_{\text{obs}}(y_t)$ penalizes collisions with the obstacles of the environment.

For the \texttt{waypoint-tracking} scenario, we seek to specify a cost function promoting
that waypoints are visited in the correct order but without specifying the reaching time of each waypoint. 
This can be done using temporal logic statements for defining the cost \cite{li2017reinforcement,leung2023backpropagation}.
Specifically, we use truncated linear temporal logic (TLTL) cost functions, as described in~\citet{li2017reinforcement}.
TLTL is a specification language leveraging a set of operators defined over finite-time trajectories. 
It allows incorporating domain knowledge and constraints (in a soft fashion) into the learning process,
such as ``always avoid obstacles'', ``eventually visit $x$'', or ``do not visit $y$ until visiting $x$''.
Then, using quantitative semantics, one can transform temporal logical formulae into real-valued reward functions that are compositions of $\min$ and $\max$ functions over a finite period of time~\cite{li2017reinforcement, leung2023backpropagation}. 
Note that TLTL costs cannot be written, in general, as the sum of stage costs like \eqref{eq:loss_CA}.
In the scenario \texttt{waypoint-tracking}, the loss formulation for the \textit{orange} agent is translated into plain English as
``\textit{%
	Visit $g_c$ then $g_b$ then $g_a$;
	and don't visit $g_b$ or $g_a$ until visiting $g_c$;
	and don't visit $g_a$ until visiting $g_b$;
	and if visited $g_c$, don't visit $g_c$ again;
	and if visited $g_b$, don't visit $g_b$ again;
	and always avoid obstacles;
	and always avoid collisions;
	and eventually 
	state at the final goal.%
}''
The implementation details and the full expression of the TLTL cost function can be found in Appendix~\ref{app:implementation}.

\subsection{Results}

\begin{figure*}
	\centering
	\begin{minipage}{0.33\linewidth}
		\includegraphics[width=\linewidth]{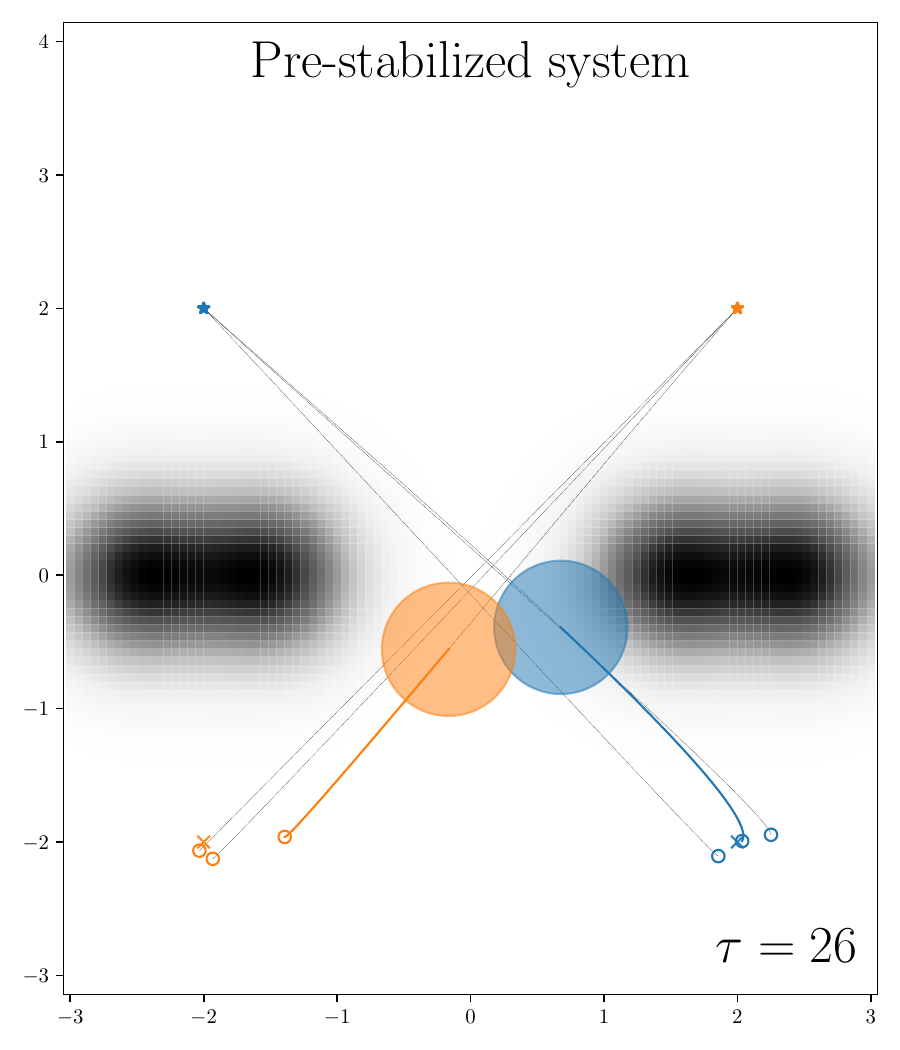}
	\end{minipage}%
	\begin{minipage}{0.33\linewidth}
		\includegraphics[width=\linewidth]{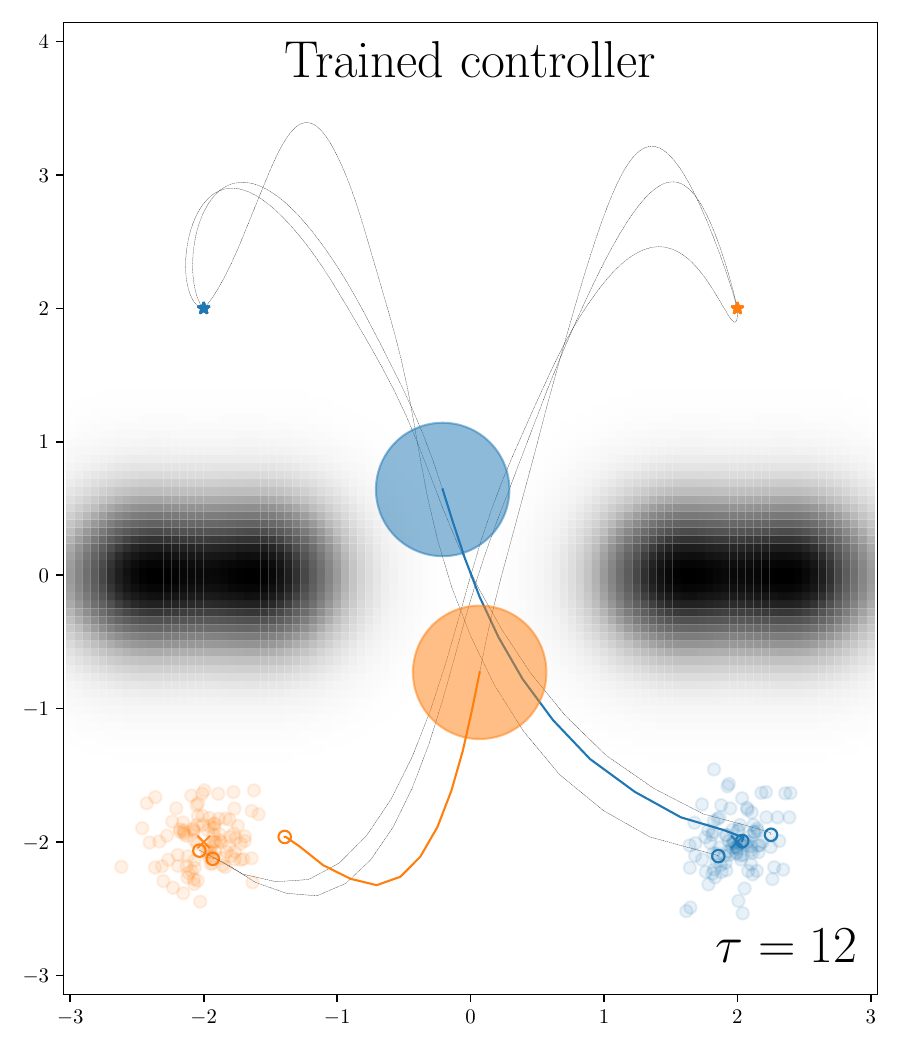}
	\end{minipage}%
	\begin{minipage}{0.33\linewidth}
		\includegraphics[width=\linewidth]{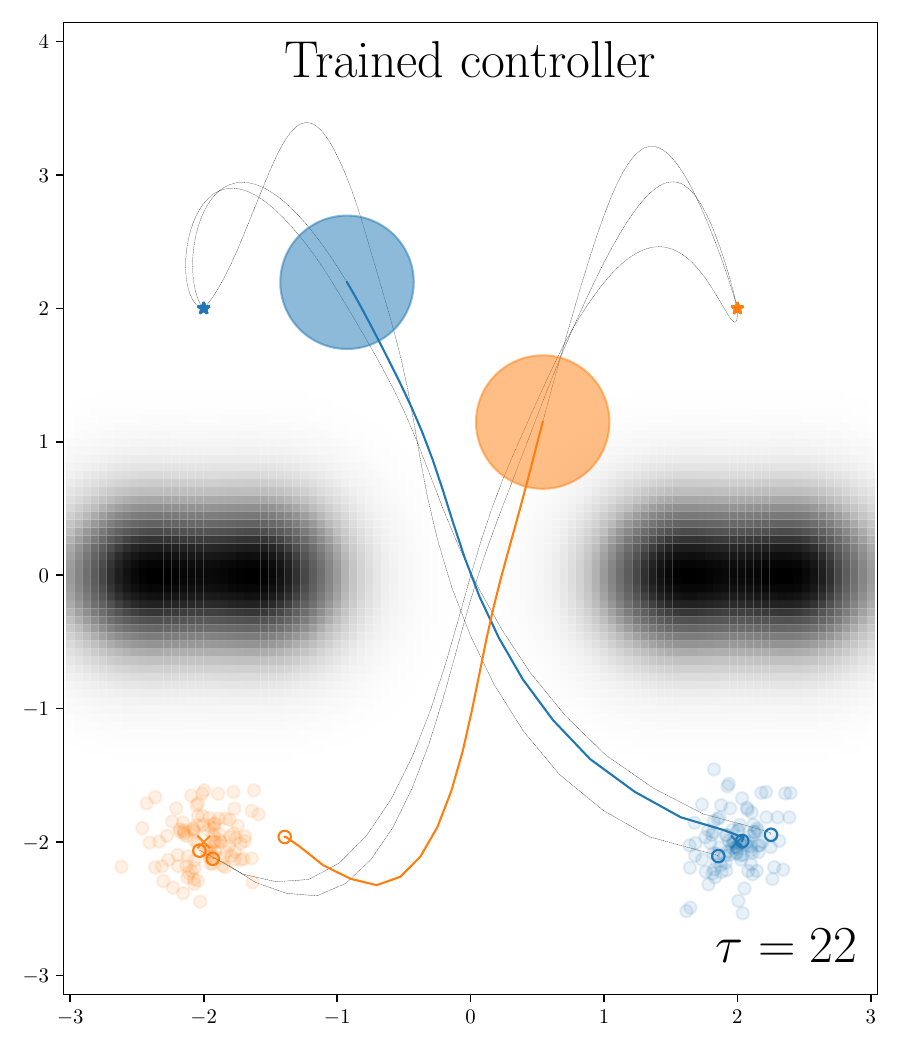}
	\end{minipage}
	\caption{\texttt{Corridor} --- Closed-loop trajectories after training over 100 randomly sampled initial conditions marked with $\circ$. Snapshots taken at instant $\tau$. Colored (gray) lines show the trajectories in $[0,\tau_i]$ ($[\tau_i,\infty)$). Colored balls (and their radius) represent the agents (and their size for collision avoidance).} 
	\label{fig:corridor}
\end{figure*}

\begin{figure*}
	\centering
	\begin{minipage}{0.33\linewidth}
		\includegraphics[width=\linewidth]{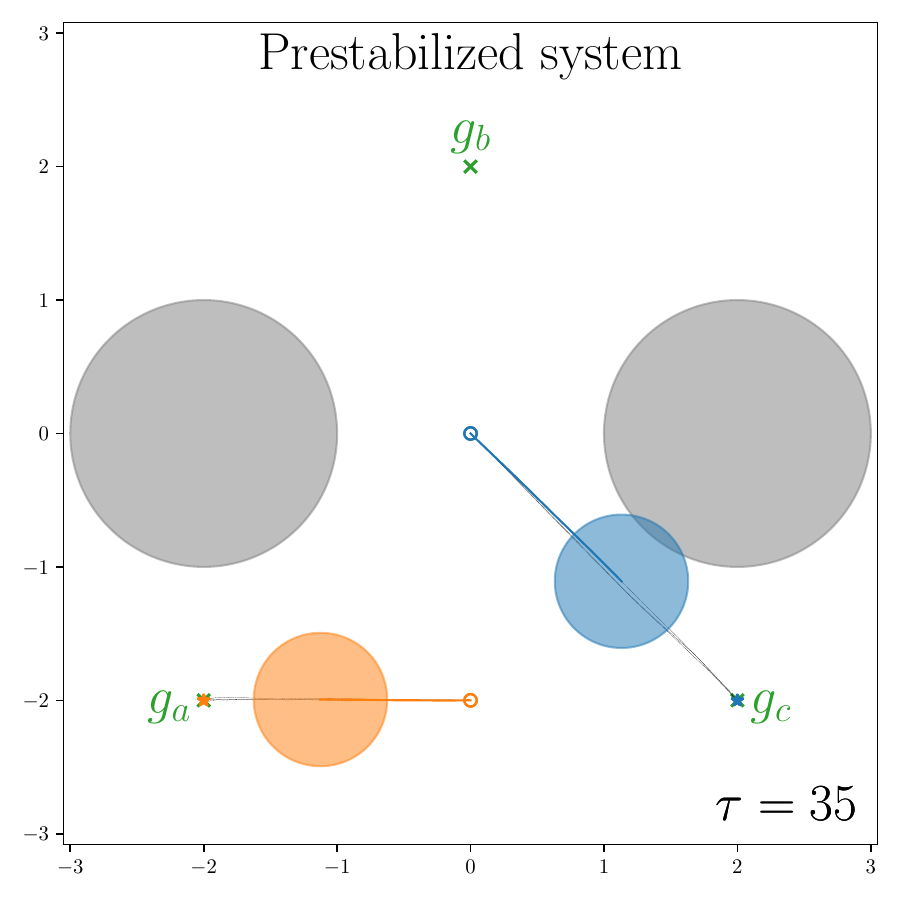}
	\end{minipage}%
	\begin{minipage}{0.33\linewidth}
		\includegraphics[width=\linewidth]{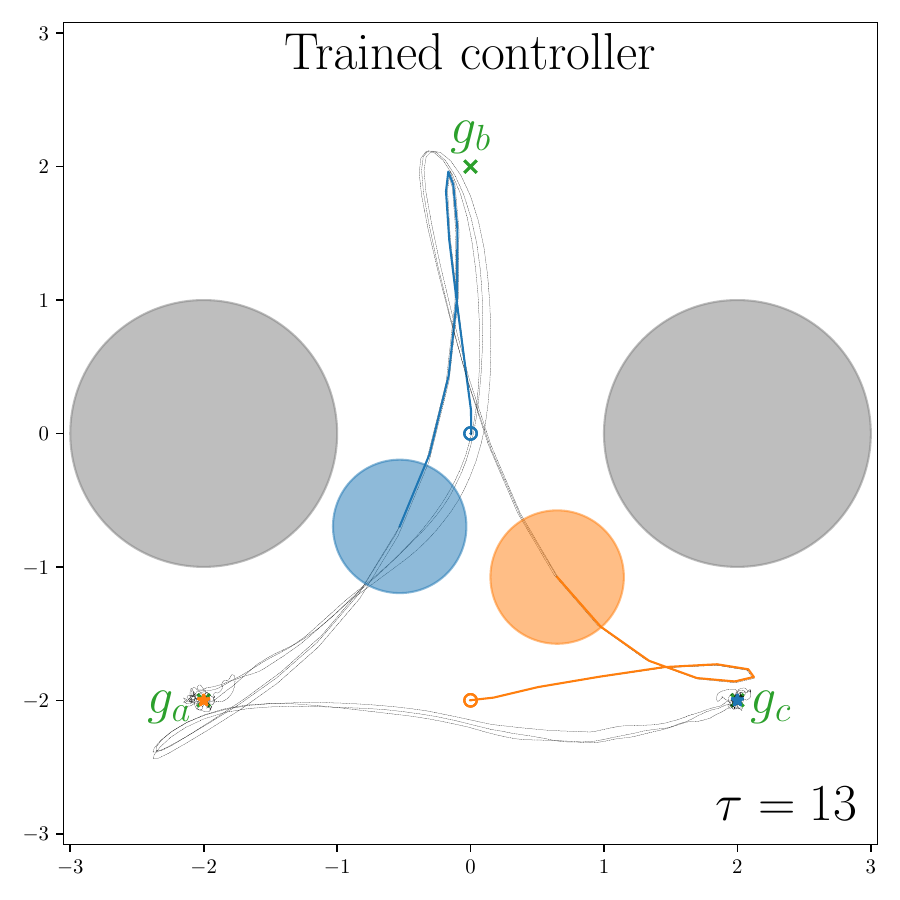}
	\end{minipage}%
	\begin{minipage}{0.33\linewidth}
		\includegraphics[width=\linewidth]{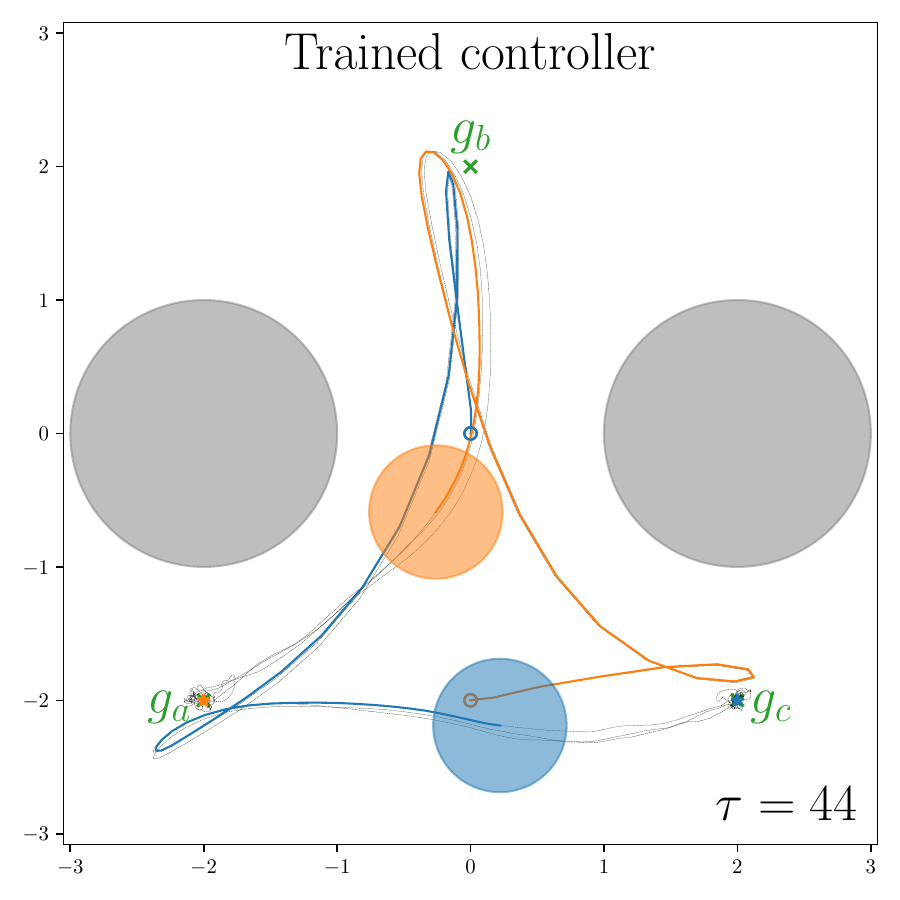}
	\end{minipage}
	\caption{\texttt{Waypoint-tracking} --- Closed-loop trajectories after training. Snapshots taken at instant $\tau$. Colored (gray) lines show the trajectories in $[0,\tau_i]$ ($[\tau_i,T]$). Colored balls (and their radius) represent the agents (and their size for collision avoidance).}
	\label{fig:waypoints}
\end{figure*}

We design control policies to optimize the performance over a horizon of $T = 100$ time-steps. 
Figures~\ref{fig:corridor} and~\ref{fig:waypoints} show the trajectories of the systems with only the prestabilizing controller (left), and the trajectories after training (middle and right).
It can be seen that while the base controller allows the stabilization of the system around the desired equilibrium, it has poor performance, and collisions occur.
After the training process, the obtained control policies avoid collisions and achieve optimized trajectories, boosting the performance of the base controller, thanks to minimizing \eqref{eq:loss_CA} or the TLTL cost.
Note that, despite the use of finite-horizon costs, Theorem~\ref{th:stabilizing_SLS} guarantees 
that targets are asymptotically reached and the system is $\ell_2$-stable around them.

\begin{table*}
\caption{Comparison with other works}
\label{tab:comparison}
\setlength{\tabcolsep}{4pt}
\renewcommand{\arraystretch}{1.2}
\begin{center}
\begin{tabular}{p{3.2cm}|c c c c c c}
\hline
& \multirow{2}{*}{Nonlinear} & \multirow{2}{*}{CT or DT?} & Architecture & Parametrization & Controller & Distributed \\
& & & as in \eqref{eq:system_control} & of $\Phiyu{\Gb}{\Kb}$ & implementation & implementation \\
\hline
\hline
\multirow{1}{*}{\citet{desoer1982}} & \cmark & CT \& DT & \cmark & (a) & \xmark & \xmark \\
\hline
\multirow{1}{*}{\citet{desoer1984simultaneous}} & \multirow{1}{*}{\cmark} & \multirow{1}{*}{CT \& DT} & \multirow{1}{*}{\cmark} & \xmark & \xmark & \xmark \\
\hline
\multirow{2}{3.2cm}{\citet{fujimoto1998youla,fujimoto2000}} & \multirow{2}{*}{\cmark} & \multirow{2}{*}{CT} & \multirow{2}{*}{\cmark} & \multirow{2}{*}{(b)} & \multirow{2}{*}{\xmark} & \multirow{2}{*}{\xmark} \\
&&&&&&\\
\hline
\multirow{2}{3.2cm}{\citet{paice1996tac}} & \multirow{2}{*}{\cmark} & \multirow{2}{*}{CT} & \multirow{2}{*}{\cmark} & \multirow{2}{*}{(b)} & \multirow{2}{*}{\xmark} & \multirow{2}{*}{\xmark} \\
&&&&&&\\
\hline
\citet{wang2019sls} & \xmark & DT & \cmark & \cmark & \cmark & \cmark \\
\hline
\citet{furieri2019input}  & \xmark & CT \& DT & \cmark & \cmark & \cmark & \cmark \\
\hline
\citet{ho2020system} & \cmark & DT & \xmark & \cmark & \cmark & \xmark \\
\hline
\citet{Furieri22b} & \cmark & DT & \xmark & \cmark & \cmark & \xmark \\
\hline
\citet{wang2023linear} & (c) & DT & \cmark & \cmark & \cmark & \xmark \\
\hline
\citet{barbara2023learning} & \cmark & DT & (d) & \xmark & \cmark & \xmark \\
\hline
\textbf{This paper} & \cmark & DT & \cmark & \cmark & \cmark & \cmark \\
\hline
\hline
\multicolumn{7}{l}{(a): Authors only parametrize the map $\db\mapsto\tyb$.} \\
\hline
\multicolumn{7}{l}{(b): Authors parameterize the closed-loop maps through kernel representations.} \\
\hline
\multicolumn{7}{l}{(c): Authors consider linear plants with nonlinear controllers.} \\
\hline
\multicolumn{7}{l}{(d): Authors consider measurable disturbances $\db$ (or equivalently, a set-point).} \\
\hline
\end{tabular}
\end{center}
\end{table*}

\section{Conclusions}
\label{sec:conclusions}

As we move towards designing nonlinear policies for addressing general optimal control problems, it is crucial to guarantee closed-loop stability during and after optimization. 
In this work, we present parametrizations of all and only stabilizing controllers for a given system, which are described in terms of a single stable operator. 
We show the compatibility of our parametrizations with scenarios where only an approximate system description is available and with distributed setups.
Importantly, these parametrizations lead to optimization problems that can be tackled by training DNNs with unconstrained gradient descent.

In the following, we discuss the limitations of the proposed approach and possible research directions to address them.
To obtain a tractable formulation of the NOC problem \eqref{eq:NOC_problem}, the true cost \eqref{eq:cost_J} has been replaced by its empirical counterpart \eqref{eq:loss}, based on sampled disturbances. This approximation raises the issue of quantifying the potential degradation of the cost when the controller is deployed, and the system is affected by out-of-sample disturbance profiles. A common heuristic approach consists of evaluating performance via simulations on a test set of disturbance sequences not used for control design.
An alternative, more rigorous approach would be to derive generalization bounds --- i.e., upper bounds on the uncomputable true cost \eqref{eq:cost_J} based on its empirical counterpart \eqref{eq:loss} and independent of the disturbance distribution. These bounds would allow for performance characterization without the need of validation experiments. Generalization bounds for stochastic NOC problems and state-feedback controllers have been studied by \citet{Ghoddousi24} using a PAC-Bayesian framework. Extending these results to output-feedback controllers represents a promising research direction.
Similarly, one could investigate the worst-case cost degradation under different disturbance models, e.g., disturbances lying in a prescribed bounded set or drawn from probability distributions belonging to an uncertainty set around a nominal one.\footnote{This setting is typically considered in distributionally robust optimization \cite{mohajerinesfahaniDatadrivenDistributionallyRobust2018a}.} An even more important direction would be to study how to design controllers that optimize these objective performance measures. As shown in \citet{Ghoddousi24}, this might require a substantial reformulation of the NOC problems considered in this paper.

This work focuses on control design methods aimed at ensuring closed-loop stability. However, in many applications, the primary challenge lies in tracking specific classes of setpoints. Furthermore, while the proposed controllers optimize system performance over a finite time horizon, many real-world applications require cost optimization over the entire operational lifespan of the system. Overcoming these limitations would require investigating generalizations to tracking problems and controller optimization in a receding-horizon fashion. The fundamental challenge would be to derive guarantees of stable closed-loop tracking under continuous updates of the controllers.

Finally, the controller synthesis methods developed in this paper are model-based. A significant research direction would be to extend these synthesis algorithms to cases where the system models are obtained from a finite number of noisy experimental data points. This problem, which is central to the field of reinforcement learning, is notoriously challenging. However, the fail-safe design methods discussed in this paper could provide inspiration for the development of controllers capable of ensuring closed-loop stability even during the online learning phase of the model.

\bibliographystyle{elsarticle-harv}        
\bibliography{references}           

\begin{thebibliography}{51}
\expandafter\ifx\csname natexlab\endcsname\relax\def\natexlab#1{#1}\fi
\providecommand{\url}[1]{\texttt{#1}}
\providecommand{\href}[2]{#2}
\providecommand{\path}[1]{#1}
\providecommand{\DOIprefix}{doi:}
\providecommand{\ArXivprefix}{arXiv:}
\providecommand{\URLprefix}{URL: }
\providecommand{\Pubmedprefix}{pmid:}
\providecommand{\doi}[1]{\href{http://dx.doi.org/#1}{\path{#1}}}
\providecommand{\Pubmed}[1]{\href{pmid:#1}{\path{#1}}}
\providecommand{\bibinfo}[2]{#2}
\ifx\xfnm\relax \def\xfnm[#1]{\unskip,\space#1}\fi
\bibitem[{Anantharam and Desoer(1984)}]{anantharam1984stabilization}
\bibinfo{author}{Anantharam, V.}, \bibinfo{author}{Desoer, C.A.}, \bibinfo{year}{1984}.
\newblock \bibinfo{title}{On the stabilization of nonlinear systems}.
\newblock \bibinfo{journal}{IEEE Transactions on Automatic Control} \bibinfo{volume}{29}, \bibinfo{pages}{569--572}.
\bibitem[{Anderson et~al.(2019)Anderson, Doyle, Low and Matni}]{anderson2019system}
\bibinfo{author}{Anderson, J.}, \bibinfo{author}{Doyle, J.C.}, \bibinfo{author}{Low, S.H.}, \bibinfo{author}{Matni, N.}, \bibinfo{year}{2019}.
\newblock \bibinfo{title}{System level synthesis}.
\newblock \bibinfo{journal}{Annual Reviews in Control} \bibinfo{volume}{47}, \bibinfo{pages}{364--393}.
\bibitem[{Astolfi(1998)}]{astolfi1998discontinuous}
\bibinfo{author}{Astolfi, A.}, \bibinfo{year}{1998}.
\newblock \bibinfo{title}{Discontinuous control of the brockett integrator}.
\newblock \bibinfo{journal}{European Journal of Control} \bibinfo{volume}{4}, \bibinfo{pages}{49--63}.
\bibitem[{Barbara et~al.(2023)Barbara, Wang and Manchester}]{barbara2023learning}
\bibinfo{author}{Barbara, N.H.}, \bibinfo{author}{Wang, R.}, \bibinfo{author}{Manchester, I.R.}, \bibinfo{year}{2023}.
\newblock \bibinfo{title}{Learning over contracting and {L}ipschitz closed-loops for partially-observed nonlinear systems}, in: \bibinfo{booktitle}{IEEE Conference on Decision and Control (CDC)}, pp. \bibinfo{pages}{1028--1033}.
\bibitem[{Bonassi et~al.(2022)Bonassi, Farina, Xie and Scattolini}]{bonassi2022}
\bibinfo{author}{Bonassi, F.}, \bibinfo{author}{Farina, M.}, \bibinfo{author}{Xie, J.}, \bibinfo{author}{Scattolini, R.}, \bibinfo{year}{2022}.
\newblock \bibinfo{title}{On recurrent neural networks for learning-based control: Recent results and ideas for future developments}.
\newblock \bibinfo{journal}{Journal of Process Control} \bibinfo{volume}{114}, \bibinfo{pages}{92--104}.
\bibitem[{Boroujeni et~al.(2024)Boroujeni, Galimberti, Krause and Ferrari-Trecate}]{Ghoddousi24}
\bibinfo{author}{Boroujeni, M.G.}, \bibinfo{author}{Galimberti, C.L.}, \bibinfo{author}{Krause, A.}, \bibinfo{author}{Ferrari-Trecate, G.}, \bibinfo{year}{2024}.
\newblock \bibinfo{title}{A {PAC}-{B}ayesian framework for optimal control with stability guarantees}, in: \bibinfo{booktitle}{IEEE Conference on Decision and Control (CDC)}, pp. \bibinfo{pages}{8237--8244}.
\bibitem[{Boyd and Vandenberghe(2004)}]{Boyd_Vandenberghe_2004}
\bibinfo{author}{Boyd, S.}, \bibinfo{author}{Vandenberghe, L.}, \bibinfo{year}{2004}.
\newblock \bibinfo{title}{Convex Optimization}.
\newblock \bibinfo{publisher}{Cambridge University Press}.
\bibitem[{D'Amico et~al.(2023)D'Amico, La~Bella, Dercole and Farina}]{damico2023data}
\bibinfo{author}{D'Amico, W.}, \bibinfo{author}{La~Bella, A.}, \bibinfo{author}{Dercole, F.}, \bibinfo{author}{Farina, M.}, \bibinfo{year}{2023}.
\newblock \bibinfo{title}{Data-based control design for nonlinear systems with recurrent neural network-based controllers}.
\newblock \bibinfo{journal}{IFAC-PapersOnLine} \bibinfo{volume}{56}, \bibinfo{pages}{6235--6240}.
\bibitem[{{de Souza} et~al.(2023){de Souza}, Girard and Tarbouriech}]{deSouza2023event}
\bibinfo{author}{{de Souza}, C.}, \bibinfo{author}{Girard, A.}, \bibinfo{author}{Tarbouriech, S.}, \bibinfo{year}{2023}.
\newblock \bibinfo{title}{Event-triggered neural network control using quadratic constraints for perturbed systems}.
\newblock \bibinfo{journal}{Automatica} \bibinfo{volume}{157}, \bibinfo{pages}{111237}.
\bibitem[{Desoer and Lin(1984)}]{desoer1984simultaneous}
\bibinfo{author}{Desoer, C.}, \bibinfo{author}{Lin, C.}, \bibinfo{year}{1984}.
\newblock \bibinfo{title}{Simultaneous stabilization of nonlinear systems}.
\newblock \bibinfo{journal}{IEEE Transactions on Automatic Control} \bibinfo{volume}{29}, \bibinfo{pages}{455--457}.
\bibitem[{Desoer and Chan(1976)}]{desoer1976feedback}
\bibinfo{author}{Desoer, C.A.}, \bibinfo{author}{Chan, W.S.}, \bibinfo{year}{1976}.
\newblock \bibinfo{title}{The feedback interconnection of multivariable systems: simplifying theorems for stability}.
\newblock \bibinfo{journal}{Proceedings of the IEEE} \bibinfo{volume}{64}, \bibinfo{pages}{139--144}.
\bibitem[{Desoer and Liu(1982)}]{desoer1982}
\bibinfo{author}{Desoer, C.A.}, \bibinfo{author}{Liu, R.W.}, \bibinfo{year}{1982}.
\newblock \bibinfo{title}{Global parametrization of feedback systems with nonlinear plants}.
\newblock \bibinfo{journal}{Systems \& Control Letters} \bibinfo{volume}{1}, \bibinfo{pages}{249--251}.
\bibitem[{Economou et~al.(1986)Economou, Morari and Palsson}]{Economou_Morari_nlIMC_1986}
\bibinfo{author}{Economou, C.G.}, \bibinfo{author}{Morari, M.}, \bibinfo{author}{Palsson, B.O.}, \bibinfo{year}{1986}.
\newblock \bibinfo{title}{Internal model control: extension to nonlinear system}.
\newblock \bibinfo{journal}{Industrial \& Engineering Chemistry Process Design and Development} \bibinfo{volume}{25}, \bibinfo{pages}{403--411}.
\bibitem[{Falkovich(2011)}]{falkovich2011fluid}
\bibinfo{author}{Falkovich, G.}, \bibinfo{year}{2011}.
\newblock \bibinfo{title}{Fluid mechanics: A short course for physicists}.
\newblock \bibinfo{publisher}{Cambridge University Press}.
\bibitem[{Fazel et~al.(2018)Fazel, Ge, Kakade and Mesbahi}]{fazel2018global}
\bibinfo{author}{Fazel, M.}, \bibinfo{author}{Ge, R.}, \bibinfo{author}{Kakade, S.}, \bibinfo{author}{Mesbahi, M.}, \bibinfo{year}{2018}.
\newblock \bibinfo{title}{Global convergence of policy gradient methods for the linear quadratic regulator}, in: \bibinfo{booktitle}{International conference on machine learning}, \bibinfo{organization}{PMLR}. pp. \bibinfo{pages}{1467--1476}.
\bibitem[{Forgione and Piga(2021)}]{forgione2021dynonet}
\bibinfo{author}{Forgione, M.}, \bibinfo{author}{Piga, D.}, \bibinfo{year}{2021}.
\newblock \bibinfo{title}{dynonet: A neural network architecture for learning dynamical systems}.
\newblock \bibinfo{journal}{International Journal of Adaptive Control and Signal Processing} \bibinfo{volume}{35}, \bibinfo{pages}{612--626}.
\bibitem[{Fujimoto and Sugie(1998)}]{fujimoto1998youla}
\bibinfo{author}{Fujimoto, K.}, \bibinfo{author}{Sugie, T.}, \bibinfo{year}{1998}.
\newblock \bibinfo{title}{Youla-{K}ucera parameterization for nonlinear systems via observer based kernel representations}.
\newblock \bibinfo{journal}{Transactions of the Society of Instrument and Control Engineers} \bibinfo{volume}{34}, \bibinfo{pages}{376--383}.
\bibitem[{Fujimoto and Sugie(2000)}]{fujimoto2000}
\bibinfo{author}{Fujimoto, K.}, \bibinfo{author}{Sugie, T.}, \bibinfo{year}{2000}.
\newblock \bibinfo{title}{Characterization of all nonlinear stabilizing controllers via observer-based kernel representations}.
\newblock \bibinfo{journal}{Automatica} \bibinfo{volume}{36}, \bibinfo{pages}{1123--1135}.
\bibitem[{Furieri et~al.(2022)Furieri, Galimberti and Ferrari-Trecate}]{Furieri22b}
\bibinfo{author}{Furieri, L.}, \bibinfo{author}{Galimberti, C.L.}, \bibinfo{author}{Ferrari-Trecate, G.}, \bibinfo{year}{2022}.
\newblock \bibinfo{title}{Neural system level synthesis: Learning over all stabilizing policies for nonlinear systems}, in: \bibinfo{booktitle}{IEEE Conference on Decision and Control (CDC)}, pp. \bibinfo{pages}{2765--2770}.
\bibitem[{Furieri et~al.(2024)Furieri, Galimberti and Ferrari-Trecate}]{furieri2024performance}
\bibinfo{author}{Furieri, L.}, \bibinfo{author}{Galimberti, C.L.}, \bibinfo{author}{Ferrari-Trecate, G.}, \bibinfo{year}{2024}.
\newblock \bibinfo{title}{Learning to boost the performance of stable nonlinear systems}.
\newblock \bibinfo{journal}{IEEE Open Journal of Control Systems} \bibinfo{volume}{3}, \bibinfo{pages}{342--357}.
\bibitem[{Furieri and Kamgarpour(2020)}]{furieri2020first}
\bibinfo{author}{Furieri, L.}, \bibinfo{author}{Kamgarpour, M.}, \bibinfo{year}{2020}.
\newblock \bibinfo{title}{First order methods for globally optimal distributed controllers beyond quadratic invariance}, in: \bibinfo{booktitle}{2020 American Control Conference (ACC)}, \bibinfo{organization}{IEEE}. pp. \bibinfo{pages}{4588--4593}.
\bibitem[{Furieri et~al.(2019)Furieri, Zheng, Papachristodoulou and Kamgarpour}]{furieri2019input}
\bibinfo{author}{Furieri, L.}, \bibinfo{author}{Zheng, Y.}, \bibinfo{author}{Papachristodoulou, A.}, \bibinfo{author}{Kamgarpour, M.}, \bibinfo{year}{2019}.
\newblock \bibinfo{title}{An input--output parametrization of stabilizing controllers: Amidst {Y}oula and system level synthesis}.
\newblock \bibinfo{journal}{IEEE Control Systems Letters} \bibinfo{volume}{3}, \bibinfo{pages}{1014--1019}.
\bibitem[{Garcia and Morari(1982)}]{Garcia_Morari_IMC_1982}
\bibinfo{author}{Garcia, C.E.}, \bibinfo{author}{Morari, M.}, \bibinfo{year}{1982}.
\newblock \bibinfo{title}{Internal model control. {A} unifying review and some new results}.
\newblock \bibinfo{journal}{Industrial \& Engineering Chemistry Process Design and Development} \bibinfo{volume}{21}, \bibinfo{pages}{308--323}.
\bibitem[{Gu et~al.(2022a)Gu, Goel and Ré}]{gu2022efficiently}
\bibinfo{author}{Gu, A.}, \bibinfo{author}{Goel, K.}, \bibinfo{author}{Ré, C.}, \bibinfo{year}{2022}a.
\newblock \bibinfo{title}{Efficiently modeling long sequences with structured state spaces}, in: \bibinfo{booktitle}{International Conference on Learning Representations}.
\bibitem[{Gu et~al.(2022b)Gu, Yin, El~Ghaoui, Arcak, Seiler and Jin}]{gu2022recurrent}
\bibinfo{author}{Gu, F.}, \bibinfo{author}{Yin, H.}, \bibinfo{author}{El~Ghaoui, L.}, \bibinfo{author}{Arcak, M.}, \bibinfo{author}{Seiler, P.}, \bibinfo{author}{Jin, M.}, \bibinfo{year}{2022}b.
\newblock \bibinfo{title}{Recurrent neural network controllers synthesis with stability guarantees for partially observed systems}, in: \bibinfo{booktitle}{Proceedings of the AAAI Conference on Artificial Intelligence}, pp. \bibinfo{pages}{5385--5394}.
\bibitem[{Ho(2020)}]{ho2020system}
\bibinfo{author}{Ho, D.}, \bibinfo{year}{2020}.
\newblock \bibinfo{title}{A system level approach to discrete-time nonlinear systems}, in: \bibinfo{booktitle}{2020 American Control Conference (ACC)}, \bibinfo{organization}{IEEE}. pp. \bibinfo{pages}{1625--1630}.
\bibitem[{Imura and Yoshikawa(1997)}]{imura1997}
\bibinfo{author}{Imura, J.i.}, \bibinfo{author}{Yoshikawa, T.}, \bibinfo{year}{1997}.
\newblock \bibinfo{title}{Parametrization of all stabilizing controllers of nonlinear systems}.
\newblock \bibinfo{journal}{Systems \& Control Letters} \bibinfo{volume}{29}, \bibinfo{pages}{207--213}.
\bibitem[{Kim et~al.(2018)Kim, Ríos~Patrón and Braatz}]{KimPatronBraatz18}
\bibinfo{author}{Kim, K.K.K.}, \bibinfo{author}{Ríos~Patrón, E.}, \bibinfo{author}{Braatz, R.D.}, \bibinfo{year}{2018}.
\newblock \bibinfo{title}{Standard representation and unified stability analysis for dynamic artificial neural network models}.
\newblock \bibinfo{journal}{Neural Networks} \bibinfo{volume}{98}, \bibinfo{pages}{251--262}.
\bibitem[{Koelewijn et~al.(2021)Koelewijn, Tóth and Weiland}]{Koelewijn_2021}
\bibinfo{author}{Koelewijn, P.J.W.}, \bibinfo{author}{Tóth, R.}, \bibinfo{author}{Weiland, S.}, \bibinfo{year}{2021}.
\newblock \bibinfo{title}{Incremental dissipativity based control of discrete-time nonlinear systems via the {LPV} framework}, in: \bibinfo{booktitle}{IEEE Conference on Decision and Control (CDC)}, \bibinfo{organization}{IEEE}. pp. \bibinfo{pages}{3281--3286}.
\bibitem[{Leung et~al.(2023)Leung, Aréchiga and Pavone}]{leung2023backpropagation}
\bibinfo{author}{Leung, K.}, \bibinfo{author}{Aréchiga, N.}, \bibinfo{author}{Pavone, M.}, \bibinfo{year}{2023}.
\newblock \bibinfo{title}{Back-propagation through signal temporal logic specifications: Infusing logical structure into gradient-based methods}.
\newblock \bibinfo{journal}{The International Journal of Robotics Research} \bibinfo{volume}{42}, \bibinfo{pages}{356--370}.
\bibitem[{Li et~al.(2017)Li, Vasile and Belta}]{li2017reinforcement}
\bibinfo{author}{Li, X.}, \bibinfo{author}{Vasile, C.I.}, \bibinfo{author}{Belta, C.}, \bibinfo{year}{2017}.
\newblock \bibinfo{title}{Reinforcement learning with temporal logic rewards}, in: \bibinfo{booktitle}{2017 IEEE/RSJ International Conference on Intelligent Robots and Systems (IROS)}, \bibinfo{organization}{IEEE}. pp. \bibinfo{pages}{3834--3839}.
\bibitem[{Lu(1995)}]{Lu1995}
\bibinfo{author}{Lu, W.M.}, \bibinfo{year}{1995}.
\newblock \bibinfo{title}{A state-space approach to parameterization of stabilizing controllers for nonlinear systems}.
\newblock \bibinfo{journal}{IEEE Transactions on Automatic Control} \bibinfo{volume}{40}, \bibinfo{pages}{1576--1588}.
\bibitem[{Martinelli et~al.(2023)Martinelli, Galimberti, Manchester, Furieri and Ferrari-Trecate}]{martinelli2023unconstrained}
\bibinfo{author}{Martinelli, D.}, \bibinfo{author}{Galimberti, C.L.}, \bibinfo{author}{Manchester, I.R.}, \bibinfo{author}{Furieri, L.}, \bibinfo{author}{Ferrari-Trecate, G.}, \bibinfo{year}{2023}.
\newblock \bibinfo{title}{Unconstrained parametrization of dissipative and contracting neural ordinary differential equations}, in: \bibinfo{booktitle}{IEEE Conference on Decision and Control (CDC)}, \bibinfo{organization}{IEEE}. pp. \bibinfo{pages}{3043--3048}.
\bibitem[{Mohajerin~Esfahani and Kuhn(2018)}]{mohajerinesfahaniDatadrivenDistributionallyRobust2018a}
\bibinfo{author}{Mohajerin~Esfahani, P.}, \bibinfo{author}{Kuhn, D.}, \bibinfo{year}{2018}.
\newblock \bibinfo{title}{Data-driven distributionally robust optimization using the {{Wasserstein}} metric: Performance guarantees and tractable reformulations}.
\newblock \bibinfo{journal}{Mathematical Programming} \bibinfo{volume}{171}, \bibinfo{pages}{115--166}.
\bibitem[{Onken et~al.(2021)Onken, Nurbekyan, Li, Fung, Osher and Ruthotto}]{onken2021neural}
\bibinfo{author}{Onken, D.}, \bibinfo{author}{Nurbekyan, L.}, \bibinfo{author}{Li, X.}, \bibinfo{author}{Fung, S.W.}, \bibinfo{author}{Osher, S.}, \bibinfo{author}{Ruthotto, L.}, \bibinfo{year}{2021}.
\newblock \bibinfo{title}{A neural network approach applied to multi-agent optimal control}, in: \bibinfo{booktitle}{IEEE European Control Conference (ECC)}, pp. \bibinfo{pages}{1036--1041}.
\bibitem[{Paice and van~der Schaft(1994a)}]{paice1994cdc}
\bibinfo{author}{Paice, A.D.B.}, \bibinfo{author}{van~der Schaft, A.J.}, \bibinfo{year}{1994}a.
\newblock \bibinfo{title}{Stable kernel representations and the {Y}oula parameterization for nonlinear systems}, in: \bibinfo{booktitle}{IEEE Conference on Decision and Control (CDC)}, pp. \bibinfo{pages}{781--786}.
\bibitem[{Paice and van~der Schaft(1994b)}]{paice1994stable}
\bibinfo{author}{Paice, A.D.B.}, \bibinfo{author}{van~der Schaft, A.J.}, \bibinfo{year}{1994}b.
\newblock \bibinfo{title}{Stable kernel representations as nonlinear left coprime factorizations}, in: \bibinfo{booktitle}{IEEE Conference on Decision and Control (CDC)}, \bibinfo{organization}{IEEE}. pp. \bibinfo{pages}{2786--2791}.
\bibitem[{Paice and van~der Schaft(1996)}]{paice1996tac}
\bibinfo{author}{Paice, A.D.B.}, \bibinfo{author}{van~der Schaft, A.J.}, \bibinfo{year}{1996}.
\newblock \bibinfo{title}{The class of stabilizing nonlinear plant controller pairs}.
\newblock \bibinfo{journal}{IEEE Transactions on Automatic Control} \bibinfo{volume}{41}, \bibinfo{pages}{634--645}.
\bibitem[{Pauli et~al.(2024)Pauli, Wang, Manchester and Allg{\"o}wer}]{pauliLipKernelLipschitzBoundedConvolutional2024}
\bibinfo{author}{Pauli, P.}, \bibinfo{author}{Wang, R.}, \bibinfo{author}{Manchester, I.}, \bibinfo{author}{Allg{\"o}wer, F.}, \bibinfo{year}{2024}.
\newblock \bibinfo{title}{{LipKernel}: {L}ipschitz-bounded convolutional neural networks via dissipative layers}.
\newblock \bibinfo{journal}{arXiv preprint arXiv:2410.22258} .
\bibitem[{Revay et~al.(2024)Revay, Wang and Manchester}]{revay2021RENs}
\bibinfo{author}{Revay, M.}, \bibinfo{author}{Wang, R.}, \bibinfo{author}{Manchester, I.R.}, \bibinfo{year}{2024}.
\newblock \bibinfo{title}{Recurrent equilibrium networks: Flexible dynamic models with guaranteed stability and robustness}.
\newblock \bibinfo{journal}{IEEE Transactions on Automatic Control} \bibinfo{volume}{69}, \bibinfo{pages}{2855--2870}.
\bibitem[{van~der Schaft(2017)}]{vanderSchaft2017}
\bibinfo{author}{van~der Schaft, A.}, \bibinfo{year}{2017}.
\newblock \bibinfo{title}{L$_2$-Gain and Passivity Techniques in Nonlinear Control}.
\newblock \bibinfo{publisher}{Springer}.
\bibitem[{Tang et~al.(2021)Tang, Zheng and Li}]{tang2021analysis}
\bibinfo{author}{Tang, Y.}, \bibinfo{author}{Zheng, Y.}, \bibinfo{author}{Li, N.}, \bibinfo{year}{2021}.
\newblock \bibinfo{title}{Analysis of the optimization landscape of linear quadratic {G}aussian ({LQG}) control}, in: \bibinfo{booktitle}{Learning for Dynamics and Control}, \bibinfo{organization}{PMLR}. pp. \bibinfo{pages}{599--610}.
\bibitem[{Wang et~al.(2023)Wang, Barbara, Revay and Manchester}]{wang2023linear}
\bibinfo{author}{Wang, R.}, \bibinfo{author}{Barbara, N.H.}, \bibinfo{author}{Revay, M.}, \bibinfo{author}{Manchester, I.R.}, \bibinfo{year}{2023}.
\newblock \bibinfo{title}{Learning over all stabilizing nonlinear controllers for a partially-observed linear system}.
\newblock \bibinfo{journal}{IEEE Control Systems Letters} \bibinfo{volume}{7}, \bibinfo{pages}{91--96}.
\bibitem[{Wang and Manchester(2023)}]{wangDirectParameterizationLipschitzBounded2023a}
\bibinfo{author}{Wang, R.}, \bibinfo{author}{Manchester, I.}, \bibinfo{year}{2023}.
\newblock \bibinfo{title}{Direct {{Parameterization}} of {{Lipschitz-Bounded Deep Networks}}}, in: \bibinfo{booktitle}{Proceedings of the 40th {{International Conference}} on {{Machine Learning}}}, \bibinfo{publisher}{PMLR}. pp. \bibinfo{pages}{36093--36110}.
\bibitem[{Wang et~al.(2018)Wang, Matni and Doyle}]{wang2018separable}
\bibinfo{author}{Wang, Y.S.}, \bibinfo{author}{Matni, N.}, \bibinfo{author}{Doyle, J.C.}, \bibinfo{year}{2018}.
\newblock \bibinfo{title}{Separable and localized system-level synthesis for large-scale systems}.
\newblock \bibinfo{journal}{IEEE Transactions on Automatic Control} \bibinfo{volume}{63}, \bibinfo{pages}{4234--4249}.
\bibitem[{Wang et~al.(2019)Wang, Matni and Doyle}]{wang2019sls}
\bibinfo{author}{Wang, Y.S.}, \bibinfo{author}{Matni, N.}, \bibinfo{author}{Doyle, J.C.}, \bibinfo{year}{2019}.
\newblock \bibinfo{title}{A system-level approach to controller synthesis}.
\newblock \bibinfo{journal}{IEEE Transactions on Automatic Control} \bibinfo{volume}{64}, \bibinfo{pages}{4079--4093}.
\bibitem[{Youla et~al.(1976)Youla, Jabr and Bongiorno}]{Youla1976}
\bibinfo{author}{Youla, D.}, \bibinfo{author}{Jabr, H.}, \bibinfo{author}{Bongiorno, J.}, \bibinfo{year}{1976}.
\newblock \bibinfo{title}{Modern {W}iener-{H}opf design of optimal controllers--part {II}: The multivariable case}.
\newblock \bibinfo{journal}{IEEE Transactions on Automatic Control} \bibinfo{volume}{21}, \bibinfo{pages}{319--338}.
\bibitem[{Zakwan and Ferrari-Trecate(2024)}]{zakwan2024neural}
\bibinfo{author}{Zakwan, M.}, \bibinfo{author}{Ferrari-Trecate, G.}, \bibinfo{year}{2024}.
\newblock \bibinfo{title}{Neural port-{H}amiltonian models for nonlinear distributed control: An unconstrained parametrization approach}.
\newblock \bibinfo{journal}{arXiv preprint arXiv:2411.10096} .
\bibitem[{Zames(1966)}]{zames1966input}
\bibinfo{author}{Zames, G.}, \bibinfo{year}{1966}.
\newblock \bibinfo{title}{On the input-output stability of time-varying nonlinear feedback systems part one: Conditions derived using concepts of loop gain, conicity, and positivity}.
\newblock \bibinfo{journal}{IEEE Transactions on Automatic Control} \bibinfo{volume}{11}, \bibinfo{pages}{228--238}.
\bibitem[{Zheng et~al.(2022)Zheng, Furieri, Kamgarpour and Li}]{zheng2022system}
\bibinfo{author}{Zheng, Y.}, \bibinfo{author}{Furieri, L.}, \bibinfo{author}{Kamgarpour, M.}, \bibinfo{author}{Li, N.}, \bibinfo{year}{2022}.
\newblock \bibinfo{title}{System-level, input-output and new parameterizations of stabilizing controllers, and their numerical computation}.
\newblock \bibinfo{journal}{Automatica} \bibinfo{volume}{140}, \bibinfo{pages}{110211}.
\bibitem[{Zhou and Doyle(1998)}]{zhou1998essentials}
\bibinfo{author}{Zhou, K.}, \bibinfo{author}{Doyle, J.C.}, \bibinfo{year}{1998}.
\newblock \bibinfo{title}{Essentials of robust control}. volume \bibinfo{volume}{104}.
\newblock \bibinfo{publisher}{Prentice Hall Upper Saddle River, NJ}.

\end{thebibliography}

\appendix

\section{Proof of Proposition~\ref{prop:phi_achievable}}\label{app:proof_phi_achievable}

\begin{pf} 
	We prove that given a $\Kb \in \Cc$, 
	the closed-loop map $\Phiyu{G}{K}$ satisfies \eqref{eq:SLSo1_phi}-\eqref{eq:SLSo3_phi}.
	
	We first prove  \eqref{eq:SLSo1_phi}.
	For any $\vb\in\lpe{r}$, $\db\in\lpe{m}$, we have 
	$\tPhiu{G}{K} (\vb;\db) = \tub = \Kb(\tyb+\vb) = \Kb(\Gb(\tub+\db)+\vb)$. 
	Since $\Gb\in\Cs$, the previous operator equation corresponds to the recursive equation $u^{\mathrm{o}}_t = K_t(G_{t:0}(u^{\mathrm{o}}_{t-1:0}+d_{t-1:0}) + v_{t:0})$
    which implies that $u^{\mathrm{o}}_t$ depends on its own past values and on $v_{t:0}$ and $d_{t-1:0}$. 
    Hence, for any $t$, $u^{\mathrm{o}}_t = \Phi^{\mathrm{o}}_{t,\mathbf{G},\mathbf{K}} (v_{t:0}; d_{t-1:0})$.
	Thus, $\tPhiu{G}{K}(\vb;\db)$ is causal on its first input and strictly causal on its second input, meaning that \eqref{eq:SLSo1_phi} holds.
	
	Second, we prove that \eqref{eq:SLSo2_phi} holds.
	Per definition of the closed-loop maps, we have that $\tyb = \tPhiy{G}{K}(\vb;\db)$ and $\ub = \Phiu{G}{K}(\vb;\db)$ satisfy the closed-loop dynamics \eqref{eq:system_control}. 
	Thus, for any $\vb\in\lpe{r}$ and $\db\in\lpe{m}$ \eqref{eq:SLSo2_phi} holds, i.e. $\tPhiy{G}{K}(\vb;\db) = \tyb = \Gb\ub = \Gb\Phiu{G}{K}(\vb;\db)$.

	Next, we prove that \eqref{eq:SLSo3_phi} holds.
	Note that, from Definition~\ref{def:closed_loop_maps}, we have
	$\Kb\yb = \tub$ or equivalently $\Kb \Phiy{G}{K} (\vb;\db) = \tPhiu{G}{K}(\vb;\db)$.
	Thus, it holds that 
	$\Kb\IO\Phiyu{G}{K} = \tPhiu{G}{K}$. Since the inverse $\Phiyu{G}{K}^{-1}$ exists due to Proposition~\ref{prop:phi_invertible}, we can state that
	\begin{equation}
		\Kb\IO = \tPhiu{G}{K}(\Phiyu{G}{K})^{-1} \,.
	\end{equation}
	Moreover, it is clear that the operator $\Kb\IO$ is invariant to its second input, i.e.,
	for any signal $\ab \in\lpe{r}$ 
    and any $\bb,\bb'\in\lpe{m}$, we have
	$\Kb \IO \bvec{\ab}{\bb} = \Kb \ab = \Kb \IO \bvec{\ab}{\bb'}$.
	Thus,
	$\tPhiu{G}{K} (\Phiyu{G}{K} )^{-1} = \tPhiu{G}{K} (\Phiyu{G}{K} )^{-1} \IOOO$.
	Then, by composing it with $\Phiyu{G}{K} = (\Phiy{G}{K} ; \Phiu{G}{K})$, we obtain \eqref{eq:SLSo3_phi}.
	
	Finally, \eqref{eq:SLSo3_phi} can be equivalently written as $\tub = \tPhiu{\Gb}{\Kb} \Phiyu{\Gb}{\Kb}^{-1} \IOv \yb$,
	implying that the controller $\Kb$ can is described by
	$\Kb = \tPhiu{\Gb}{\Kb} \Phiyu{\Gb}{\Kb}^{-1} \IOv$, 
	i.e., \eqref{eq:SLSoK_phi}. 
	This concludes the proof.
\end{pf}

\section{Proof of Proposition~\ref{prop:invertibility}}\label{app:proof_invertibility}
\begin{pf}
	Given $(\yb;\ub)$, we want to find $(\vb;\db)$ such that $\Psiyu(\vb;\db) = (\yb;\ub)$. 
	Equivalently, we can write $(\vb;\db) = (\yb;\ub) - \tPsiyu(\vb;\db)$ and split it into two relations
	$\vb = \yb - \tPsiy(\vb;\db)$ and $\db = \ub - \tPsiu(\vb;\db)$.
	Since $\tPsiy\in\Css$ and $\tPsiu\in\Ccs$ by assumption, we can rewrite the 
	two expressions in the recursive form
	\begin{align}
		v_t &= y_t - \Psi^{y^\mathrm{o}}_t(v_{t-1:0}; d_{t-1:0}) \,,
		\label{eq:prop1_vt} \\
		d_t &= u_t - \Psi^{u^\mathrm{o}}_t(v_{t:0}; d_{t-1:0}) \,,
		\label{eq:prop1_dt}
	\end{align}
	where, in \eqref{eq:prop1_dt}, $d_t$ depends on $v_t$ whose expression is given by \eqref{eq:prop1_vt}.
	Then, \eqref{eq:prop1_vt} and \eqref{eq:prop1_dt} prove the existence and uniqueness of $(\vb;\db)$, 
	along with providing an algorithm for their computation.	
\end{pf}

\section{Proof of Theorem~\ref{th:achiev_of}}\label{app:proof_th_achiev_of}

\begin{pf} 
	We split the proof in three parts: necessity, sufficiency and uniqueness.

    \noindent
    (1) \emph{Necessity:} 
    We prove that given a $\Kb \in \Cc$, 
    the closed-loop map $\Phiyu{G}{K}$ satisfies \eqref{eq:SLSo} for 
    \begin{equation*}
        \Psiy = \Phiy{G}{K} \text{ and } \Psiu = \Phiu{G}{K} \,.
    \end{equation*}
    This is straightforward since it follows from Proposition~\ref{prop:phi_achievable}.

    \noindent
    (2) \emph{Sufficiency:}
    We prove that given the operators $(\Psiy,\Psiu)$ that satisfy \eqref{eq:SLSo}, there exists a $\Kb \in \Cc$ such that $(\Psiy,\Psiu)$ are the induced closed-loop maps $(\Phiy{G}{K},\Phiu{G}{K})$ of the plant $\Gb$.

    First, note that from \eqref{eq:SLSo1}, \eqref{eq:SLSo2} and since $\Gb\in\Cs$, we have that $\tPsiy\in\Css$.
    Then, using Proposition~\ref{prop:invertibility}, $\Psiyu^{-1}$ exists and it is causal.
    Let us now set
    \begin{equation}
        \Kb' = \tPsiu \Psiyu^{-1}\IOv\,,
        \label{eq:K_choice}
    \end{equation}
    and note that $\Kb' \in \Cc$ since 
    \eqref{eq:SLSo1} holds and
    $\Psiyu^{-1} \in \Cc$. 
    Since \eqref{eq:SLSo3} is equivalent to 
    $\tPsiu \Psiyu^{-1} \IOv \IO = \tPsiu \Psiyu^{-1}$, then, we have that 
    \begin{equation}\label{eq:K-IO}
        \Kb' \IO = \tPsiu  \Psiyu^{-1}\,.
    \end{equation}
    
    It remains to prove that \eqref{eq:K_choice} is such that the resulting control policy 
    achieves the closed-loop maps $(\Phiy{G}{K'},\Phiu{G}{K'}) = (\Psiy,\Psiu)$.
    
    Given any $\vb\in\lpe{r}$ and $\db\in\lpe{m}$, let
    $(\yb, \ub)$ be the signals obtained when considering the feedback loop of $\Gb$ and $\Kb'$ defined in \eqref{eq:K_choice}. In other words, we have that $\yb = \Gb\ub +\vb$ and $\ub = \Kb'\yb +\db$.
    Then, stacking the equations together, we have
    \begin{equation}\label{eq:CL_choice}
        \bvec{\yb}{\ub} = \bvec{\Gb\ub + \vb}{\Kb'\yb + \db}\,.
    \end{equation}
    Multiplying the left hand side of \eqref{eq:CL_choice} by $\I = \Psiyu\, \Psiyu^{-1}$, we have
    \begin{equation*}
        \bvec{\Psiy \Psiyu^{-1} (\yb;\ub)}{\Psiu \Psiyu^{-1} (\yb;\ub)} = \bvec{\Gb\ub + \vb}{\Kb'\yb+\db}\,.
    \end{equation*}
    Using the closed-loop dynamics \eqref{eq:CL_choice} and the fact that $\Kb'\yb = \Kb'\IO (\yb;\ub)$, 
    we can equivalently write the previous equation as
    \begin{align}
        \Psiy \Psiyu^{-1} (\yb;\ub) &= \Gb(\Kb'\IO (\yb;\ub) + \db) + \vb \,,
        \label{eq:line1} \\
        \Psiu \Psiyu^{-1} (\yb;\ub) &= \Kb'\IO (\yb;\ub) + \db\,.
        \label{eq:line2}
    \end{align}
    Then, from \eqref{eq:line2},
     \eqref{eq:K-IO}, and since $\Psiu = \tPsiu + \OI$, we have:
    \begin{align}
        \Psiu \Psiyu^{-1} (\yb;\ub) &= \Kb'\IO (\yb;\ub) + \db \,,\\
        (\tPsiu+\OI) \Psiyu^{-1}  (\yb;\ub) &= \tPsiu\Psiyu^{-1}(\yb;\ub) + \db \,,\nonumber\\
        \OI\Psiyu^{-1}(\yb;\ub) &= \db \,.
        \label{eq:proof_d}
    \end{align}
    Moreover, from \eqref{eq:line1}, \eqref{eq:K-IO} and since $\Psiy = \tPsiy + \IO$, we have
    \begin{equation*}
        (\tPsiy + \IO) \Psiyu^{-1} (\yb;\ub) 
        = \Gb(\tPsiu  \Psiyu^{-1} (\yb;\ub) + \db) + \vb \,,
    \end{equation*}
    and replacing $\db$ from \eqref{eq:proof_d}, we have
    \begin{multline*}
        (\tPsiy + \IO) \Psiyu^{-1} (\yb;\ub) 
        = \\
        \Gb(\tPsiu  \Psiyu^{-1} (\yb;\ub) + \OI\Psiyu^{-1}(\yb;\ub)) + \vb \,,
    \end{multline*}
    and therefore
    \begin{multline*}
        \tPsiy  \Psiyu^{-1} (\yb;\ub)  + \IO \Psiyu^{-1} (\yb;\ub)
        = \\
        \Gb(\tPsiu  + \OI) \Psiyu^{-1} (\yb;\ub) + \vb \,.
    \end{multline*}
    Since \eqref{eq:SLSo2} holds, we have that $\tPsiy = \Gb(\tPsiu  + \OI)$. Then, 
    \begin{equation}\label{eq:proof_v}
        \IO \Psiyu^{-1} (\yb;\ub)
        =  \vb \,.
    \end{equation}
    
    Finally,  stacking together \eqref{eq:proof_v} and \eqref{eq:proof_d}, one obtains
    \begin{align*}
        \bvec{\IO \Psiyu^{-1}(\yb;\ub)}{\OI \Psiyu^{-1}(\yb;\ub)} &= \bvec{\vb}{\db} \,,\\
        \bvec{\IO}{\OI} \Psiyu^{-1}(\yb;\ub) &= \bvec{\vb}{\db} \,,\\
        \bvec{\yb}{\ub} &= \Psiyu \bvec{\vb}{\db} \,.
    \end{align*}
    Thus, $\Psiyu$ is indeed the closed-loop map $\Phiyu{G}{K'}$.
    
    \noindent
    (3) \emph{Uniqueness:}
    Assume that $\Psiyu = \Phiyu{G}{K'} = \Phiyu{G}{K''}$ for some $\Kb', \Kb''$. Then, for any $\vb\in\lpe{r}$ and any $\db\in\lpe{m}$, 
    it holds
    \begin{equation*}
        \tPsiu(\vb;\db) = \Kb' \IO \Psiyu (\vb;\db) = \Kb'' \IO \Psiyu (\vb;\db)\,.
    \end{equation*}
    Since $\Psiyu$ is invertible, it implies $\Kb' \IO = \Kb'' \IO$.
    Then, we have that for any $\ab \in\lpe{r}$ and any $\bb\in\lpe{m}$
    \begin{equation*}
        \Kb' \IO \bvec{\ab}{\bb} = \Kb'' \IO \bvec{\ab}{\bb}\,.
    \end{equation*}
    Thus, $\Kb' \ab = \Kb'' \ab $ and then $\Kb' = \Kb''$.
\end{pf}

\section{Proof of Theorem~\ref{th:SLS_M_2}}\label{app:proof_SLS_M}

Before stating the proof, we define an operator, called $\boldsymbol{\mathcal{S}}$, which will be useful for the proof and 
we highlight a property it enjoys that plays a key role in the proof.
\begin{definition}\label{def:S}
	Consider two operators $\Gb\in\Cs(\lpe{m} , \lpe{r})$ and  $\Emme\in\Ccs(\lpe{r}\times\lpe{m} ,  \lpe{m})$. 
	Define 
	$\boldsymbol{\mathcal{S}} \in\Ccs(\lpe{r}\times\lpe{m} ,  \lpe{r}\times\lpe{m})$ 
	as the operator given by 
	$(\betab,\deltab) = \boldsymbol{\mathcal{S}} (\vb,\db)$, 
	where 
	$\betab = \vb + \Gb(\db-\deltab) -\Gb(\Ob)$ 
	and 
	$\deltab = -\Emme (\betab,\deltab)$.
\end{definition}
Note that, since $\Gb\in\Cs$ and $\Emme\in\Ccs$, one has that the operator $\boldsymbol{\mathcal{S}}$ belongs to $\Ccs$ and can be recursively implemented as follows
\begin{align}
	\beta_t &= v_t + G_t(d_{t-1:0} - \delta_{t-1:0}) -G_t(0_{t-1:0})\,, \label{eq:app_beta}\\
	\delta_t &= -\mathcal{M}_t(\beta_{t:0};\delta_{t-1:0})\,, \label{eq:app_delta}
\end{align}
for $t=0,1,\dots$.
\begin{proposition}\label{prop:S_invariant}
	Consider a dynamical system $\Gb$ and an operator $\Emme\in\Ccs$. Then the operator $\boldsymbol{\mathcal{S}}$  in Definition~\ref{def:S} satisfies 
	$\boldsymbol{\mathcal{S}}\boldsymbol{\mathcal{S}}(\vb;\db) = \boldsymbol{\mathcal{S}}(\vb;\db)$,
	for any signals $\vb\in\lpe{r}$ and $\db\in\lpe{m}$.
\end{proposition}
\begin{pf}
	For any $\vb\in\lpe{r}$ and $\db\in\lpe{m}$,
	set $(\hat{\betab}; \hat{\deltab}) = \boldsymbol{\mathcal{S}}\boldsymbol{\mathcal{S}}(\vb;\db)$
	and 
	$(\betab;\deltab) = \boldsymbol{\mathcal{S}} (\vb;\db)$. 
	We need to show that $(\hat{\betab}; \hat{\deltab}) = (\betab;\deltab)$.
	We will prove it by induction over each component.
	\begin{itemize}
		\item 
		Base case:
        We have that $\beta_0 = v_0 + g_0 - g_0 = v_0$. Moreover, $\hat{\beta}_0 = \beta_0 + g_0 - g_0 = \beta_0$. Thus, $\hat{\beta}_0 = \beta_0 = v_0$.
        Then, 
		$\hat{\delta}_0 = -\mathcal{M}_0(\hat{\beta}_0) = -\mathcal{M}_0({\beta}_0) = \delta_0$.
		\item 
		Inductive step:
		Assume that $\hat{\beta}_{t:0} = \beta_{t:0}$ and $\hat{\delta}_{t:0} = \delta_{t:0}$. Then,
		$\hat{\beta}_{t+1} 
		= 
		\beta_{t+1} + G_{t+1}(\delta_{t:0}-\hat{\delta}_{t:0}) - G_{t+1}(0_{t:0})
		=
		\beta_{t+1} + G_{t+1}(\delta_{t:0}-{\delta}_{t:0}) - G_{t+1}(0_{t:0})
		=
		\beta_{t+1} + G_{t+1}(0_{t:0}) - G_{t+1}(0_{t:0})
		=
		\beta_{t+1}
		$
		and 
		$\hat{\delta}_{t+1} 
		=
		-\mathcal{M}_{t+1}(\hat{\beta}_{t+1:0},\hat{\delta}_{t:0}) 
		=
		-\mathcal{M}_{t+1}({\beta}_{t+1:0},{\delta}_{t:0}) 
		=
		\delta_{t+1}
		$.
	\end{itemize}	
\end{pf}

We are now ready to prove Theorem~\ref{th:SLS_M_2}.

\begin{pf}
	We start by proving point (1) for a given system $\Gb$.
	We need to show that the operator $\Psiyu$ constructed through \eqref{eq:SLSo_psi_2}
	satisfies $\Psiyu \in \CL{\Gb}$, i.e., \eqref{eq:SLSo} holds.
	
	Note that \eqref{eq:SLSo_psiu_2} can be re-written as 
	\begin{equation*}
		\tPsiu(\vb;\db) = \Emme(\betab;\deltab)  \,,
	\end{equation*}
	with
	\begin{align}
		\betab &= \vb+\Gb(\db + \tPsiu(\vb;\db)) \,, \label{eq:SLSo_beta}  \\ 
		\deltab &= -\tPsiu(\vb;\db) \,, \label{eq:SLSo_delta} 
	\end{align}
	or equivalently:
	\begin{align*}
		\tPsiu(\vb;\db) &= -\deltab \,,\\
		& \betab = \vb+\Gb(\db - \deltab) \,,\\
		& \deltab = -\Emme(\betab;\deltab) \,.
	\end{align*}
	Moreover, using Definition~\ref{def:S}, \eqref{eq:SLSo_psi_2} is equivalent to
	\begin{align}
		\tPsiu(\vb;\db) &= -\deltab \,, \label{eq:app_psiu_2}\\
		\tPsiy(\vb;\db) &= \Gb (\db-\deltab) \,, \label{eq:app_psiy_2} \\
		\text{with }& (\betab;\deltab) = \boldsymbol{\mathcal{S}} (\vb;\db) \,.\label{eq:app_beta_delta_2}
	\end{align}

	Then, we need to show that $(\tPsiy, \tPsiu)$ given by \eqref{eq:app_psiu_2}-\eqref{eq:app_beta_delta_2} satisfies \eqref{eq:SLSo}.
	
	We start by showing that \eqref{eq:SLSo1} holds. 
	Since $\boldsymbol{\mathcal{S}} \in\Ccs$ (by definition), then $\tPsiu(\vb;\db) = -\deltab$ given by \eqref{eq:app_beta_delta_2} satisfies $\tPsiu \in\Ccs$, i.e., \eqref{eq:SLSo1} holds true.
	It is also clear that \eqref{eq:SLSo2} holds since it is the same as \eqref{eq:SLSo_psiy_2}. 
	Note that \eqref{eq:SLSo2} can also be retrieved from \eqref{eq:app_psiy_2} and \eqref{eq:app_psiu_2} as follows 
	\begin{align}
		\tPsiy (\vb;\db) &= \Gb (\db-\deltab) = \Gb (\db+\tPsiu (\vb;\db)) \,, \nonumber \\
        &= \Gb \Psiu (\vb;\db) \,.
        \label{eq:aux1}
	\end{align}

	Finally, we show that \eqref{eq:SLSo3} holds.
	Define $(\hat{\betab},\hat{\deltab}) = \boldsymbol{\mathcal{S}} ({\betab},{\deltab})$, where $({\betab},{\deltab})$ are given by \eqref{eq:app_beta_delta_2}.
	From Proposition~\ref{prop:S_invariant}, we have that $(\hat{\betab},\hat{\deltab}) = ({\betab},{\deltab})$.
	Moreover, plugging $({\betab},{\deltab})$ in \eqref{eq:SLSo_psiu_2} gives
	\begin{align*}
		\tPsiu(\betab;\deltab) &= -\hat{\deltab} \,,\\
		\tPsiy(\betab;\deltab) &= \Gb (\deltab-\hat{\deltab}) \,, \\
		\text{with }& (\hat{\betab};\hat{\deltab}) = \boldsymbol{\mathcal{S}} (\betab;\deltab) \,.
	\end{align*}
	From Proposition~\ref{prop:S_invariant}, we have that $(\hat{\betab},\hat{\deltab}) = ({\betab},{\deltab})$. Then, 
	\begin{align}
		\tPsiu(\betab;\deltab) &= -\hat{\deltab} = -\deltab = \tPsiu(\vb;\db) \,, \label{eq:aux2}\\
		\tPsiy(\betab;\deltab) &= \Gb (\deltab-\hat{\deltab}) = \Gb (\deltab-\deltab) = \Gb(\Ob) \,. \label{eq:aux3}
	\end{align}
	
	Thus, replacing with \eqref{eq:aux1}, \eqref{eq:aux2} and \eqref{eq:aux3} in \eqref{eq:SLSo_beta}-\eqref{eq:SLSo_delta},
	it holds that
	\begin{align*}
		\betab &= \vb + \tPsiy(\vb;\db) - \tPsiy(\betab;\deltab)\,, \\
		\deltab &= \Ob - \tPsiu(\betab;\deltab)\,,
	\end{align*}
	whose equivalent recursive form is given by
	\begin{align*}
		\beta_t &= (v_t + \Psi^{y^\mathrm{o}}_t(v_{t-1:0};d_{t-1:0})) - \Psi^{y^\mathrm{o}}_t(\beta_{t-1:0};\delta_{t-1:0})\,, \\
		\delta_t &= 0 - \Psi^{u^\mathrm{o}}_t (\beta_{t:0};\delta_{t-1:0})\,.
	\end{align*}
	By using Proposition~\ref{prop:invertibility}, this is equal to
	\begin{align*}
		(\betab;\deltab) &= \Psiyu^{-1}\bvec{\vb + \tPsiy(\vb;\db)}{\Ob} \,,\\
		&= \Psiyu^{-1}\bvec{\Psiy(\vb;\db)}{\Ob} \,, \\
		&= \Psiyu^{-1}\bvec{\I}{\Ob} \Psiy(\vb;\db) \,.
	\end{align*}
	And since $\tPsiu(\vb;\db) = \tPsiu(\betab;\deltab)$,
	we obtain
	\begin{equation*}
		\tPsiu(\vb;\db) = \tPsiu\Psiyu^{-1}\bvec{\I}{\Ob} \Psiy(\vb;\db) \,,
	\end{equation*}
	which is \eqref{eq:SLSo3}.
	This concludes the proof of point~(1). 
	
	
	We now prove the converse, i.e., point~(2) for a given system $\Gb$.
	We need to show that for any operator $\Psiyu\in\CL{\Gb}$ there exists an operator $\Emme\in\Ccs$ such that \eqref{eq:SLSo_psi_2} is verified. 
	First, \eqref{eq:SLSo_psiy_2} holds true because it is the same as \eqref{eq:SLSo1}.
	
	Set $\Emme = \tPsiu$.
	Define $(\betab;\deltab) = \boldsymbol{\mathcal{S}}(\vb;\db)$, as per Definition~\ref{def:S} with $\Emme = \tPsiu$.
	Define $(\hat{\betab};\hat{\deltab}) = \boldsymbol{\mathcal{S}} ({\betab};{\deltab})$. 
	Using Proposition~\ref{prop:S_invariant}, we have that $(\hat{\betab};\hat{\deltab}) = ({\betab};{\deltab})$. 
	Thus, 
	$\tPsiu (\vb;\db) = -\deltab = -\hat{\deltab} = \Emme (\betab;\deltab)$, hence \eqref{eq:SLSo_psiu_2} holds true.

    The uniqueness of $\Emme$ can be proved by contradiction.
    Assume that there exists $\Emme^1\in\Ccs$ and $\Emme^2\in\Ccs$ verifying \eqref{eq:SLSo_psi_2} such that $\Emme^1 \neq \Emme^2$.
    For any $\vb\in\lpe{r}$ and $\db\in\lpe{m}$,
    we will show by induction that this is a contradiction.
    For $t=0$, one has
    $\Psi_0^{u^\mathrm{o}}(v_0) = \mathcal{M}_0^1(v_0) = \mathcal{M}_0^2(v_0)$ which implies that $\mathcal{M}_0^1 = \mathcal{M}_0^2$.
    For $t=\tau$, assume that $\mathcal{M}^1_s = \mathcal{M}^2_s$ for $0\leq s < \tau$. Then,
    \begin{multline}
        \label{eq:emme_unique1}
        \Psi_\tau^{u^\mathrm{o}}(v_{\tau:0},d_{\tau-1:0}) 
        = \mathcal{M}_\tau^1(v_{\tau:0} + G_{\tau:0}(d_{\tau-1:0} + \\
        \Psi_{\tau-1:0}^{u^\mathrm{o}}(v_{\tau-1:0}, d_{\tau-2:0} )) - G_{\tau:0}(0_{\tau-1:0}) ; \\
        -\Psi_{\tau-1:0}^{u^\mathrm{o}}(v_{\tau-1:0}, d_{\tau-2:0} ) ) ,
    \end{multline}
    and
    \begin{multline}
        \label{eq:emme_unique2}
        \Psi_\tau^{u^\mathrm{o}}(v_{\tau:0},d_{\tau-1:0}) 
        = \mathcal{M}_\tau^2(v_{\tau:0} + G_{\tau:0}(d_{\tau-1:0} + \\
        \Psi_{\tau-1:0}^{u^\mathrm{o}}(v_{\tau-1:0}, d_{\tau-2:0} )) - G_{\tau:0}(0_{\tau-1:0}) ; \\
        -\Psi_{\tau-1:0}^{u^\mathrm{o}}(v_{\tau-1:0}, d_{\tau-2:0} ) ) .
    \end{multline}
    Since the left-hand side of~\eqref{eq:emme_unique1} and~\eqref{eq:emme_unique2} are the same, and the arguments of $\mathcal{M}_\tau^1$ and $\mathcal{M}_\tau^2$ coincide. Thus, the input-output map is the same for both functions.
    Therefore $\Emme^1 = \Emme^2$ which concludes the proof.
\end{pf}

\section{Proof of Proposition~\ref{prop:rec_implememntation_M}}
\label{app:proof_prop_recursivity_M}
\begin{pf}
    For a given $\Emme$, \eqref{eq:SLSo_psiu_2} allows us to explicitly compute the operator $\tPsiu$ at every time instant, because $\Emme \in \Ccs$ and $\Gb\in\Cs$. In particular, for any sequence $(\vb;\db)$, we have the following recursive formulae for $t=0,1,\dots$:
    \begin{multline}
    	\label{eq:emme_recursive}
    	\Psi^{u^{\mathrm{o}}}_t(v_{t:0},d_{t-1:0}) 
    	= \\
    	\mathcal{M}_{t}
        \left(v_{t:0}  + 
    		  G_{t:0} \left(d_{t-1:0} + \Psi^{u^{\mathrm{o}}}_{t-1:0}(v_{t-1:0},d_{t-2:0}) \right) 
        \right.  \\ \left.
            -G_{t:0}(0_{t-1:0})
        \,;\,
    	  -\Psi^{u^{\mathrm{o}}}_{t-1:0}(v_{t-1:0},d_{t-2:0}) \right)
    	     \,.
    \end{multline}
    Observe that
    since $\Psiyu\in\CL{\Gb}$, 
    one has $u^{\mathrm{o}}_{t-1:0} = \Psi^{u^{\mathrm{o}}}_{t-1:0}(v_{t-1:0};d_{t-2:0})$,
    and, from \eqref{eq:system_dyn}, one has
    $y_{t:0} = v_{t:0}  + G_{t:0} \left(d_{t-1:0} + u^{\mathrm{o}}_{t-1:0} \right)$.
    Thus, one can notice that the operator $\Emme$ takes as argument the sequence 
    $(\yb - \yb^{\text{free}}  ; \tub)$, i.e., 
    \begin{equation}\label{eq:psiu_emme}
        \Psi^{u^{\mathrm{o}}}_t(v_{t:0} ; d_{t-1:0}) = \mathcal{M}_{t} (y_{t:0} - y^{\text{free}}_{t:0} ; -u^{\mathrm{o}}_{t-1:0} )\,.
    \end{equation}
    This implies that
    the output $\tub = \Kb \yb$ can be computed recursively through \eqref{eq:K_implementation}.
\end{pf}

\section{Proof of Theorem~\ref{th:SLS_stable}}\label{app:proof_SLS_stable}	
We start by introducing a preliminary result that will be useful for the proof of  Theorem~\ref{th:SLS_stable}.
This result is introduced in Theorem~1 by \citet{desoer1976feedback}, and we state it here for completeness.

\begin{proposition}{(Theorem~1 in~\citet{desoer1976feedback})}\label{prop:app_th_desoer}
	Consider the closed-loop system given by \eqref{eq:system_control}. Assume that 
	$\Gb\in\Lp$ and $\Gb$ is i.f.g. $\ell_p$-stable.
	Then, 
	\begin{equation*}
		\Phiyu{G}{K}  \in \Lp \Leftrightarrow \Kb(\I-\Gb\Kb)^{-1} \in \Lp \,.
	\end{equation*}
\end{proposition}

\begin{pf}
	We first prove $\Kb(\I-\Gb\Kb)^{-1} \in \Lp \Rightarrow \Phiyu{G}{K} \in \Lp$. Recall that $(\yb;\ub) = \Phiyu{G}{K} (\vb;\db)$.
	
	Consider  any $\db\in\ell_p^m$ and any $\yb\in\ell_p^r$. 
    Let 
    \begin{equation}
        \label{eq:v_tilde_def_app}
        \tilde{\vb} = \Gb(\db+\Kb\yb) - \Gb\Kb\yb\,. 
    \end{equation}
    Then, 
	\begin{align}
		\norm{\tilde{\vb}}_p &= \norm{\Gb(\db+\Kb\yb) - \Gb(\Kb\yb)}_p \,,\nonumber\\
        &\leq \gamma_{\Gb} \norm{\db+\Kb\yb-\Kb\yb}_p = \gamma_{\Gb} \norm{\db}_p\,,
        \label{eq:v_tilde_app}
	\end{align}
    where $\gamma_{\Gb}$ is the incremental $\ell_p$-gain of $\Gb$.
	Since $\db\in\ell_p^m$, we have that $\tilde{\vb}\in\ell_p^r$.
	From the system equation \eqref{eq:system_control}, for any $\vb\in\ell_p^r$,  we have that
		$\vb = \yb - \Gb\ub = \yb - \Gb(\Kb\yb+\db)$.
	Then,
	\begin{align*}
		\vb +\tilde{\vb} 
		&= \yb - \Gb(\Kb\yb+\db) + \Gb(\db + \Kb\yb) - \Gb\Kb\yb \,, \\
		&= \yb - \Gb\Kb\yb \,, \\
		&= (\I - \Gb\Kb) \yb \,. 
	\end{align*}
	Since $\Gb\Kb\in\Cs$, then $(\I - \Gb\Kb)^{-1}$ exists and is causal from Corollary~\ref{cor:invertibility_ho}.
	Thus, $(\I - \Gb\Kb)^{-1}(\vb +\tilde{\vb}) = \yb$ and, hence $\Kb(\I - \Gb\Kb)^{-1}(\vb +\tilde{\vb}) = \Kb\yb$.
	Note that the left hand side is an $\Lp$ operator, since $\vb,\tilde{\vb}\in\ell_p^r$ and $\Kb(\I - \Gb\Kb)^{-1}\in\Lp$ by assumption. Thus, $\Kb\yb\in\ell_p^m$.
	Then, $\ub = \Kb\yb+\db \in \ell_p^m$.
	Finally, since $\Gb\in\Lp$ by assumption, we have that $\yb = \Gb\ub + \vb\in\ell_p^r$.
	Thus, $\Phiyu{G}{K}\in\Lp$.
	
	We now prove that $\Phiyu{G}{K} \in \Lp \Rightarrow \Kb(\I-\Gb\Kb)^{-1} \in \Lp$.
	
	For any $\db\in\ell_p^m$ and any $\vb\in\ell_p^r$, consider $\tilde{\vb}$ as defined in \eqref{eq:v_tilde_def_app}. Then, $\tilde{\vb}\in\ell_p^r$.
	
	Let us consider the case where $\db=\Ob$. Then, from \eqref{eq:v_tilde_app} $\tilde{\vb} = \Ob$.
	Then, the mapping $\vb\mapsto\ub$ is given by 
	\begin{equation*}
		\ub = \Kb\yb + \Ob = \Kb(\I - \Gb\Kb)^{-1}(\vb +\Ob ) = \Kb(\I - \Gb\Kb)^{-1}(\vb) \,.
	\end{equation*}
	Finally, since $\Phiyu{G}{K}\in\Lp$, we have that $\Phiu{G}{K} (\vb,\Ob)\in\ell_p^m$. 
	Thus, for any $\vb\in\ell_p^r$, we have that $\ub = \Kb(\I - \Gb\Kb)^{-1} \,\vb \in \ell_p^m$. Hence, $\Kb(\I - \Gb\Kb)^{-1}\in\Lp$.
\end{pf}

We are now ready to prove Theorem~\ref{th:SLS_stable}.

\begin{pf}
	We assume that an i.f.g. $\ell_p$-stable plant $\Gb\in\Lp$ is given.
	We split the proof into two parts: sufficiency and necessity.
	
	\noindent
    (1)
	\textit{Sufficiency}: We show that the operator $\Psiyu$ given 
	in Theorem~\ref{th:SLS_M_2} with $\Emme$ as per \eqref{eq:M_from_Q}
	is $\Lp$ for any $\Qb \in\Lp$.
	
	In other words, we want to show that for any $\vb\in\ell_p^r$ and any $\db\in\ell_p^m$, we have that $\yb = \tyb+\vb = \Psiy(\vb;\db) \in \ell_p^r$ and $\ub = \tub+\db = \Psiu(\vb;\db) \in \ell_p^m$.
	
	Using 
	Theorem~\ref{th:SLS_M_2}, one has $\tub = \Emme (\betab; \deltab)$
    with 
    $\betab = \vb + \Gb(\db-\deltab) -\Gb(\Ob)$ 
	and 
	$\deltab = -\Emme (\betab;\deltab)$.
    Using the definition of the operator $\Emme$ in \eqref{eq:M_from_Q}, one has
    \begin{align}
		\tub &= \Qb (\betab - \Gb(-\deltab) +\Gb(\Ob) ) \,, \nonumber \\
        &= \Qb(\vb +\Gb(\db-\deltab) - \Gb(-\deltab) +\Gb(\Ob) )\,.
        \label{eq:prop_SLS_stable_eq1}
    \end{align}
	Moreover, due to $\Gb$ being i.f.g. $\ell_p$-stable,  there exists $\gamma_{\Gb}>0$ such that
	\begin{equation}\label{eq:prop_SLS_stable_eq2}
		\norm{\Gb(\db-\deltab) - \Gb(-\deltab)}_p \leq \gamma_{\Gb}\norm{\db-\deltab + \deltab}_p = \gamma_{\Gb}\norm{\db}_p\,,
	\end{equation} 
    and since $\Gb\in\Lp$, one has $\Gb(\Ob)\in\ell_p^r$.
	Then, for any $\vb\in\ell_p^r$ and any $\db\in\ell_p^m$, one has 
	\begin{equation*}
	 	\vb + \Gb(\db-\deltab) - \Gb(-\deltab) +\Gb(\Ob) \in \ell_p^r \,.
	\end{equation*}
 	From \eqref{eq:prop_SLS_stable_eq1}, it follows that, for any $\Qb\in\Lp$, we have $\tub \in\ell_p^m$ which implies $\tPsiu\in\Lp$ and $\Psiu\in\Lp$.
 	Moreover, from Theorem~\ref{th:SLS_M_2}, one has that $\tPsiy = \Gb\Psiu$. Thus $\tPsiy \in \Lp$ and hence $\Psiy \in \Lp$.

 	\noindent
    (2)
 	\textit{Necessity}: We show that given an $\ell_p$-stable closed-loop map $\hat{\Psiyu}$, there exists a $\Qb\in\Lp$ such that 
    constructing $\Psiyu$ using Theorem~\ref{th:SLS_M_2} with $\Emme$ as per \eqref{eq:M_from_Q}, one has $\hat{\Psiyu} = \Psiyu$.
 	
 	Since $\hat{\Psiyu}$ is an $\ell_p$-stable closed-loop map for the plant $\Gb$, i.e. $\hat{\Psiyu}\in\CLp{\Gb}$, then \eqref{eq:SLSo}-\eqref{eq:SLSo3} are satisfied for $\hat{\Psiyu}$ and we can obtain the corresponding $\Kb$ from \eqref{eq:SLSoK}.
 	Then, for any $\vb\in\ell_p^r$ and any $\db\in\ell_p^m$, one has
    \begin{equation}\label{eq:phi_app_MQ}
        (\yb;\ub) = \Phiyu{\Gb}{\Kb} (\vb;\db) = \hat{\Psiyu}(\vb;\db)\,.
    \end{equation}
 	Next, set 
 	\begin{equation}\label{eq:prop_SLS_stable_eq4}
 		\Qb = \Kb(\I-\Gb\Kb)^{-1}\,,
 	\end{equation}
 	where Corollary~\ref{cor:invertibility_ho} ensures the existence of the inverse since $\Gb\Kb\in\Cs$.
    Note that, from Proposition~\ref{prop:app_th_desoer}, we have that $\Qb \in \Lp$.  
    With $\Qb$ as defined in \eqref{eq:prop_SLS_stable_eq4}, one can obtain $\Psiyu$ using Theorem~\ref{th:SLS_M_2} with $\Emme$ as per \eqref{eq:M_from_Q}.
    It reads
    \begin{equation} \label{eq:emme_from_Q_app}
        \Emme = \Kb(\I-\Gb\Kb)^{-1} \left( \IO - \Gb \OnegI + \Gb \OO \right) \,.
    \end{equation}
    Moreover, from Theorem~\ref{th:SLS_M_2}, point (1), we have that, $\Emme$ and $\Gb$ define the operator $\Psiyu = (\Psiy;\Psiu) = (\tPsiy;\tPsiu) + \I$ as per
    \begin{subequations}
        \label{eq:psiy_app_MQ}
        \begin{align}
        \tPsiu(\vb;\db) 
        &= 
        \begin{multlined}[t]
        \Kb(\I-\Gb\Kb)^{-1} \\  
        \left( \IO - \Gb \OnegI + \Gb \OO \right) (\betab;\deltab) \,,
        \end{multlined} 
        \\
        \tPsiy(\vb;\db) 
        &= 
        \Gb \Psiu(\vb;\db) \,, 
        \end{align}
    \end{subequations}
    where
    \begin{align}
        \betab &= \vb + \Gb(\db + \tPsiu(\vb;\db)) - \Gb(\Ob) \,, \label{eq:beta_app_MQ}\\
        \deltab &= - \tPsiu(\vb;\db) \label{eq:delta_app_MQ}\,.
    \end{align}
    Denote 
    $\boldsymbol{\eta} = (\I-\Gb\Kb)^{-1}(\betab-\Gb(-\deltab) +\Gb(\Ob))$. 
    Then, 
    \begin{equation}\label{eq:psiu_app_MQ}
        \tPsiu(\vb;\db) = \Kb\boldsymbol{\eta}\,.
    \end{equation}
    We use the recursive algorithm provided in Corollary~\ref{cor:invertibility_ho} for the inverse computation appearing in the definition of $\boldsymbol{\eta}$.
    It reads
    $\eta_t = \beta_t-G_t(-\delta_{t-1:0}) + G_t(0_{t-1:0}) + G_t (K_{t-1:0}(\eta_{t-1:0})) $, for $t=0,1,\dots$.
    Moreover, from \eqref{eq:delta_app_MQ} and \eqref{eq:psiu_app_MQ}, one has $-\delta_{t-1:0} = \Psi^{u^\mathrm{o}}_{t-1:0}(v_{t-1:0};d_{t-2:0}) = K_{t-1:0}(\eta_{t-1:0})$.
    Thus, $\eta_t = \beta_t + G_t(0_{t-1:0}) $ for all $t=0,1,\dots$, implying that $\boldsymbol{\eta} = \betab + \Gb(\Ob)$.

    Thus, \eqref{eq:psiu_app_MQ} reads $\tPsiu(\vb;\db) = \Kb\boldsymbol{\eta}$, and using \eqref{eq:beta_app_MQ}, one has
    \begin{align*}
        \tPsiu(\vb;\db) 
        &= \Kb ( \vb + \Gb(\db + \tPsiu(\vb;\db)) \hspace{-1pt}-\hspace{-1pt} \Gb(\Ob) \hspace{-1pt}+\hspace{-1pt} \Gb(\Ob) ) ,\\
        &= \Kb \left( \vb + \Gb(\Psiu(\vb;\db))  \right) \,.
    \end{align*}
    Moreover, using \eqref{eq:psiy_app_MQ}, one obtains
    \begin{equation}
    \label{eq:psiu2_app_MQ}
        \tPsiu(\vb;\db) = \Kb \left( \vb + \tPsiy(\vb;\db)  \right) = \Kb \left(  \Psiy(\vb;\db)  \right) \,.
    \end{equation}
    
    Finally, note that \eqref{eq:psiy_app_MQ} and \eqref{eq:psiu2_app_MQ} are indeed the definition of the closed-loop map $\Phiyu{G}{K}$. Thus, 
    the choice of $\Qb$ as per \eqref{eq:prop_SLS_stable_eq4} leads to the closed-loop map $\Phiyu{\Gb}{\Kb}$, i.e.,
    $\Psiyu =\Phiyu{G}{K} = \hat{\Psiyu}$.
\end{pf}

\section{Proof of Proposition~\ref{prop:rec_implemenatation_KQ}}
\label{app:proof_rec_implementation_KQ}
\begin{pf}
To retrieve the controller $\Kb$ associated with a given $\Qb\in\Lp$ one can use the recursive implementation \eqref{eq:K_implementation}, where, according to Theorem~\ref{th:SLS_stable}, $\Emme$ is given in \eqref{eq:M_from_Q}.
Then, we have that
$u^\mathrm{o}_t 
= \mathcal{Q}_t(y_{t:0} - G_{t:0}( u^\mathrm{o}_{t-1:0})) $, since $ y^{\text{free}}_{t:0} = G_{t:0}(0_{t-1:0})$.
By defining $\omega_t = y_{t} - G_{t}( u^\mathrm{o}_{t-1:0})$, one has 
$u^\mathrm{o}_t =  \mathcal{Q}_t(\omega_{t:0})$ and $\omegab$ can be computed recursively through \eqref{eq:recursive_implementation_omega_1}.
\end{pf}

\section{Proof of Theorem~\ref{th:stabilizing_SLS}}\label{app:proof_stabilizing}
\begin{pf}
    We start by proving point (1).
    We assume that $\Phiyu{G}{K'}\in\Lp$. Thus, for any $\vb\in\ell_p^r$ and any $\db,\boldsymbol{\nu} \in \ell_p^m$, one has
    $(\yb; \ub) = \Phiyu{G}{K'}(\vb;\db+\boldsymbol{\nu}) \in\ell_p$ and 
    $(\tyb; \tub) = \tPhiyu{G}{K'}(\vb;\db+\boldsymbol{\nu}) \in\ell_p$.

    Choose $\boldsymbol{\nu} = \Qb(\tomegab)$, 
    with $\tomegab$ as defined in \eqref{eq:omega_prestabilized}.
    Observe that $\Qb$ and $\Kb'$ defines a new controller $\Kb$ as in \eqref{eq:stabilizing_input}, whose closed-loop map is $\Phiyu{G}{K}:(\vb;\db)\mapsto (\yb;\ub)$.
    We will show that for any $\Qb\in\Lp$, one has $\Phiyu{G}{K}\in\Lp$.

    Notice that, 
	\begin{align*}
		\norm{\tomegab}_p 
		&= \norm{\yb-\Gb(\tub)}_p \,,\\
		&= \norm{\vb + \Gb(\tub+\db) - \Gb(\tub)}_p \,,\\
		&\leq \norm{\vb}_p + \gamma_G \norm{\tub+\db-\tub}_p \,,\\
		&= \norm{\vb}_p + \gamma_G \norm{\db}_p\,,
	\end{align*}
    where $\gamma_G$ is the incremental finite gain of $\Gb$, that exists since $\Gb$ is i.f.g $\ell_p$-stable.
	Then, for any $\vb\in\ell_p^r$ and any $\db\in\ell_p^m$, we have that $\tomegab\in\ell_p^r$. 
	Thus, for any $\Qb\in\Lp$, we have that $\boldsymbol{\nu}\in\ell_p^m$, hence the closed-loop system satisfies $\Phiyu{\Gb}{\Kb} \in \CLp{\Gb}$.

	We prove point (2).
    We are given $\Psiyu \in \CLp{G}$ and $\Kb'\in\Lp$ such that $\Phiyu{G}{K'}\in\Lp$.
    For any $\vb\in\ell_p$ and any $\db\in\ell_p$,
    consider the signal $(\vb;\db+\boldsymbol{\nu})$ 
    with 
    $\boldsymbol{\nu} = (\tPsiu - \Kb'\Psiy) (\vb;\db)$. 
    Since $\Kb', \Psiy, \Psiu \in \Lp$ and $\vb,\db\in\ell_p$, one has that $\boldsymbol{\nu}\in\ell_p$.
    Consider now the map $\hat{\Psiyu}: (\vb;\db) \mapsto (\yb;\ub)$ such that $(\yb;\ub) = \Phiyu{G}{K'}(\vb;\db+\boldsymbol{\nu}) = \Phiyu{G}{K'}(\vb;\db+(\tPsiu - \Kb'\Psiy) (\vb;\db))$. 
    This map satisfies $\hat{\Psiyu}\in\Lp$.
    We will first show that this choice of $\boldsymbol{\nu}$ leads to $\hat{\Psiyu} = \Psiyu$. 
    Then, we will show that $\boldsymbol{\nu}$ can be equivalently retrieved as $\boldsymbol{\nu} = \Qb (\tomegab)$, with $\tomegab$ as defined in \eqref{eq:omega_prestabilized}.
    
    From the definition of $\hat{\Psiyu}$, one has 
    \begin{align*}
        \hat{\boldsymbol{\Psi}}^{\yb}(\vb;\db) 
        &= \Gb(\hat{\boldsymbol{\Psi}}^{\ub}(\vb;\db)) + \vb \,,  
        \\
        \hat{\boldsymbol{\Psi}}^{\ub}(\vb;\db) 
        &= 
        \begin{multlined}[t]
            \Kb'(\hat{\boldsymbol{\Psi}}^{\yb}(\vb;\db)) + \db  \\
                + \tPsiu(\vb;\db) - \Kb'(\Psiy (\vb;\db)) \,,
        \end{multlined}
    \end{align*}
    and, since $\Psiyu \in \CLp{G}$, there exists a $\Kb\in\Cc$ such that 
    \begin{align*}
        \Psiy(\vb;\db) &= \Gb(\Psiu(\vb;\db)) + \vb \,,\\
        \Psiu(\vb;\db) &=\Kb(\Psiy (\vb;\db)) + \db \,.
    \end{align*}
    We prove that $\hat{\Psiyu} = \Psiyu$ by induction.
    The base case is
    \begin{equation*}
        \hat{\Psi}^y_0(v_{0}) =  g_0 + v_0 = \Psi^y_0(v_{0}) \,.
    \end{equation*}
    Moreover,
    \begin{align*}
        \hat{\Psi}^u_0(v_{0};d_0) 
        &= K_0'(\hat{\Psi}^y_0(v_{0})) \hspace{-1pt}+\hspace{-1pt} d_0 + \Psi^{u^\mathrm{o}}_0(v_{0}) \hspace{-1pt}-\hspace{-1pt} K_0'({\Psi}^y_0(v_{0}))  \,,\\
        &= \Psi^{u}_0(v_{0};d_0) \,.
    \end{align*}
    As for the inductive step,
    assume that $\hat{\Psi}_i = \Psi_i$ for $0 \leq i \leq j$. 
    We aim to show that $\hat{\Psi}_{j+1} = \Psi_{j+1}$.
    One has
    \begin{align*}
        \hat{\Psi}^y_{j+1}(v_{j+1:0};d_{j:0}) 
        &= G_{j+1}(\hat{\Psi}^u_{j:0}(v_{j:0};d_{j:0})) + v_{j+1} \,,\\
        &= G_{j+1}({\Psi}^u_{j:0}(v_{j:0};d_{j:0})) + v_{j+1} \,,\\
        &= {\Psi}^y_{j+1}(v_{j+1:0};d_{j:0}) \,.
    \end{align*}
    Moreover,
    \begin{align*}
        &\hat{\Psi}^u_{j+1}(v_{j+1:0};d_{j+1:0}) \\
        &= 
        \begin{multlined}[t]
        K_{j+1}'(\hat{\Psi}^y_{j+1:0}(v_{j+1:0};d_{j:0})) + d_{j+1} \\
        + \Psi^{u^\mathrm{o}}_{j+1}(v_{j+1:0};d_{j:0}) - K_{j+1}'({\Psi}^y_{j+1:0}(v_{j+1:0};d_{j:0})) , 
        \end{multlined}
        \\
        &= 
        \begin{multlined}[t]
        K_{j+1}'(\hat{\Psi}^y_{j+1:0}(v_{j+1:0};d_{j:0})) + \Psi^{u}_{j+1}(v_{j+1:0};d_{j+1:0}) \\
        - K_{j+1}'(\hat{\Psi}^y_{j+1:0}(v_{j+1:0};d_{j:0}))\,, 
        \end{multlined}
        \\
        &= \Psi^{u}_{j+1}(v_{j+1:0};d_{j+1:0}) \,.
    \end{align*}
    Thus, $\hat{\Psiyu} = \Psiyu$, i.e., we have shown that 
    $\hat{\Psiyu}: (\vb;\db) \mapsto (\yb;\ub) = \Phiyu{G}{K'}(\vb;\db+(\tPsiu - \Kb'\Psiy) (\vb;\db))$ achieves the closed-loop maps $\Psiyu$.

    Now, let us define $\tomegab = (\Psiy - \Gb\tPsiu)(\vb;\db)$. Note that, since $\Gb, \Psiy, \tPsiu \in\Lp$, then, for any $\vb\in\ell_p$ and any $\db\in\ell_p$, we have that $\tomegab\in\ell_p$.

    We now show that the signal $\boldsymbol{\nu}$ can be equivalently reconstructed using a new operator $\Qb$, as per  $\boldsymbol{\nu} = \Qb(\tomegab)$.
    Define $\Qb$ as the operator that for $t=0,1,\dots$ satisfies $\hat{\nu}_t = \mathcal{Q}_t(\tilde{\omega}_{t:0})$ as per:
    \begin{align*}
        \hat{y}_t &= \tilde{\omega}_t + G_t(\hat{u}^{\mathrm{o}}_{t-1:0}) \,, \\
        \hat{u}^{\mathrm{o}}_t &= \Psi^{u^\mathrm{o}}_t(\hat{y}_{t:0} - G_{t:0}(0_{t-1:0}) ; -\hat{u}^{\mathrm{o}}_{t-1:0}) \,, \\
        \hat{\nu}_t &= \hat{u}^{\mathrm{o}}_t - K_{t}'(\hat{y}_{t:0}) \,, 
    \end{align*}
    where $\hat{y}_t$ and $\hat{u}^{\mathrm{o}}_t$ represent the \textit{internal states} of the operator. 
    Next, we show that, for $t=0,1,\dots$, one has $\hat{\nu}_t = \nu_t$.
    We prove it by induction. 
    As for the base case,
    note that 
    \begin{equation*}
        \hat{y}_0 
        = \tilde{\omega}_0 + g_0 
        = \Psi^y_0(v_0) - g_0 + g_0 
        = \Psi^y_0(v_0) \,,
    \end{equation*}
    where
    $\Psi^y_0(v_0) = g_0 + v_0$,
    and
    \begin{equation*}
        \hat{u}^{\mathrm{o}}_0 
        = \Psi^{u^\mathrm{o}}_0(\hat{y}_{0} - g_0) 
        = \Psi^{u^\mathrm{o}}_0(g_0 + v_0 - g_0) 
        = \Psi^{u^\mathrm{o}}_0(v_0) \,.
    \end{equation*}
    Then,
    \begin{align*}
        \hat{\nu}_0 
        &= \hat{u}^{\mathrm{o}}_0 - K_{0}'(\hat{y}_{0}) \,, \\
        &= \Psi^{u^\mathrm{o}}_0(v_0) - K_{0}'(\Psi^y_0(v_0)) \,, \\
        &= \nu_0 \,.
    \end{align*}
    The inductive step proceeds as follows. 
    Assume that, for $0 \leq i \leq j$, 
    $\hat{y}_i = \Psi^y_i(v_{i:0};d_{i-1:0})$,
    $\hat{u}^{\mathrm{o}}_i = \Psi^{u^\mathrm{o}}_i(v_{i:0};d_{i-1:0})$ 
    and  
    $\hat{\nu}_i = \nu_i$.
    We show that 
    $\hat{y}_{j+1} = \Psi^y_{j+1}(v_{j+1:0};d_{j:0})$,
    $\hat{u}^{\mathrm{o}}_{j+1} = \Psi^{u^\mathrm{o}}_{j+1}(v_{j+1:0};d_{j:0})$
    and $\hat{\nu}_{j+1} = \nu_{j+1}$.

    Let us start with $\hat{y}_{j+1}$.
    \begin{align*}
        \hat{y}_{j+1} 
        &= \tilde{\omega}_{j+1} + G_{j+1}(\hat{u}^{\mathrm{o}}_{j:0})\,, \\
        &= 
        \begin{multlined}[t]
        \Psi^{y}_{j+1}(v_{j+1:0};d_{j:0}) - G_{j+1}(\Psi^{u^\mathrm{o}}_{j:0}(v_{j:0};d_{j-1:0})) \\
        + G_{j+1}(\Psi^{u^\mathrm{o}}_{j:0}(v_{j:0};d_{j-1:0})) \,,
        \end{multlined}
        \\
        &= \Psi^{y}_{j+1}(v_{j+1:0};d_{j:0}) \,.
    \end{align*}
    Then,
    \begin{align*}
        \hat{u}^{\mathrm{o}}_{j+1} 
        &= \Psi^{u^\mathrm{o}}_{j+1}(\hat{y}_{j+1:0} - G_{j+1:0}(0_{j:0}) ; -\hat{u}^{\mathrm{o}}_{j:0}) \,, \\
        &= 
        \begin{multlined}[t]
        \Psi^{u^\mathrm{o}}_{j+1}(\Psi^{y}_{j+1:0}(v_{j+1:0};d_{j:0}) - G_{j+1:0}(0_{j:0}) ; \\
        -\Psi^{u^\mathrm{o}}_{j:0}(v_{j:0};d_{j-1:0})) \,,
        \end{multlined}
        \\
        &= \Psi^{u^\mathrm{o}}_{j+1}(v_{j+1:0};d_{j:0})\,.
    \end{align*}
    where the last equality holds since $\Psiyu\in\CLp{G}$, thus \eqref{eq:SLSo_psiu_2} holds, and we can use Theorem~\ref{th:SLS_M_2}, point (2) with $\Emme = \tPsiu$ (as shown in the proof --- see also \eqref{eq:psiu_emme}).
    Moreover,
    \begin{align*}
        \hat{\nu}_{j+1} 
        &= \hat{u}^{\mathrm{o}}_{j+1} - K_{j+1}'(\hat{y}_{j+1:0}) \,,\\
        &= \Psi^{u^\mathrm{o}}_{j+1}(v_{j+1:0};d_{j:0}) 
        - K_{j+1}'(\Psi^{y}_{j+1}(v_{j+1:0};d_{j:0}) ) ,\\
        &= {\nu}_{j+1} \,.
    \end{align*}

    Finally, note that $\Qb\in\Lp$ since, for any $\tomegab\in\ell_p$ there exists a $(\vb;\db)\in\ell_p$ that allows reconstructing it.
    In particular, choose any $\db\in\ell_p$ and for any $t=0,1,\dots$, choose:
    $v_t = \tilde{\omega}_t - \Psi^{y^\mathrm{o}}_t(v_{t-1:0};d_{t-1:0}) + G_t(\Psi^{u^\mathrm{o}}_{t-1}(v_{t-1:0};d_{t-2:0}))$.
\end{pf}

\section{Proof of Proposition~\ref{prop:rec_implemenatation_KQ_with_base}}
\label{app:proof_rec_implementation_KQ_base}
\begin{pf}
Equation~\eqref{eq:K_stable_output} is equivalent to  \eqref{eq:stabilizing_input} at each time instant $t=0,1,\dots$.
Then, using \eqref{eq:omega_prestabilized}, we can obtain the expression for $\tilde{\omega}_t$, which reads
$\tilde{\omega}_t = y_t + G_t(u^\mathrm{o}_{t-1:0})$. 
By replacing $u^\mathrm{o}_{t-1:0}$ using \eqref{eq:K_stable_output}, we obtain \eqref{eq:K_stable_omega}, concluding the proof.
\end{pf}

\section{Proof of Proposition~\ref{prop:robust}}\label{app:proof_robust}

\begin{pf}
	Since $\omegab = \tilde{\Gb}(\tub+\db)+\vb-\Gb\tub$, by adding and subtracting on the right-hand side the term $\tilde{\Gb}\tub$, and taking the norms, we have
	\begin{align}
		\norm{\omegab}_p 
		&\hspace{-1pt}\leq\hspace{-1pt} \norm{\tilde{\Gb}(\tub+\db) -\tilde{\Gb}\tub}_p \hspace{-2pt}+\hspace{-1pt} \norm{\tilde{\Gb}\tub-\Gb\tub}_p \hspace{-2pt}+\hspace{-1pt} \norm{\vb}_p \hspace{-1pt}, \nonumber \\
		&\leq \gamma_{\tilde{G}} \norm{\db}_p + \norm{(\tilde{\Gb}-\Gb)\tub}_p  + \norm{\vb}_p \,, \label{eq:robust_ineq}
	\end{align}
	where $\gamma_{\tilde{G}}>0$ is the incremental finite gain of $\tilde{\Gb}$ that exists and is finite since $\tilde{\Gb}$ is i.f.g. $\ell_p$-stable. 
	
	We prove point {(1)}.
	Since $(\tilde{\Gb} - \Gb)\in\Lp$, there exists $0<\gamma_{\Delta} \leq \infty$ such that for any $\tub\in\ell_p$, we have that 
	$\norm{(\tilde{\Gb} - \Gb) \tub }_p  \leq \gamma_{\Delta}\norm{\tub}_p$.
	Then, one can close the loop using the operator $\Qb: \omegab \mapsto \tub$, which has some finite gain $\gamma_Q>0$.
	We have that 
	\begin{align*}
		\norm{\omegab}_p &\leq \gamma_{\tilde{G}} \norm{\db}_p + \gamma_{\Delta}\gamma_Q \norm{\omegab}_p  + \norm{\vb}_p \,,\\
		(1 - \gamma_{\Delta}\gamma_Q) \norm{\omegab}_p &\leq \gamma_{\tilde{G}} \norm{\db}_p  + \norm{\vb}_p \,.
	\end{align*}
	By selecting $\gamma_Q<\frac{1}{\gamma_{\Delta}}$, we have 
	\begin{equation*}
		\norm{\omegab}_p \leq \frac{\gamma_{\tilde{G}}}{(1 - \gamma_{\Delta}\gamma_Q)} \norm{\db}_p + \frac{1}{(1 - 	\gamma_{\Delta}\gamma_Q)} \norm{\vb}_p\,.
	\end{equation*}
	Thus, the closed-loop system is $\ell_p$-stable.
	
	We prove point {(2)}.
	Since $(\tilde{\Gb} - \Gb)\ub\in\ell_p$ for any $\ub$, it holds for the case $\ub = \tub$, i.e., $(\tilde{\Gb} - \Gb)\tub\in\ell_p$.
    Using \eqref{eq:robust_ineq}, we have that for any $\db,\vb\in\ell_p$, it holds that $\omegab \in \ell_p$.
	Thus, any choice of $\Qb\in\Lp$ will lead to stable closed-loop maps.
\end{pf}

\section{Proof of Proposition~\ref{prop:distributed}} 
\label{app:proof_prop_distributed}

For completeness, we start by introducing a property of binary matrices that will be useful for the proof of Proposition~\ref{prop:distributed}.

\begin{proposition}
    \label{prop:props_of_binary_matrix_inequalities}
    Given $S_1,S_2,S_3\in\{0,1\}^{M\times M}$, one has
    $S_1 \leq S_3$ and $S_2 \leq S_3$, if and only if $S_1 + S_2 \leq S_3$.
\end{proposition}

\begin{pf}
    We first prove that if $S_1 \leq S_3$ and $S_2 \leq S_3$, then $S_1 + S_2 \leq S_3$.
    By definition of binary matrices, the assumptions are equivalent to $[S_1]_{i,j} \leq [S_3]_{i,j}$ and $[S_2]_{i,j} \leq [S_3]_{i,j}$, for all $i,j=1,\dots,M$.
    Then, for each $(i,j)\in[1,\dots,M]$, one has:
    \begin{itemize}
        \item 
        If $[S_1]_{i,j} = 0$ and $[S_2]_{i,j} = 0$, then $[S_1]_{i,j} + [S_2]_{i,j} = 0 \leq [S_3]_{i,j}$.
        \item 
        If at least one of $[S_1]_{i,j}$ or $[S_2]_{i,j}$ is $1$, then $[S_3]_{i,j} = 1$, and $[S_1]_{i,j} + [S_2]_{i,j} = 1 \leq [S_3]_{i,j}$, by definition of summation of binary matrices.
    \end{itemize}
    Thus,  $S_1 + S_2 \leq S_3$.
    
    We now prove that if $S_1 + S_2 \leq S_3$, then $S_1 \leq S_3$ and $S_2 \leq S_3$.
    By definition of binary matrices, the assumption is equivalent to $[S_1]_{i,j} + [S_2]_{i,j} \leq [S_3]_{i,j}$. Since $[S_1]_{i,j},[S_2]_{i,j}\in\{0,1\}$, then $[S_1]_{i,j} \leq [S_3]_{i,j}$ and $[S_2]_{i,j} \leq [S_3]_{i,j}$.
\end{pf}

We now prove Proposition~\ref{prop:distributed}.
\begin{pf}
    We split the proof into two parts. 
    First, we show that given an operator $\Gb$ (respectively, $\Qb$) described by $M$ operators $\Gb^i\in\Lp$ (respectively, $\Qb^i\in\Lp$) then, $\Gb$ (respectively, $\Qb$) is also an operator in $\Lp$. Thus, Theorem~\ref{th:SLS_stable} can be applied and one has that the closed-loop system with $\Kb$ as in \eqref{eq:controller_youla} is $\ell_p$-stable.
    Second, we show that the interconnection structure of the controller verifies \eqref{eq:sparsity_controller_distributed} and \eqref{eq:sparsity_controller_distributed_GK} when implemented using Algorithm~\ref{alg:dist_implementation}.
    \begin{enumerate}
    \item
    We show that if $\Gb^i\in\Lp$ then $\Gb\in\Lp$.
    Consider a signal $\ub\in\ell_p$ split into $M$ signals of suitable dimensions, such that $\Gb\ub = (\Gb^1\ub^1;\dots; \Gb^M\ub^M)$. Since $\Gb^i\in\Lp$, one has that $(\Gb^1\ub^1;\dots; \Gb^M\ub^M)\in\ell_p$, thus $\Gb\ub\in\ell_p$, verifying that $\Gb\in\Lp$.
    By using similar arguments, one can show that $\Qb^i\in\Lp$ implies that $\Qb\in\Lp$.
    
    Then, Theorem~\ref{th:SLS_stable} ensures that the controller \eqref{eq:controller_youla} is $\ell_p$-stabilizing for system $\Gb$ 
    Furthermore, note that \eqref{eq:recursive_implementation_omega} gives a recursive implementation of \eqref{eq:controller_youla}. 
    When $\Gb$ (respectively, $\Qb$) is composed by $M$ subsystems $\Gb^i$ (respectively $\Qb^i$) for $i=1,\dots,M$ with coupling topology indicated by $\mathcal{D}(\Gb)$ (respectively $\mathcal{D})(\Qb)$), then \eqref{eq:recursive_implementation_omega} is equivalent to \eqref{eq:omega_algorithm}-\eqref{eq:algorithm_u} due to the definition of in-neighbors given in \eqref{eq:in_neighbors}.
    \item
    We now need to show that the controller given by \eqref{eq:controller_youla} and implemented in Algorithm~\ref{alg:dist_implementation} verifies \eqref{eq:sparsity_controller_distributed} and \eqref{eq:sparsity_controller_distributed_GK}.
    
    At each iteration of Algorithm~\ref{alg:dist_implementation}, each local controller $i$, for $i=1,\dots,M$, receives information from its neighbors $j\in \bar{\mathcal{N}}^-_{\mathcal{Q}}(i)$ in step 5, and from its neighbors $j\in \bar{\mathcal{N}}^-_{G}(i)$ in step 8.
    Thus, it receives information according to the communication links encoded by the adjacent matrix $\mathcal{D}(\Qb) + \mathcal{D}(\Gb)$, i.e., one has that 
    \begin{equation}
    \label{eq:proof_distributed_eq1}
        \mathcal{D}(\Kb) = \mathcal{D}(\Qb) + \mathcal{D}(\Gb) \,.
    \end{equation}
    Furthermore, by assumption, one has that $\mathcal{D}(\Qb) \leq \mathcal{T}$ and $\mathcal{D}(\Gb) \leq \mathcal{T}$.
    Using Proposition~\ref{prop:props_of_binary_matrix_inequalities}, one obtains 
    \begin{equation}
    \label{eq:proof_distributed_eq2}
        \mathcal{D}(\Qb) + \mathcal{D}(\Gb) \leq \mathcal{T} \,.
    \end{equation}
    Thus, \eqref{eq:proof_distributed_eq1} and \eqref{eq:proof_distributed_eq2} are equivalent to \eqref{eq:sparsity_controller_distributed}.
    Finally, from \eqref{eq:sparsity_controller_distributed} and using Proposition~\ref{prop:props_of_binary_matrix_inequalities}, one obtains \eqref{eq:sparsity_controller_distributed_GK}.%
    \end{enumerate}%
\end{pf}

\section{Proof of Theorem~\ref{th:achievability_measured_disturbances}}
\label{app:proof_achievability_measured_disturbances}
\begin{pf}
We split the proof into three parts: necessity, sufficiency, and uniqueness.
	
	\noindent
    (1)
	\emph{Necessity:} 
	We prove that given a $\Kb \in \Ccs$,
	the closed-loop map $\Phiyu{G}{K}$ satisfies \eqref{eq:SLSio1}-\eqref{eq:SLSio2} for 
	\begin{equation*}
		\Psiy = \Phiy{G}{K} \text{ and } \Psiu = \Phiu{G}{K} \,.
	\end{equation*}
	
	We first prove that $\tPsiu = \tPhiu{G}{K} \in \Ccs$. 
	For any $\vb\in\lpe{r}$, $\db\in\lpe{m}$, we have 
	$\tPhiu{G}{K} (\vb;\db) = \tub = \Kb(\tyb+\vb ; \tub+\db) = \Kb(\Gb(\tub+\db)+\vb ; \tub+\db)$. 
	Since $\Gb\in\Cs$ and $\Kb \in \Ccs$, 
	the previous operator equation corresponds to the recursive formulae 
	$u^{\mathrm{o}}_t = K_t(G_{t:0}(u^{\mathrm{o}}_{t-1:0}+d_{t-1:0}) + v_{t:0} ; u^{\mathrm{o}}_{t-1:0}+d_{t-1:0})$ 
    for $t=0,1,\dots$,
	which implies that $u^{\mathrm{o}}_t$ depends on its own past values and on $v_{t:0}$ and $d_{t-1:0}$. 
	Thus, $\tPhiu{G}{K}(\vb;\db)$ is causal with respect to its first input and strictly causal with respect to its second input, i.e., \eqref{eq:SLSio1} holds true.
	
	Next, we prove that \eqref{eq:SLSio2} is verified.
	As per the definition of the closed-loop maps, we have that $\tyb = \tPhiy{G}{K}(\vb;\db) = \tPsiy(\vb;\db)$ and $\ub = \Phiu{G}{K}(\vb;\db) = \Psiu(\vb;\db)$ satisfy the closed-loop dynamics given by  \eqref{eq:system_control_Kyu}. 
	Thus, for any $\vb\in\lpe{r}$ and $\db\in\lpe{m}$, \eqref{eq:SLSio2} holds true, i.e., $\tPsiy(\vb;\db) = \tyb = \Gb\ub = \Gb\Psiu(\vb;\db)$.
	
	
	\noindent
    (2)
	\emph{Sufficiency:}
	We prove that given the operators $(\Psiy,\Psiu)$ that satisfy \eqref{eq:SLSio1}-\eqref{eq:SLSio2}, there exists $\Kb \in \Ccs$ such that $(\Psiy,\Psiu)$ are the induced closed-loop maps $(\Phiy{G}{K},\Phiu{G}{K})$ of the plant $\Gb$.
	
	First, note that from \eqref{eq:SLSio1}, \eqref{eq:SLSio2} and $\Gb\in\Cs$, we have that $\tPsiy\in\Css$.
	Then, using Proposition~\ref{prop:invertibility}, $\Psiyu^{-1}$ exists and it is causal.
	Let us now set
	\begin{equation}
		\Kb' = \tPsiu \Psiyu^{-1}\,,
		\label{eq:K_choice_io}
	\end{equation}
	and note that $\Kb' \in \Ccs$ since 
	\eqref{eq:SLSio1} holds and
	$\Psiyu^{-1} \in \Ccscc$.

	It remains to prove that \eqref{eq:K_choice_io} is such that the resulting control policy 
	achieves the closed-loop map $(\Phiy{G}{K'};\Phiu{G}{K'}) = (\Psiy;\Psiu)$.
	
	Given any $\vb\in\lpe{r}$ and $\db\in\lpe{m}$, let
	$(\yb; \ub)$ be the signals obtained when considering the feedback loop of $\Gb$ and $\Kb'$ defined in \eqref{eq:K_choice_io}.
	In other words, we have that $\yb = \Gb\ub +\vb$ and $\ub = \Kb'(\yb; \ub) +\db$.
	Then, stacking the equations together, we have
	\begin{equation}\label{eq:CL_choice_io}
		\bvec{\yb}{\ub} = \bvec{\Gb\ub + \vb}{\Kb'(\yb;\ub) + \db}\,.
	\end{equation}
	By multiplying the left hand side of \eqref{eq:CL_choice_io} by $\I = \Psiyu\, \Psiyu^{-1}$, we have
	\begin{align}
		\Psiy \Psiyu^{-1} (\yb;\ub) &= \Gb(\ub) + \vb \,,
		\label{eq:line1_io} \\
		\Psiu \Psiyu^{-1} (\yb;\ub) &= \Kb'(\yb;\ub) + \db \,.
		\label{eq:line2_io}
	\end{align}
	Then,  using the definition of $\Kb'$ \eqref{eq:K_choice_io} and the fact that $\Psiu = \tPsiu + \OI$, 
	\eqref{eq:line2_io} can be equivalently written as
	\begin{align}
		\Psiu \Psiyu^{-1} (\yb;\ub) 
        &= \Kb' (\yb;\ub) + \db \,, \nonumber 
        \\
		(\tPsiu+\OI) \Psiyu^{-1}  (\yb;\ub) 
        &= \tPsiu\Psiyu^{-1}(\yb;\ub) + \db \,,\nonumber
        \\
        \begin{multlined}[b]
          \tPsiu \Psiyu^{-1} (\yb;\ub) \\
          + \OI \tPsiu \Psiyu^{-1} (\yb;\ub) 
        \end{multlined}
        &= \tPsiu\Psiyu^{-1}(\yb;\ub) + \db \,, \nonumber
        \\
		\OI\Psiyu^{-1}(\yb;\ub) 
        &= \db \,.
		\label{eq:proof_d_io}
	\end{align}

	Moreover, from \eqref{eq:line1_io} and since $\ub = \Kb(\yb;\ub) + \db$, one has
	\begin{equation*}
		\Psiy \Psiyu^{-1} (\yb;\ub) = \Gb(\Kb(\yb;\ub) + \db) + \vb \,.
	\end{equation*}
	Then, using the definition of $\Kb'$, i.e. \eqref{eq:K_choice_io}, and the fact that $\Psiy = \tPsiy + \IO$, 
	the previous equation can be equivalently written as
	\begin{equation*}
		(\tPsiy + \IO) \Psiyu^{-1} (\yb;\ub) 
		= \Gb(\tPsiu  \Psiyu^{-1} (\yb;\ub) + \db) + \vb \,,
	\end{equation*}
	and replacing $\db$ from \eqref{eq:proof_d_io}, we have
	\begin{multline*}
		(\tPsiy + \IO) \Psiyu^{-1} (\yb;\ub) 
		= \\
        \Gb(\tPsiu  \Psiyu^{-1} (\yb;\ub) + \OI\Psiyu^{-1}(\yb;\ub)) + \vb \,.
    \end{multline*}
    The last equation is equivalent to
    \begin{multline*}
		\tPsiy  \Psiyu^{-1} (\yb;\ub)  + \IO \Psiyu^{-1} (\yb;\ub)
		= \\
        \Gb(\tPsiu  + \OI) \Psiyu^{-1} (\yb;\ub) + \vb \,.
	\end{multline*}
	Since \eqref{eq:SLSio2} holds, we have that $\tPsiy = \Gb(\tPsiu  + \OI)$. Then, 
	\begin{equation}
		\label{eq:proof_v_io}
		\IO \Psiyu^{-1} (\yb;\ub)
		=  \vb \,.
	\end{equation}
	
	Finally,  stacking together \eqref{eq:proof_v_io} and \eqref{eq:proof_d_io}, one obtains
	\begin{align*}
		\bvec{\IO \Psiyu^{-1}(\yb;\ub)}{\OI \Psiyu^{-1}(\yb;\ub)} &= \bvec{\vb}{\db} \,, \\
		\Psiyu^{-1}(\yb;\ub) &= \bvec{\vb}{\db} \,,\\
		\bvec{\yb}{\ub} &= \Psiyu \bvec{\vb}{\db} \,.
	\end{align*}
    These relations show that
	$\Psiyu$ is the closed-loop map $\Phiyu{G}{K'}$.
	
	
	\noindent
    (3)
	\emph{Uniqueness:}
	Assume that $\Psiyu = \Phiyu{G}{K'} = \Phiyu{G}{K''}$ for some $\Kb', \Kb''$. Then, for any $\vb\in\lpe{r}$ and any $\db\in\lpe{m}$, 
	it holds
	\begin{equation*}
		\tPsiu(\vb;\db) = \Kb'  \Psiyu (\vb;\db) = \Kb'' \Psiyu (\vb;\db)\,.
	\end{equation*}
	Since $\Psiyu$ is invertible, this implies that $\Kb' = \Kb'' $.
\end{pf}

\section{Proof of Proposition~\ref{prop:rec_implemenatation_Kio}}
\label{app:proof_rec_implementation_Kio}

\begin{pf}
    We evaluate \eqref{eq:K_io} over signals $(\yb;\ub)$, obtaining
    $\tub = \Kb(\yb;\ub) = \tPsiu \Psiyu^{-1} (\yb;\ub)$. 
    Then, using Proposition~\ref{prop:invertibility}, one obtains \eqref{eq:rec_impl_disturbances_1}-\eqref{eq:rec_impl_disturbances_2}, which are the recursive calculation of $(\boldsymbol{\beta};\boldsymbol{\delta}) = \Psiyu^{-1} (\yb;\ub)$.
    Finally, $u^\mathrm{o}_t$ for $t=0,1,\dots$ is calculated
    from $\tub = \tPsiu (\boldsymbol{\beta};\boldsymbol{\delta})$, resulting in \eqref{eq:rec_impl_disturbances_3}.
\end{pf}

\section{Proof of Theorem~\ref{th:stability_measured_disturbances}}
\label{app:proof_stability_measured_disturbances}

\begin{pf}
Assume that $\Gb\in\Lp$.
By using the definition of $\CL{\Gb}$ in \eqref{eq:SLSio}
the set $\hCLp{\Gb}$ defined in \eqref{eq:CLp_measured_disturbances}
can be written as
\begin{align*}
	\hCLp{\Gb} 
    = 
    & \{\Psiyu = (\Psiy; \Psiu) = (\tPsiy; \tPsiu) + \I ~~| \\
	&\quad \tPsiu  \in \Ccs\,, \\
	&\quad \tPsiy = \Gb \Psiu \,, \\
    &\quad \Psiyu \in \Lp
	\}\,.
\end{align*}
In other words, we must show that 
given $\Psiyu = (\Psiy; \Psiu) = (\tPsiy; \tPsiu) + \I$ with $\tPsiu  \in \Ccs$ and $\tPsiy = \Gb \Psiu$, 
then
a necessary and sufficient condition for $\Psiyu$ to be in $\Lp$ is that $\tPsiu\in\Lp$. 
\begin{enumerate}
    \item 
    \emph{Necessity:} 
    We need to show that  $\Psiyu\in\Lp$ implies $\tPsiu\in\Lp$.
    This is straightforward because 
    ($i$) $\Psiyu$ is assumed to be in $\Lp$ and hence $\Psiu\in\Lp$;  
    ($ii$) $\tPsiu = \Psiu - \OI$; and 
    ($iii$) $\Lp$ is closed under sum and $\OI\in\Lp$.
    \item
    \emph{Sufficiency:}
    We prove that 
    $\tPsiu\in\Lp$ implies $\Psiyu\in\Lp$.
    Since $\Lp$ is closed under summation and composition, 
    and $\OI\in\Lp$, one has that
    $\Psiu = \tPsiu +  \OI \in\Lp$ and hence $\Psiy = \Gb \Psiu + \IO \in\Lp$.
    Thus, $\Psiyu = (\Psiy; \Psiu)\in\Lp$.
\end{enumerate}
\end{pf}

\section{Properties of exponentially stable operators and  i.f.g \texorpdfstring{$\ell_p$}{lp}-stable operators}
\label{app:exponential_stability}

For the sake of completeness, in this Appendix, we show that the
set of operators in $\Lexp$ and the set of i.f.g $\ell_p$-stable operators are closed under the summation and composition.

\subsection{Properties of exponentially stable operators}
\begin{proposition}\label{app:prop_exp_stable_sum}
	If $\Ab_1\in\Lexp$ and  $\Ab_2\in\Lexp$, then $(\Ab_1+\Ab_2)\in\Lexp$.
\end{proposition}
\begin{pf}
	Consider any signal $\xb \in\ell_{exp}$. 
	Then, since $\Ab_1,\Ab_2\in\Lexp$, we have that $\Ab_1\xb,\Ab_2\xb \in\ell_{exp}$. 
    Then, by means of Proposition~\ref{app:prop_sum_exp_signals}, one has that $\Ab_1\xb + \Ab_2\xb = (\Ab_1 + \Ab_2)\xb \in\ell_{exp}$.
\end{pf}

\begin{proposition}\label{app:prop_exp_stable_composition}
	If $\Ab_1\in\Lexp$ and  $\Ab_2\in\Lexp$, then $\Ab_1\Ab_2\in\Lexp$.
\end{proposition}
\begin{pf}
	Consider any signal $\xb \in\ell_{exp}$. 
	Then, since $\Ab_2\in\Lexp$, we have that $\Ab_2\xb \in\ell_{exp}$. 
	Moreover, since $\Ab_1\in\Lexp$, we have that $\Ab_1\Ab_2\xb \in\ell_{exp}$.
\end{pf}

The proof of Proposition~\ref{app:prop_exp_stable_sum} relies on the following result.
\begin{proposition}\label{app:prop_sum_exp_signals}
	If $\xb\in\ell_{exp}$ and  $\yb\in\ell_{exp}$, then $\xb+\yb\in\ell_{exp}$.
\end{proposition}
\begin{pf}
	By assumption, there exists $\alpha_x,\alpha_y\in(0,1)$ and $K_x,K_y>0$, such that $\lvert x_t\rvert \leq K_x\alpha_x^t$ and $\lvert y_t\rvert \leq K_y\alpha_y^t$, for $t=0,1, \dots$. 
 Then, 
	\begin{align*}
		\lvert x_t + y_t\rvert &\leq \lvert x_t\rvert + \lvert y_t\rvert \,,\\
		&\leq K_x\alpha_x^t  + K_y\alpha_y^t \,, \\
		&\leq  K_{xy}\alpha_{xy}^t \,,
	\end{align*}
	with $\alpha_{xy} = \max(\alpha_x,\alpha_y)$ and $K_{xy} = K_x + K_y$.
\end{pf}

\subsection{Properties of i.f.g. \texorpdfstring{$\ell_p$}{lp}-stable operators}

\begin{proposition}\label{app:prop_ifg_sum}
	If $\Ab_1$ and  $\Ab_2$ are i.f.g. $\ell_p$-stable operators, then $(\Ab_1+\Ab_2)$ is also an i.f.g. $\ell_p$-stable operator.
\end{proposition}
\begin{pf}
	Since $\Ab_1$ and $\Ab_2$ are i.f.g. $\ell_p$-stable operators,
    we have that for any $\xb_1,\xb_2$, 
    $\norm{\Ab_1\xb_1 -\Ab_1\xb_2}_p \leq \gamma_1 \norm{\xb_1 - \xb_2}_p$ 
    and
    $\norm{\Ab_2\xb_1 -\Ab_2\xb_2}_p \leq \gamma_2 \norm{\xb_1 - \xb_2}_p$. 
    Then,
    \begin{align*}
        &\norm{(\Ab_1+\Ab_2)\xb_1 -(\Ab_1+\Ab_2)\xb_2}_p  \\
        &\qquad\qquad\qquad= 
        \norm{\Ab_1\xb_1 +\Ab_2 \xb_1 - \Ab_1\xb_2 - \Ab_2\xb_2}_p ,\\
        &\qquad\qquad\qquad\leq
        \norm{\Ab_1\xb_1  - \Ab_1\xb_2}_p
        +
        \norm{\Ab_2 \xb_1 - \Ab_2\xb_2}_p ,\\
        &\qquad\qquad\qquad\leq
        \gamma_1 \norm{\xb_1 - \xb_2}_p
        +
        \gamma_2 \norm{\xb_1 - \xb_2}_p ,\\
        &\qquad\qquad\qquad= 
        (\gamma_1 + \gamma_2) \norm{\xb_1 - \xb_2}_p.
    \end{align*}
\end{pf}
\begin{proposition}\label{app:prop_ifg_composition}
	If $\Ab_1$ and  $\Ab_2$ are i.f.g. $\ell_p$-stable operators, then $\Ab_1 \Ab_2$ is also an i.f.g. $\ell_p$-stable operator.
\end{proposition}
\begin{pf}
	Since $\Ab_1$ and $\Ab_2$ are i.f.g. $\ell_p$-stable operators,
    we have that for any $\xb_1,\xb_2$, 
    $\norm{\Ab_1\xb_1 -\Ab_1\xb_2}_p \leq \gamma_1 \norm{\xb_1 - \xb_2}_p$ 
    and
    $\norm{\Ab_2\xb_1 -\Ab_2\xb_2}_p \leq \gamma_2 \norm{\xb_1 - \xb_2}_p$. 
    Then,
    \begin{align*}
        \norm{\Ab_1\Ab_2\xb_1 -\Ab_1\Ab_2\xb_2}_p 
        & \leq
        \gamma_1\norm{\Ab_2\xb_1 -\Ab_2\xb_2}_p \,,\\
        & \leq
        \gamma_1\gamma_2\norm{\xb_1 -\xb_2}_p\,.
    \end{align*}
\end{pf}

\section{Proof of Theorem~\ref{th:exp_stab}}
\label{app:proof_exp_stab}

\begin{pf}
Assume that $\Gb\in\Lexp$.
By using the definition of $\CL{\Gb}$ in \eqref{eq:SLSio}
the set $\CL{\Gb}_{exp}$ defined in \eqref{eq:set_exponentially_stable_short}
can be written as
\begin{align*}
	\CL{\Gb}_{exp} 
    = 
    & \{\Psiyu = (\Psiy; \Psiu) = (\tPsiy; \tPsiu) + \I ~~| \\
	&\quad \tPsiu  \in \Ccs\,, \\
	&\quad \tPsiy = \Gb \Psiu \,, \\
    &\quad \Psiyu \in \Lexp
	\}\,.
\end{align*}
In other words, we must show that 
given $\Psiyu = (\Psiy; \Psiu) = (\tPsiy; \tPsiu) + \I$ with $\tPsiu  \in \Ccs$ and $\tPsiy = \Gb \Psiu$, 
then
a necessary and sufficient condition for $\Psiyu$ to be in $\Lexp$ is that $\tPsiu\in\Lexp$. 
\begin{enumerate}
    \item 
    \emph{Necessity:} 
    We need to show that  $\Psiyu\in\Lexp$ implies $\tPsiu\in\Lexp$.
    This is straightforward because 
    ($i$) $\Psiyu$ is assumed to be in $\Lexp$ and hence $\Psiu\in\Lexp$;  
    ($ii$) $\tPsiu = \Psiu - \OI$; and 
    ($iii$) the set of operators in $\Lexp$ is closed under sum (see Proposition~\ref{app:prop_exp_stable_sum}) and $\OI\in\Lexp$.
    \item
    \emph{Sufficiency:}
    We prove that 
    $\tPsiu\in\Lexp$ implies $\Psiyu\in\Lexp$.
    Since the set of operators in $\Lexp$ is closed under summation and composition (see Proposition~\ref{app:prop_exp_stable_sum} and Proposition~\ref{app:prop_exp_stable_composition}), 
    and $\OI\in\Lexp$, one has that
    $\Psiu = \tPsiu +  \OI \in\Lexp$ and hence $\Psiy = \Gb \Psiu + \IO \in\Lexp$.
    Thus, $\Psiyu = (\Psiy; \Psiu)\in\Lexp$.
\end{enumerate}
\end{pf}

\section{Proof of Theorem~\ref{th:lipschitzness}}
\label{app:proof_lipschitzness}

\begin{pf}
Assume that $\Gb$ is i.f.g. $\ell_p$-stable.
By using the definition of $\CL{\Gb}$ in \eqref{eq:SLSio}
the set $\CL{\Gb}_{i.f.g}$ defined in \eqref{eq:set_ifg_stable_short}
can be written as
\begin{align*}
	\CL{\Gb}_{i.f.g} 
    = 
    & \{\Psiyu = (\Psiy; \Psiu) = (\tPsiy; \tPsiu) + \I ~~| \\
	&\quad \tPsiu  \in \Ccs\,, \\
	&\quad \tPsiy = \Gb \Psiu \,, \\
    &\quad \Psiyu \text{ is } i.f.g.\, \ell_p\text{-stable}
	\}\,.
\end{align*}
In other words, we must show that 
given $\Psiyu = (\Psiy; \Psiu) = (\tPsiy; \tPsiu) + \I$ with $\tPsiu  \in \Ccs$ and $\tPsiy = \Gb \Psiu$, 
then
a necessary and sufficient condition for $\Psiyu$ to be i.f.g. $\ell_p$-stable is that $\tPsiu$ is i.f.g. $\ell_p$-stable. 

\begin{enumerate}
    \item 
    \emph{Necessity:} 
    We need to show that  $\Psiyu$ being i.f.g. $\ell_p$-stable implies $\tPsiu$ being i.f.g. $\ell_p$-stable.
    This is straightforward because 
    ($i$) $\Psiyu$ is assumed to be i.f.g. $\ell_p$-stable and hence $\Psiu$ has the same property;  
    ($ii$) $\tPsiu = \Psiu - \OI$; and 
    ($iii$) the set of i.f.g. $\ell_p$-stable operators is closed under sum (see Proposition~\ref{app:prop_ifg_sum}) and $\OI$ is an i.f.g. $\ell_p$-stable operator.
    \item 
    \emph{Sufficiency:} 
    We prove that 
    $\tPsiu$ being i.f.g. $\ell_p$-stable implies $\Psiyu$ being i.f.g. $\ell_p$-stable.
    Since the set of i.f.g. $\ell_p$-stable operators is closed under summation and composition (see Proposition~\ref{app:prop_ifg_sum} and Proposition~\ref{app:prop_ifg_composition}), 
    and $\OI$ is an i.f.g. $\ell_p$-stable operator, one has that
    $\Psiu = \tPsiu +  \OI$ is also an i.f.g. $\ell_p$-stable operator and so it is the operator $\Psiy = \Gb \Psiu + \IO$.
    Thus, $\Psiyu = (\Psiy; \Psiu)$ is also an i.f.g. $\ell_p$-stable operator.
\end{enumerate}
\end{pf}

\section{Proof of Proposition~\ref{prop:disturbance_localization}}
\label{app:proof_disturbance_localization}

\begin{pf}
	Observe that the mapping $(\vb;\db)\mapsto \ub$ is given by $\mathbf{\tPsiu} + \OI$. 
	Since $\mathcal{D}(\tPsiu) = S_u$
	and 
	$[S_u]_{i,i} = 1$ for $i=1,\dots,M$, 
	the map $(\vb;\db)\mapsto \ub$ has the same sparsity pattern, i.e. $\mathcal{D}(\mathbf{\Psiu}) = S_u$. 
	
	We now study the mapping $(\vb;\db)\mapsto \yb$, which is given by $\tPsiy + \IO$. 
	Since $\tPsiy = \Gb \Psiu$, one has $\mathcal{D}(\tPsiy) = \mathcal{D}(\Gb) S_u$. 
 Since both $[S_u]_{i,i} = 1$ and  $[\mathcal{D}(\Gb)]_{i,i}=1$ for $i=1,\dots,M$, it follows that $\mathcal{D}(\tPsiy) = \mathcal{D}(\Gb) S_u$.
\end{pf}

\section{Comparison of Theorem~\ref{th:achiev_of} with the LTI case}\label{app:generalization}

Let us consider the closed-loop system given by \eqref{eq:system_control} and assume now that $\Gb$ and $\Kb$ are transfer function matrices, i.e., LTI operators. 
The corresponding closed-loop map $\Phiyu{\Gb}{\Kb}$ is also a transfer function matrix, which can be written in the block form 
\begin{equation*}
	\Phiyu{\Gb}{\Kb} 
	=
	\begin{bmatrix}
		\mathbf{X} & \mathbf{W} \\
		\mathbf{Y} & \mathbf{Z}
	\end{bmatrix}\,.
\end{equation*}
Moreover, we have $\Phiy{\Gb}{\Kb} = \begin{bmatrix}\mathbf{X} & \mathbf{W}\end{bmatrix}$ and $\Phiu{\Gb}{\Kb} = \begin{bmatrix}\mathbf{Y} & \mathbf{Z} \end{bmatrix}$.

In this setup, we will show the relationship of Theorem~\ref{th:achiev_of} with the existing results in the LTI case.

\begin{theorem}\label{th:achiev_LTI}
    \emph{\textbf{(Achievability constraints of Theorem~1 by \citet{furieri2019input})}}
	Consider the system \eqref{eq:system_control}, where $\Gb\in\Cs$ and $\Kb\in\Cc$ are transfer matrices. The following statements hold.
	\begin{enumerate}
		\item 
		For any transfer matrix $\Kb \in \Cc$, there exists four corresponding transfer matrices $(\mathbf{X},\mathbf{Y},\mathbf{W},\mathbf{Z})$ that lie in the affine subspace defined by the equations
		\begin{equation}
			\begin{bmatrix}
				\I & -\Gb
			\end{bmatrix}
			\begin{bmatrix}
				\mathbf{X} & \mathbf{W} \\
				\mathbf{Y} & \mathbf{Z}
			\end{bmatrix}
			= 
			\begin{bmatrix}
				\I & \Ob
			\end{bmatrix}\,,
			\label{eq:achiev_LTI1}
		\end{equation}
		\begin{equation}
			\begin{bmatrix}
				\mathbf{X} & \mathbf{W} \\
				\mathbf{Y} & \mathbf{Z}
			\end{bmatrix}
			\begin{bmatrix}
				-\Gb \\ \I
			\end{bmatrix}
			= 
			\begin{bmatrix}
				\Ob \\ \I
			\end{bmatrix}\,,
			\label{eq:achiev_LTI2}
		\end{equation}
		\begin{equation}
			(\mathbf{X},\mathbf{Y},\mathbf{W},\mathbf{Z}) \in \Cc \,.
			\label{eq:achiev_LTI3}
		\end{equation}
		\item
		For any transfer matrices $(\mathbf{X},\mathbf{Y},\mathbf{W},\mathbf{Z})$ that lie in the affine subspace \eqref{eq:achiev_LTI1}-\eqref{eq:achiev_LTI3}, the controller $\Kb = \mathbf{YX}^{-1}$ generates the closed-loop maps $(\mathbf{X},\mathbf{Y},\mathbf{W},\mathbf{Z})$ for system $\Gb$.
	\end{enumerate}
\end{theorem}

Theorem~\ref{th:achiev_LTI} corresponds to the achievability constraints of Theorem~1 in \citet{furieri2019input}.
The latter characterizes the space of all and only achievable closed-loop maps that are stabilizing for a given system $\Gb$. 
This parametrization involves constraints \eqref{eq:achiev_LTI1}-\eqref{eq:achiev_LTI3} plus a constraint imposing the stability of the closed-loop maps. Stability is imposed by constraining the matrices $(\mathbf{X}, \mathbf{W}, \mathbf{Y},\mathbf{Z})$ to be proper and have all their poles inside the unit circle.

Since the stability constraint affects only the set of transfer matrices from where the closed-loop maps belong, in Theorem~\ref{th:achiev_LTI}, we remove this constraint to parametrize all achievable closed-loop maps.

Now, we are ready to show that, when restricting to LTI operators, Theorem~\ref{th:achiev_of} in our paper is equivalent to the one in Theorem~\ref{th:achiev_LTI}.

We start by re-writing the parametrization of Theorem~\ref{th:achiev_of} in the LTI case:

\begin{align}
	& \CL{\Gb}_{LTI} 
	= 
	\{\begin{bmatrix}
		\mathbf{X} & \mathbf{W} \\
		\mathbf{Y} & \mathbf{Z}
	\end{bmatrix}
	~|~ 
	\begin{bmatrix}\mathbf{Y} & \mathbf{Z}-\I \end{bmatrix} \in \Ccs\,, \label{eq_app:SLS1}\\
	&~~ 
	\begin{bmatrix}\mathbf{X}-\I & \mathbf{W}\end{bmatrix} = \Gb (\begin{bmatrix}\mathbf{Y} & \mathbf{Z}-\I \end{bmatrix} + \OI) \,,\label{eq_app:SLS2}\\
	&~~ 
    \begin{multlined}[b]
    \begin{bmatrix}\mathbf{Y} & \mathbf{Z}-\I \end{bmatrix} = \begin{bmatrix}\mathbf{Y} & \mathbf{Z}-\I \end{bmatrix} \begin{bmatrix}
		\mathbf{X} & \mathbf{W} \\
		\mathbf{Y} & \mathbf{Z}
	\end{bmatrix}^{-1} \IOv 
    \\
    (\begin{bmatrix}\mathbf{X}-\I & \mathbf{W}\end{bmatrix}+\IO) \label{eq_app:SLS3}
	\}\,.
    \end{multlined}
\end{align}

Next, we show that the parametrization given by \eqref{eq:achiev_LTI1}-\eqref{eq:achiev_LTI3} is equivalent to \eqref{eq_app:SLS1}-\eqref{eq_app:SLS3}.

First, we show that constraints \eqref{eq_app:SLS2} and \eqref{eq:achiev_LTI1} are the same.
Start from \eqref{eq_app:SLS2}. Since all operators are linear, we apply the distributive property and reorder the terms to obtain
\begin{equation}
	\label{eq_app:SLS_LTI_syn_2}
	\left\{
	\begin{array}{r l}
		\mathbf{X}-\I &= \Gb \mathbf{Y} \\
		\mathbf{W} &= \Gb \mathbf{Z}
	\end{array}
	\right.
	\text{ and then }
	\left\{
	\begin{array}{r l}
		\mathbf{X} - \Gb \mathbf{Y} &= \I\\
		\mathbf{W} - \Gb \mathbf{Z} &= \Ob
	\end{array}
	\right. \,,
\end{equation}
which is the same as \eqref{eq:achiev_LTI1}.

Second, we rewrite \eqref{eq_app:SLS3}, in a more compact form.
Using the formula for the inverse of a block matrix, one has
\begin{equation}\label{eq:LTI_inverse}
	\begin{bmatrix}
		\mathbf{X} & \mathbf{W} \\
		\mathbf{Y} & \mathbf{Z}
	\end{bmatrix} ^{-1}
	\hspace{-5pt}=\hspace{-2pt}
	\begin{bmatrix}
		\mathbf{X}^{-1} \hspace{-2pt}+\hspace{-1pt} \mathbf{X}^{-1} \mathbf{W} \boldsymbol{\Delta}^{-1} \mathbf{Y} \mathbf{X}^{-1}
		& 
		-\mathbf{X}^{-1} \mathbf{W} \boldsymbol{\Delta}^{-1}
		\\
		-\boldsymbol{\Delta}^{-1} \mathbf{Y} \mathbf{X}^{-1}
		& 
		\boldsymbol{\Delta}^{-1}
	\end{bmatrix}\hspace{-2pt},
\end{equation}
with $\boldsymbol{\Delta} = \mathbf{Z} - \mathbf{Y} \mathbf{X}^{-1} \mathbf{W}$.
By using \eqref{eq:LTI_inverse},
\eqref{eq_app:SLS3} is the same as
\begin{align*}
	\begin{bmatrix}\mathbf{Y} & \mathbf{Z}-\I \end{bmatrix} 
	&= 
	\begin{bmatrix}\mathbf{Y} & \mathbf{Z}-\I \end{bmatrix} 
	\begin{bmatrix}
		\mathbf{X} & \mathbf{W} \\
		\mathbf{Y} & \mathbf{Z}
	\end{bmatrix} ^{-1} 
	\begin{bmatrix}
		\mathbf{X} & \mathbf{W} \\
		\mathbf{0} & \mathbf{0}
	\end{bmatrix} \,,
\end{align*}
obtaining
\begin{equation*}
	\left\{
	\begin{array}{r l}
		  \mathbf{Y} &=  (\I + (\mathbf{Y} \mathbf{X}^{-1} \mathbf{W} - \mathbf{Z} + \I) \boldsymbol{\Delta}^{-1}) \mathbf{Y} \\
		  \mathbf{Z}-\I &= (\I + (\mathbf{Y} \mathbf{X}^{-1} \mathbf{W} - \mathbf{Z} + \I) \boldsymbol{\Delta}^{-1}) 
		\mathbf{Y} \mathbf{X}^{-1} \mathbf{W} \,.
	\end{array}
	\right. 
\end{equation*}
By using the definition of $\boldsymbol{\Delta}$ we have for any $\mathbf{Y}$ and any $\mathbf{Z}$ that
\begin{equation*}
	\left\{
	\begin{array}{r l}
		\mathbf{Y} &= \boldsymbol{\Delta}^{-1}  \mathbf{Y}  \\
		\mathbf{Z} &= \boldsymbol{\Delta}^{-1} \mathbf{Z}
	\end{array}
	\right.
	\,,
\end{equation*}
which reveals that $\boldsymbol{\Delta} = \I$, i.e., 
\begin{equation}
	\label{eq_app:Delta_2}
	\mathbf{Z} - \mathbf{Y} \mathbf{X}^{-1} \mathbf{W} = \I\,.
\end{equation}
In the sequel, since they are the same, we will replace \eqref{eq_app:SLS3} with \eqref{eq_app:Delta_2}.

Third, we show the sufficiency and necessity of the remaining constraints.
I.e., assume that \eqref{eq:achiev_LTI1} (or, equivalently \eqref{eq_app:SLS2} holds). Then, we prove that \eqref{eq:achiev_LTI2} and \eqref{eq:achiev_LTI3} hold true if and only if \eqref{eq_app:SLS1} and \eqref{eq_app:Delta_2} hold true.
\begin{enumerate}
    \item 
	We first prove that \eqref{eq:achiev_LTI2} and \eqref{eq:achiev_LTI3} imply \eqref{eq_app:SLS1} and \eqref{eq_app:Delta_2}.
	
	We start by showing that \eqref{eq_app:SLS1} is verified.
	From \eqref{eq:achiev_LTI2}, we have that $-\mathbf{Y}\Gb = \mathbf{Z}-\I$. Since $\Gb \in\Cs$, one has $\mathbf{Y}\Gb\in\Cs$. Then $(\mathbf{Z}-\I) \in\Cs$. Moreover, from \eqref{eq:achiev_LTI3}, one has $\mathbf{Y}\in\Cc$. Thus, \eqref{eq_app:SLS1} is satisfied.
	
	We now show that \eqref{eq_app:Delta_2} is verified.
	From \eqref{eq:achiev_LTI1}, one has
	\begin{equation*}
		\left\{
		\begin{array}{r l}
			-\mathbf{X} \Gb + \mathbf{W} &= \Ob \\
			-\mathbf{Y} \Gb + \mathbf{Z} &= \I
		\end{array}
		\right.
		\text{ and then }
		\left\{
		\begin{array}{r l}
			\mathbf{W} &= \mathbf{X} \Gb \\
			\mathbf{Z} - \I &= \mathbf{Y} \Gb
		\end{array}
		\right.
        \,.
	\end{equation*}
	We can write the latter system of equations as
	\begin{equation*}
		\left\{
		\begin{array}{r l}
			\mathbf{X}^{-1} \mathbf{W} &=  \Gb \\
			\mathbf{Z} - \I &= \mathbf{Y} \Gb
		\end{array}
		\right.
		\,,
	\end{equation*}
    which implies $\mathbf{Z} - \I = \mathbf{Y} \mathbf{X}^{-1} \mathbf{W}$.
	Thus, \eqref{eq_app:Delta_2} is satisfied, and so it is \eqref{eq_app:SLS2},
	which completes the proof.

	\item
	Now, we prove that \eqref{eq_app:SLS1} and \eqref{eq_app:Delta_2} imply \eqref{eq:achiev_LTI2} and \eqref{eq:achiev_LTI3}.
	
	We see that the \eqref{eq_app:SLS1} is equivalent to $\mathbf{Y}\in\Cc$ and $\mathbf{Z}-\I\in\Cs$. Thus, one has  $\mathbf{Y},\mathbf{Z}\in\Cc$.
	Moreover, from \eqref{eq_app:SLS2} and $\Gb\in\Cs$, we have that $\mathbf{X}-\I, \mathbf{W} \in\Cs$. Hence $\mathbf{X}, \mathbf{W} \in\Cc$ showing that the causality conditions in \eqref{eq:achiev_LTI3} hold.
	
	We now need to show that \eqref{eq:achiev_LTI2} is verified.
	Consider \eqref{eq_app:Delta_2}, which is the same as \eqref{eq_app:SLS3}.
	Then, premultiplying \eqref{eq_app:Delta_2} by $\Gb$ and using \eqref{eq_app:SLS_LTI_syn_2} (which is the same as \eqref{eq_app:SLS2}), we have:
	\begin{align}
		\Gb \mathbf{Y}\mathbf{X}^{-1}\mathbf{W} &= \Gb\mathbf{Z} - \Gb  \,,\nonumber\\
		(\mathbf{X} - \I)\mathbf{X}^{-1}\mathbf{W} &= \mathbf{W} - \Gb \,,\nonumber\\
		\mathbf{W} - \mathbf{X}^{-1}\mathbf{W} &= \mathbf{W} - \Gb \,,\nonumber\\
		\mathbf{X}^{-1}\mathbf{W} &= \Gb \,. \label{eq_app:SLS3_part1}
	\end{align} 
    Furthermore, recall that $\mathbf{Z} - \mathbf{Y} \mathbf{X}^{-1} \mathbf{W} = \I$ from \eqref{eq_app:Delta_2}. Then, by replacing $\mathbf{X}^{-1}\mathbf{W}$ using \eqref{eq_app:SLS3_part1}, one obtains
	\begin{equation}
		\label{eq_app:SLS3_part2}
		\mathbf{Z} - \mathbf{Y} \Gb = \I \,.
	\end{equation}
	Finally, \eqref{eq_app:SLS3_part1} and \eqref{eq_app:SLS3_part2} together are the same as \eqref{eq:achiev_LTI2}, which concludes the proof.

\end{enumerate}

\section{Implementation details}
\label{app:implementation}

The parameters of the robot models in \eqref{eq:mechanical_system} are $m^i=1$, $b^i_1=2$ and $b^i_2=0.5$, for $i=1,2$. Moreover, the pre-stabilizing controllers' parameters are $\bar{k}^i_1=\bar{k}^i_2=1$, for $i=1,2$.

\subsection{Corridor scenario}
As shown in Figure~\ref{fig:corridor}, the robots start at $x^1_{1,0} = (-2,-2)$ and $x^2_{1,0} = (-2,2)$, and must reach the target positions $\bar{x}^1_1 = (2,2)$ and $\bar{x}^2_1 = (-2,2)$, respectively.
The training data consists of $100$ initial conditions sampled from a Gaussian distribution around the initial position with a standard deviation of $0.2$.

The terms of the cost function~\eqref{eq:loss_CA} are
\begin{align*}
    l_{\text{traj}}(y_t,u_t) = (y_t-\bar{y})^\top Q (y_t-\bar{y}) + \alpha_u u_t^\top u_t
\end{align*}
\begin{equation*}
l_{\text{ca}}(y_t) =
\begin{cases}
\alpha_{\text{ca}}
\sum_{i=0}^{N}
\sum_{j,\,i\ne j} (d^{i,j}_{t} + \epsilon)^{-2} 
&
\text{if} \quad d^{i,j}_{t} \leq D\,, \\ 
0 
&
\text{otherwise}\,,
\end{cases}
\end{equation*}
where 
$Q\succ 0$ and $\alpha_u,\alpha_{\text{ca}}>0$ are hyperparameters,
$\bar{y} = \bar{x}_1$ (i.e., the target positions of the robots),
$d^{i,j}_{t} = |y^i_t-y^j_t|_2 \geq 0$ denotes the distance between agent $i$ and $j$ at instant $t$,
and $\epsilon>0$ is a fixed positive small constant such that the loss remains bounded for all possible distance values.

Motivated by \citet{onken2021neural}, we represent each obstacle by using a Gaussian density function 
\begin{equation*}
\eta(z; \mu, \Sigma) = \frac{1}{2\pi \sqrt{\text{det}(\Sigma)}} \exp \hspace{-2pt}
 \left( \hspace{-1pt}
-\frac{1}{2} \left(z \hspace{-1pt}-\hspace{-1pt} \mu\right)^\top \hspace{-1pt}\Sigma^{-1}\hspace{-1pt} \left(z \hspace{-1pt}-\hspace{-1pt} \mu\right) 
 \right) \hspace{-2pt},
\end{equation*}
with mean $\mu \in \mathbb{R}^2$ and covariance $\Sigma \in \mathbb{R}^{2\times 2}$ with $\Sigma\succ 0$.
The term $l_{\text{obs}}(y_t)$ is given by
\begin{equation*}
l_{\text{obs}}(y_t) 
= 
\alpha_{\text{obs} }
\sum_{i=0}^{2} \sum_{j=0}^{4}
\eta\left(y^i_t; \mu_j, 0.2\,I\right) 
\,,
\end{equation*}
with 
$\mu_{1,2} = (\pm2.5;0)$,
and
$\mu_{3,4} = (\pm1.5;0)$.

The used REN is a deep neural network with depth $q_2 = 8$ layers, internal state $\xi$ of dimension $q_1 = 8$
and activation function $\sigma(\cdot) = \tanh(\cdot)$.\footnote{See Section~\ref{sec:intro_REN} for the meaning of $q_1$, $q_2$ and $\sigma$.} 

For the hyperparameters, we set 
$\alpha_u = 2.5\times 10^{-4}$, 
$\alpha_{\text{ca}} = 100$, 
$\alpha_{\text{obs}} = 5 \times 10^{3}$ and 
$Q=I_4$.

We use stochastic gradient descent with Adam to minimize the loss function, setting a learning rate of $0.005$.
We optimize for $1.2\times 10^{4}$ epochs with a single trajectory per batch size.

\subsection{Waypoint-tracking scenario}

As shown in Figure~\ref{fig:corridor}, the robots start at $x^1_{1,0} = (-2,0)$ and $x^2_{1,0} = (0,0)$.
The goal points $g_a$, $g_b$ and $g_c$ are located at $(-2,-2)$, $(0,2)$ and $(2,-2)$, respectively.
To describe the TLTL loss, let us define, for each robot, the following functions of time:
\begin{itemize}
    \item $d^{g_i}_t$, for $i=1,2,3$, is the distance between the robot and the goal point $g_i$;
    \item $d^{o_i}_t$, for $i=1,2$, is the distance between the robot and the $i^{\text{th}}$ obstacle;
    \item $d^{rob}_t$ is the distance between the two robots;
\end{itemize}
where $g_1$, $g_2$ and $g_3$ are the waypoints in the correct visiting order, for each robot.
Next, we define the predicates $\psi$ as described in Table~\ref{tab:predicatesTLTL},
where
$r_{\text{obs}} = 1.7$ and $r_{\text{rob}} = 0.5$ are the radii of the obstacles and vehicles, respectively, and $\tau\in[0,T]$ indicates the time at which we evaluate the predicate.

For instance, one would like to avoid collisions between the agents at all times.
This is encoded as $\square\psi_{coll}$, where the $\square$ stands for `always'.
Mathematically, $\square\psi_{coll}$ translates into
\begin{equation*}
    \min_{t\in[0,T]}  (d^{\text{rob}}_t-2r_{\text{rob}}).
\end{equation*}
Similarly, avoiding collisions with the obstacles is a desired behavior that can be encoded in the formula $\square\psi_{o_1} \wedge \square\psi_{o_2}$. This translates into
\begin{equation*}
    \min_{t\in[0,T]} \big( \min(d^{\text{obs}_1}_t-r_{\text{obs}} \,,\, d^{\text{obs}_2}_t-r_{\text{obs}}) \big).
\end{equation*}

Following the notation of \citet{li2017reinforcement}, 
the temporal logic form of the cost function for each robot is
\begin{multline}
    \label{eq:TLTL_cost}
    \left(\psi_{g_1} \,\mathcal{T}\,\psi_{g_2} \,\mathcal{T}\,\psi_{g_3} \right) 
    \wedge 
    \left(\lnot\left(\psi_{g_2} \vee \psi_{g_3}\right) \,\mathcal{U}\, \psi_{g_1}\right) \\
    \wedge 
    \left(\lnot\psi_{g_3}\,\mathcal{U}\,\psi_{g_2}\right) 
    \wedge \left(\bigwedge_{i=1,2,3}\square\left(\psi_{g_i}\Rightarrow\bigcirc\square\lnot\psi_{g_i}\right)\right) \\
    \wedge 
    \left(\bigwedge_{i=1,2}\square\psi_{o_i}\right) 
    \wedge
    \square\psi_{coll}
    \wedge
    \lozenge\square\psi_{g_3} \,.
\end{multline}
The Boolean operators $\lnot$, $\vee$, and $\wedge$ stand for negation (not), disjunction (or), and conjunction (and). The temporal operators $\mathcal{T}$, $\mathcal{U}$, $\Rightarrow$, $\bigcirc$, $\lozenge$, and $\square$  stand for `then', `until', `implies', `next', `eventually', and `always'. 

\begin{table}[!bt]
    \centering
    \begin{tabular}{c|c}
        Predicates & Expression \\
        \hline
        $\psi_{g_1}$ & $d^{g_1}_{\tau} < 0.05$ \\
        $\psi_{g_2}$ & $d^{g_2}_{\tau} < 0.05$ \\
        $\psi_{g_3}$ & $d^{g_3}_{\tau} < 0.05$ \\
        $\psi_{o_1}$ & $d^{o_1}_{\tau} > r_{obs}$ \\
        $\psi_{o_2}$ & $d^{o_2}_{\tau} > r_{obs}$ \\
        $\psi_{coll}$ & $d^{rob}_{\tau} > 2\,r_{rob}$
    \end{tabular}
    \caption{Predicates used in the TLTL formulation.} 
    \label{tab:predicatesTLTL}
\end{table}

The full mathematical expression of~\eqref{eq:TLTL_cost} for robot $i$ is given by
\begin{align*}
    &L^i_{\text{wp}} = -\min \Bigg( \\
    &~~
    \max_{t\in[0,T]} \bigg( 
    \min \Big( 
    \begin{aligned}[t]
        &\max_{\tilde{t}\in[0,t]} \big( \min( 0.05-d^{g_3}_{\tilde{t}} \,,\, \max_{\hat{t}\in[0,\tilde{t}]} 0.05-d^{g_2}_{\hat{t}} )\big) \,,\, 
        \\
        &\max_{\tilde{t}\in[0,t]} 0.05-d^{g_1}_{\tilde{t}} \Big) 
        \bigg), 
    \end{aligned}
    \\
    &~~
    \max_{t\in[0,T]} \bigg( 
    \min \Big( 
    \begin{aligned}[t]
        &\max (0.05-d^{g_2}_t \,,\, 0.05-d^{g_3}_t) \,,\,  \\
        &\max_{\tilde{t}\in[0,t]} 0.05-d^{g_1}_{\tilde{t}} \Big) 
        \bigg), 
    \end{aligned}
    \\
    &~~
    \max_{t\in[0,T]} \left( \min \Big( -(0.05-d^{g_3}_t) \,,\, \max_{\tilde{t}\in[0,t]} 0.05-d^{g_2}_{\tilde{t}} \Big) \right), \\
    & \min_{i\in\{1,2,3\}} \min_{t\in[0,T]} 
    \max \Big( 
    \begin{aligned}[t]
        &-(0.05-d^{g_i}_t) \,,\, \\
        &\min_{\tilde{t}\in[t+1,T]} -(0.05-d^{g_i}_t) 
        \Big)  , 
    \end{aligned}
    \\
    &~~
    \min_{t\in[0,T]} \big( \min(d^{obs_1}_t-r_{obs} \,,\, d^{obs_2}_t-r_{obs}) \big), \\
    &~~
    \min_{t\in[0,T]}  (d^{rob}_t-2r_{rob}), \\
    &~~
    \max_{t\in[0,T]} \left( \min_{\tilde{t}\in[t,T]}(0.05-d^{g_3}_{\tilde{t}}) \right) 
    \Bigg)\,.
\end{align*}

We also add a regularization term 
$\alpha_{\text{reg}} |y_t-\bar{y}|^2$
for promoting that the robots stay close to their target point.
The used REN is a deep neural network with depth $q_2 = 32$ layers, 
internal state $\xi$ of dimension $q_1 = 32$
and activation function $\sigma(\cdot) = \tanh(\cdot)$.\footnote{See Section~\ref{sec:intro_REN} for the meaning of $q_1$, $q_2$ and $\sigma$.} 
For the hyperparameters, we set $\alpha_{\text{reg}}=1\times 10^{-4}$.
We use stochastic gradient descent with Adam to minimize the loss function, setting a learning rate of $0.001$.
We optimize for 1500 epochs with a batch size of 5 trajectories.

\end{document}